\documentclass[aps, prd, twocolumn, superscriptaddress, floatfix, longbibliography, nofootinbib, bibnotes]{revtex4-2}

\pdfoutput=1
\usepackage{amsmath}
\usepackage{amssymb}
\usepackage{graphicx}
\usepackage{booktabs}
\usepackage{multirow}
\usepackage{placeins}

\usepackage[colorlinks=true,breaklinks=true]{hyperref}
\usepackage[dvipsnames]{xcolor}
\definecolor{darkred}{rgb}{0.5,0,0}
\definecolor{darkblue}{rgb}{0,0,0.5}
\definecolor{firebrick}{rgb}{0.75,0.125,0.125}
\definecolor{darkgreen}{rgb}{0,0.5,0}
\hypersetup{urlcolor=darkblue,
	    citecolor=darkgreen,
	    linkcolor=firebrick}

\usepackage{fontawesome5}
\newcommand{\repourl}{https://github.com/mbustama/nu-decay-geo-reactor}

\graphicspath{{figures/}}

\newcommand{\ie}{{\it i.e.}}

\newcommand{\eq}{Eq.}

\newcommand{\fig}{Fig.}

\providecommand{\Ref}{}
\AtBeginDocument{\renewcommand{\Ref}{Ref.}}
\newcommand{\Refs}{Refs.}
\newcommand{\Tab}{Table}

\newcommand{\equ}[1]{\eq~(\ref{equ:#1})}
\newcommand{\figu}[1]{\fig~\ref{fig:#1}}
\newcommand{\tabl}[1]{\Tab~\ref{tab:#1}}
\newcommand{\U}{{}^{238}\mathrm{U}}
\newcommand{\Th}{{}^{232}\mathrm{Th}}
\newcommand{\Pee}{P_{\bar\nu_e\to\bar\nu_e}}

\begin{document}

\title{Geoneutrinos (and reactor neutrinos) do not visibly decay\,\href{\repourl}{\faGithub}}

\author{Mauricio Bustamante}
\email{mbustamante@nbi.ku.dk}
\affiliation{Niels Bohr International Academy, Niels Bohr Institute,\\University of Copenhagen, 2100 Copenhagen, Denmark}

\date{\today}

\begin{abstract}
Neutrinos that decay on their way to a detector arrive depleted, more so at low energies, where time dilation lengthens their lifetime the least. Geoneutrinos---electron antineutrinos from the radioactive decay of uranium and thorium inside the Earth---have energies of a few MeV and travel up to the diameter of the Earth. We use them to test neutrino decay for the first time. We compute the geoneutrino signal at Borexino, KamLAND, SNO+, and JUNO for visible decay, in which the daughter neutrinos are themselves detected, and compare it against their published data. We find no evidence of decay that holds across Earth models. Combining the four experiments, we limit the decay $\nu_1 \to \nu_3$ in the inverted mass ordering at 90\%~C.L.: $\tau_1/m_1 > 3.7 \times 10^{-9}$~s~eV$^{-1}$ when the radiogenic power of the Earth is left free and $\tau_1/m_1 > 4.4 \times 10^{-9}$ to $1.6 \times 10^{-8}$~s~eV$^{-1}$ when it is taken from Earth models. With an Earth model assumed, the geoneutrinos set the limit; with the power free, the antineutrinos from nuclear reactors that the same experiments detect count as much as the geoneutrinos. These are the first terrestrial limits on the visible decay of $\nu_1$.
\end{abstract}

\maketitle

\section{Introduction}
\label{sec:intro}

Neutrinos have mass, so the heavier neutrinos could decay into the lighter ones. If neutrino masses are the only extension of the Standard Model, the decays proceed only through rare channels, with lifetimes that exceed the age of the Universe by many orders of magnitude~\cite{Petcov:1976ff, Marciano:1977wx, Zatsepin:1978iy, Pal:1981rm, Hosotani:1981mq, Shrock:1982sc, Nieves:1982bq}. Faster decays arise when neutrinos couple to a new light boson, such as the Majoron of models where lepton number is spontaneously broken~\cite{Bahcall:1972my, Pakvasa:1972gz, Chikashige:1980qk, Gelmini:1980re, Schechter:1981cv, Gelmini:1983ea, Valle:1983ua, Bahcall:1986gq, Nussinov:1987pc, Frieman:1987as, Gonzalez-Garcia:1988okv, Kim:1990km, Acker:1992eh}. Since neutrinos were first proposed to be unstable~\cite{Bahcall:1972my, Pakvasa:1972gz}, their decay has been sought in neutrinos from reactors, accelerators, the Sun, supernovae, and cosmic sources, and in cosmology.

\begin{figure*}[t!]
 \centering
 \includegraphics[width=\textwidth]{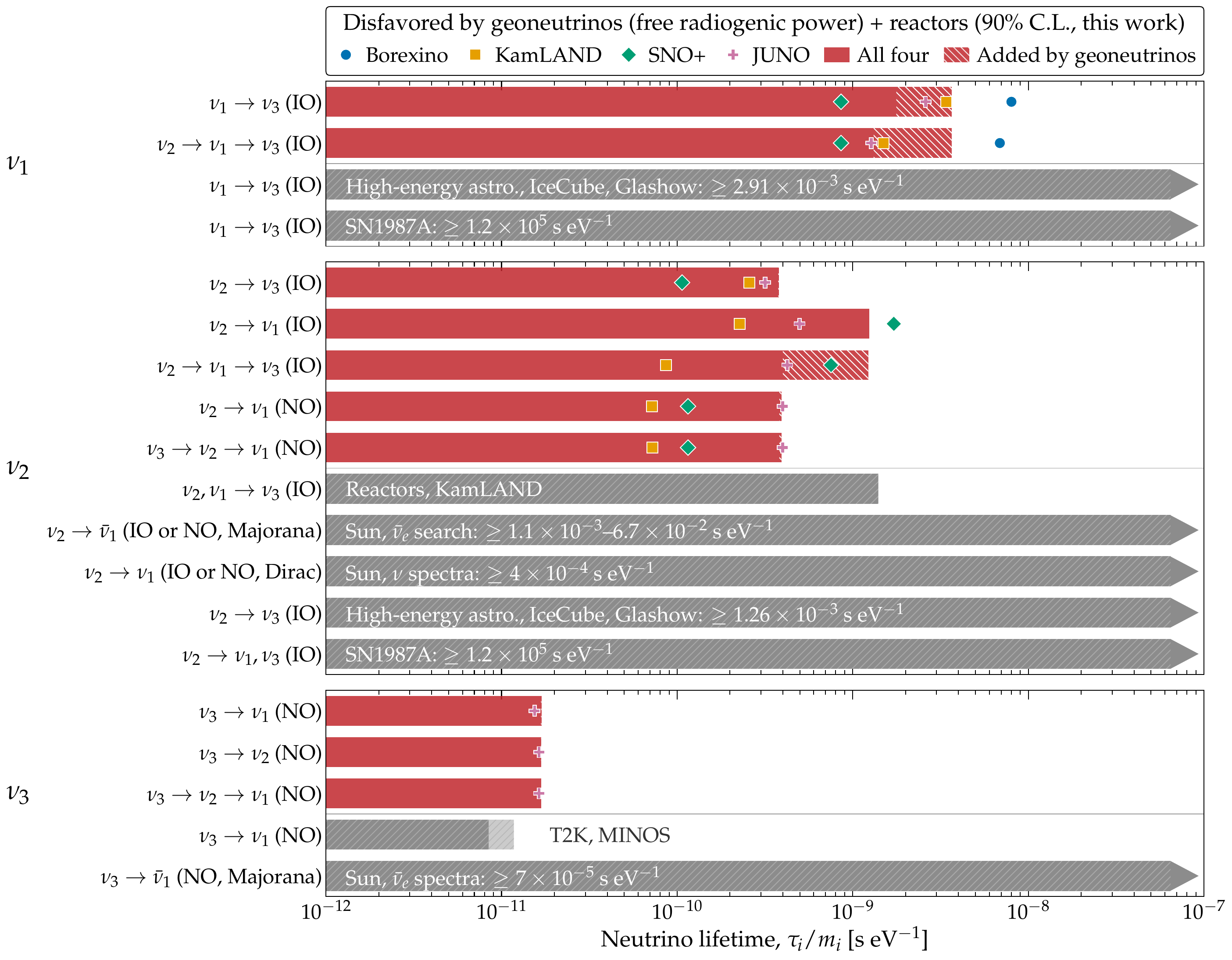}
 \caption{\textbf{Lower limits on the neutrino lifetime-to-mass ratio from visible decay.} Our limits come from the combined fit of the data of Borexino, KamLAND, SNO+, and JUNO. Each row is one decay channel, grouped by the decaying mass eigenstate. On the rows of $\nu_1$, Borexino alone sets a stronger limit than the four combined, because one lifetime must fit both Borexino's excess of geoneutrinos and KamLAND's deficit (Sec.~\ref{sec:nu1}). On the row of $\nu_2 \to \nu_1$ (IO), where decay raises the signal, SNO+ alone does. Limits from terrestrial and astrophysical neutrinos are from \Refs~\cite{Porto-Silva:2020gma, Gago:2017zzy, KamLAND:2003gfh, Picoreti:2021yct, Ivanez-Ballesteros:2023lqa, Bustamante:2020niz, deGouvea:2023jxn}. Limits from astrophysical neutrinos extend beyond the range. Our limits are at 90\% confidence level (C.L.); \tabl{published} gives the statistics of the others. See Sec.~\ref{sec:results} for details.}
 \label{fig:money}
\end{figure*}

A neutrino mass eigenstate $\nu_i$ ($i = 1, 2, 3$), with mass $m_i$ and rest-frame lifetime $\tau_i$, that travels a distance $L$ with energy $E$ survives with probability $\exp[-L m_i/(E\tau_i)]$, in natural units. Time dilation lengthens its lifetime by the Lorentz factor $E/m_i$, so experiments constrain the ratio $\tau_i/m_i$. The largest value of $\tau_i/m_i$ that an experiment can test grows with $L/E$: over a given distance, low-energy neutrinos decay more. An experiment is sensitive to $\tau_i/m_i = 3.3\times10^{-9}\,[L/(10^3~\mathrm{km})](\mathrm{MeV}/E)$~s~eV$^{-1}$. Existing limits on visible decay, in which the daughter neutrinos are detectable, span about sixteen orders of magnitude (\figu{money}). Terrestrial experiments, with reactor~\cite{Porto-Silva:2020gma} and accelerator~\cite{Gago:2017zzy} neutrinos, reach $10^{-11}$ to $10^{-9}$~s~eV$^{-1}$. Astrophysical neutrinos, from the Sun~\cite{KamLAND:2003gfh, SNO:2004eru, Picoreti:2021yct}, supernova SN1987A~\cite{Ivanez-Ballesteros:2023lqa}, and cosmic sources~\cite{Bustamante:2020niz}, reach up to $10^{5}$~s~eV$^{-1}$. Limits on invisible decay, in which the daughters escape detection, come from the same sources~\cite{Beacom:2002cb, Joshipura:2002fb, Bandyopadhyay:2002qg, Berryman:2014qha, Gomes:2014yua, Picoreti:2015ika, Choubey:2018cfz, SNO:2018pvg, Ternes:2024qui, Beccaria:2026ous, Beccaria:2026gnb, Valera:2024buc, Frieman:1987as}. Cosmology limits neutrino decay indirectly~\cite{Hannestad:2005ex, Escudero:2019gfk, Barenboim:2020vrr, Chacko:2020hmh, Chen:2022idm} (Sec.~\ref{sec:comparison}).

Geoneutrinos combine low energies with long baselines. They are the electron antineutrinos, $\bar{\nu}_e$, emitted in the beta decays of the uranium and thorium chains inside the Earth, with energies below about 3.3~MeV. They reach a detector from everywhere in the Earth, from a few kilometers away to its full diameter, 12\,742~km. With $L/E$ of a few thousand km~MeV$^{-1}$, they are sensitive to $\tau/m \sim 10^{-8}$~s~eV$^{-1}$, beyond the reach of reactor and accelerator neutrinos. 

Geoneutrinos were proposed as a probe of the Earth's interior in the 1960s~\cite{Eder:1966, Marx:1969be} and of its radiogenic heat in the decades after~\cite{Krauss:1983zn, Kobayashi:1991, Raghavan:1997gw, Rothschild:1997dd}. KamLAND detected them in 2005~\cite{Araki:2005qa}; Borexino followed in 2010~\cite{Borexino:2010dli}. Both have since measured them with growing precision~\cite{KamLAND:2011ayp, KamLAND:2013rgu, KamLAND:2022vbm, Borexino:2013tnm, Borexino:2015ucj, Borexino:2019gps}. SNO+~\cite{SNO:2025chx} and JUNO~\cite{JUNO:2025gmd} joined them recently. Their main use is geophysical: they measure how much uranium and thorium the Earth holds and, through that, how much of its heat is radiogenic~\cite{Fiorentini:2007te, Dye:2011mc, Bellini:2013wsa, Ludhova:2013hna, Sramek:2012nk, Bellini:2021sow, Leyton:2026gev}.

Yet, geoneutrinos have rarely been used to test the properties of neutrinos themselves. Previous studies computed how standard oscillations---the transitions between neutrinos of different flavors---shape their flux~\cite{Fiorentini:2003ww, Mohanty:2005fd, Wan:2016nhe, Mao:2019hbf}. One estimated the effect of sterile neutrinos and nonstandard interactions on it~\cite{Yan:2017fqm}. To our knowledge, geoneutrino data have not been used before to test physics beyond the Standard Model.

We use geoneutrinos to test neutrino decay. We compute the geoneutrino signal at Borexino, KamLAND, SNO+, and JUNO for multiple possible decay channels, including the regeneration of daughter neutrinos at lower energies. We fit the published spectrum of each experiment with a model that reproduces the experiment's own analysis and then combine the four. We find no evidence of decay that holds across Earth models, so we set limits on the neutrino lifetimes.

Our strongest limits are on the decay of $\nu_1$, possible in the inverted mass ordering (where $\nu_3$ is the lightest neutrino). Which neutrinos set these limits depends on what we assume about the Earth. The experiments detect two kinds of antineutrinos: geoneutrinos and antineutrinos from nuclear reactors. Each reactor's power fixes how many reactor antineutrinos arrive. The number of geoneutrinos that arrive is set by the uranium and thorium that the Earth holds. If we take that amount from a model of the Earth, the geoneutrinos set the limits on $\nu_1$. If we leave it free, more uranium and thorium can compensate for the geoneutrinos lost to decay, so the reactor antineutrinos count as much as the geoneutrinos. Reactor antineutrinos alone set our limits on the decay of $\nu_3$, to which geoneutrinos are insensitive. With the amount of uranium and thorium free, they also set our limits on the decay of $\nu_2$ alone, in either mass ordering.

Figure~\ref{fig:money} shows our limits beside the published limits on visible decay. We show the limits with the amount of uranium and thorium free, which rest on the fewest assumptions about the Earth. \textbf{Geoneutrinos and reactor antineutrinos set the first terrestrial limits on the visible decay of $\nu_1$,} $\tau_1/m_1 > 3.7\times10^{-9}$~s~eV$^{-1}$ at 90\%~C.L. With an Earth model (\figu{overview}), the limit is up to 4.3 times tighter. For $\nu_2$ and $\nu_3$, our limits are weaker than or comparable to terrestrial limits~\cite{Porto-Silva:2020gma, Gago:2017zzy, Ternes:2024qui, Beccaria:2026ous}. Astrophysical neutrinos reach further, over cosmological distances, but their limits rest on incomplete knowledge of the neutrino emission and of the distribution of sources~\cite{Bustamante:2016ciw}. Geoneutrinos and reactor antineutrinos test decay over terrestrial distances, from sources that are constrained independently, the Earth by geochemistry and geophysics and the reactors by their power.

Section~\ref{sec:decay} introduces neutrino decay. Section~\ref{sec:geoneutrinos} computes the flux of geoneutrinos and their decay. Section~\ref{sec:reactor_antineutrinos} describes the reactor antineutrinos and their decay. Section~\ref{sec:detection} describes the experiments and their event rates. Section~\ref{sec:stats} presents the statistical analysis. Section~\ref{sec:results} gives our limits, compared to published ones. Section~\ref{sec:summary} summarizes. The appendices detail the calculation of the geoneutrino signal (Appendix~\ref{app:signal}) and how we model each experiment (Appendix~\ref{app:reproduction}), test the limits (Appendices~\ref{app:brazil} and~\ref{app:coverage}), describe the fit parameters and their impact (Appendix~\ref{app:parameters}), and collect published limits (Appendix~\ref{app:published}) and the constraint on the solar mixing parameters that we use (Appendix~\ref{app:nufit}).

\section{Neutrino decay}
\label{sec:decay}

\subsection{Decay model}
\label{sec:decays}

We test two-body decays of a heavier mass eigenstate into a lighter one and a massless spin-zero boson, $\nu_i \to \nu_j + \phi$, such as the Majoron~\cite{Chikashige:1980qk, Gelmini:1980re, Schechter:1981cv}. The decay proceeds through the interaction $\mathcal{L} \supset \tfrac{1}{2}\, \overline{\nu_j}\, (g + i g^\prime \gamma_5)\, \nu_i\, \phi + \mathrm{h.c.}$, with a scalar coupling, $g$, and a pseudoscalar coupling, $g^\prime$; we take $g^\prime = g$~\cite{Abdullahi:2020rge}. In every channel we test, each decaying mass eigenstate has a single daughter (Sec.~\ref{sec:channels}), so the lifetime of $\nu_i$ in its rest frame is~\cite{Abdullahi:2020rge}
\begin{equation}
 \tau_i = \frac{8\pi}{g^2 m_i \left(1 - x_{ij}^{-4}\right)} \;,
 \label{equ:lifetime}
\end{equation}
where $x_{ij} = m_i/m_j$ is the ratio of the parent and daughter masses. A lower limit on $\tau_i/m_i$ is an upper limit on $g$, for given masses. Through the Hermitian-conjugate term, antineutrinos decay with the same couplings and, as CPT symmetry requires, the same lifetime as neutrinos. Our limits hold for both.

Since geoneutrinos and reactor antineutrinos are $\bar\nu_e$, the parent is an antineutrino mass eigenstate, $\bar\nu_i$. Detectors observe both through inverse beta decay (IBD), $\bar\nu_e + p \to e^+ + n$, which only $\bar\nu_e$ trigger. A decay that conserves helicity yields an antineutrino daughter, $\bar\nu_j$, which IBD can detect in proportion to its electron-flavor content, $|U_{ej}|^2$ ($U$ is the lepton mixing matrix; Sec.~\ref{sec:survival}). A decay that flips helicity yields a daughter that IBD cannot detect. Our calculation includes both kinds of decay at the same time, each at its own rate: together they set the lifetime, \equ{lifetime}, and how fast parents disappear, while only the helicity-conserving decays add daughters. With $g^\prime = g$, helicity-conserving decays make up a fraction $(x_{ij}^2 + 4 \ln x_{ij} - x_{ij}^{-2}) / [2 (x_{ij}^2 - x_{ij}^{-2})]$ of all decays~\cite{Abdullahi:2020rge}: nearly all when parent and daughter have nearly the same mass, as in $\bar\nu_2 \to \bar\nu_1$ in the inverted ordering, and about half when the daughter is much lighter. Hereafter, a decay is \textit{visible} when its daughters are included in the calculation, even if few of them are detectable.

The daughter keeps between a fraction $1/x_{ij}^2$ of its parent's energy and all of it. Within that band, the number of helicity-conserving daughters per unit energy, per decay of a parent of energy $E_\nu^\prime$, is~\cite{Abdullahi:2020rge}
\begin{equation}
 W_{ij}(E_\nu \vert E_\nu^\prime)
 =
 \frac{x_{ij}^2}{x_{ij}^4 - 1}\, \frac{1}{E_\nu^\prime}
 \left( \frac{E_\nu^\prime}{E_\nu} + x_{ij}^2\, \frac{E_\nu}{E_\nu^\prime} \right) \;,
 \label{equ:daughter}
\end{equation}
for $g^\prime = g$. Integrated over $E_\nu$ across that band, $W_{ij}$ gives the fraction of decays that conserve helicity. The daughters of a much heavier parent spread over the whole range; those of a parent of nearly equal mass keep nearly all of its energy. Therefore, visible decay depends on the absolute neutrino masses~\cite{deGouvea:2022cmo}. The measured mass-squared differences fix them up to the mass of the lightest neutrino, which we leave free in our fits (Sec.~\ref{sec:stats}).

\subsection{Survival and regeneration}
\label{sec:survival}

A $\bar\nu_e$ is produced as a coherent superposition of the mass eigenstates $\nu_i$, with amplitudes $U_{ei}$, elements of the lepton mixing matrix that connects the flavor and mass bases. In its standard parametrization~\cite{ParticleDataGroup:2024cfk}, $|U_{e1}|^2 = \cos^2\theta_{12}\cos^2\theta_{13}$, $|U_{e2}|^2 = \sin^2\theta_{12}\cos^2\theta_{13}$, and $|U_{e3}|^2 = \sin^2\theta_{13}$, where $\theta_{12}$ and $\theta_{13}$ are two of its mixing angles; the third angle ($\theta_{23}$) and the CP-violation phase ($\delta_{\rm CP}$) do not affect the survival of a $\bar\nu_e$. During propagation, the eigenstates acquire different phases, depending on their masses, so their interference makes the probability of detecting a $\bar\nu_e$ oscillate with distance.

The oscillation length associated with the mass-squared splitting $\Delta m^2_{ij} = m_i^2 - m_j^2$ is $4\pi E_\nu/\vert\Delta m^2_{ij}\vert$. For illustration, we take $\theta_{12}$, $\theta_{13}$, $\Delta m^2_{21}$, and $\Delta m^2_{31}$ from the global fit NuFIT~6.1, with the atmospheric-neutrino data of Super-Kamiokande included~\cite{Esteban:2024eli, NuFIT61}, at their best fit for each mass ordering. Later, in our fits to data, they vary under NuFIT's constraints (Sec.~\ref{sec:stats}). For these values, the oscillation length is 60--110~km for $\Delta m^2_{21}$ and 2--3~km for $\vert\Delta m^2_{31}\vert$ for the geoneutrinos that IBD detects, with energies from 1.8 to 3.3~MeV.

Here and in Sec.~\ref{sec:channels}, we write the survival probability in vacuum, where it has a closed form that allows us to illustrate what decay does. Inside the Earth, the electrons of the rock change the survival probability~\cite{Wolfenstein:1977ue, Mikheyev:1985zog}, by about 1\% along paths through the mantle~\cite{Borexino:2019gps}. Our analysis includes this change, which moves our limits by at most 2.3\% (Appendix~\ref{sec:matter}). The probability amplitude for an antineutrino emitted as a $\bar\nu_e$ to be detected as one, after a distance $L$ in vacuum, is~\cite{Abdullahi:2020rge}
\begin{equation}
 \mathcal{A}(L, E_\nu)
 =
 \sum_{i=1}^{3} |U_{ei}|^2
 \exp\left(-i\,\frac{m_i^2 L}{2E_\nu}\right)
 \exp\left(-\frac{L m_i}{2E_\nu \tau_i}\right) \;.
 \label{equ:amplitude}
\end{equation}
In each term in the sum, $|U_{ei}|^2$ is the product of the amplitudes of $\nu_i$ at production and at detection; the first exponential is the phase that $\nu_i$ acquires; the second is the damping of its amplitude by decay, whose square is the probability that $\nu_i$ survives (Sec.~\ref{sec:intro}). The survival probability, $\Pee = |\mathcal{A}|^2$, is then~\cite{Abdullahi:2020rge, Porto-Silva:2020gma}
\begin{multline}
 \Pee(L, E_\nu)
 =
 \sum_{i,j=1}^{3} |U_{ei}|^2 |U_{ej}|^2
 \cos\left(\frac{\Delta m^2_{ij} L}{2E_\nu}\right)
 \\
 \times
 \exp\left[-\frac{L}{2E_\nu}\left(\frac{m_i}{\tau_i} + \frac{m_j}{\tau_j}\right)\right] \;.
 \label{equ:psurv}
\end{multline}
The terms with $i = j$ count each eigenstate on its own, weighted by its electron-flavor content, $|U_{ei}|^2$, at production and again at detection. The terms with $i \neq j$ are the interference between the amplitudes of $\nu_i$ and $\nu_j$, damped by the decay of both. Averaged over many oscillation lengths, the interference terms vanish. Without decay, the survival probability is then $\sum_i |U_{ei}|^4 = 0.549$.

\begin{figure}[t!]
 \centering
 \includegraphics[width=\columnwidth]{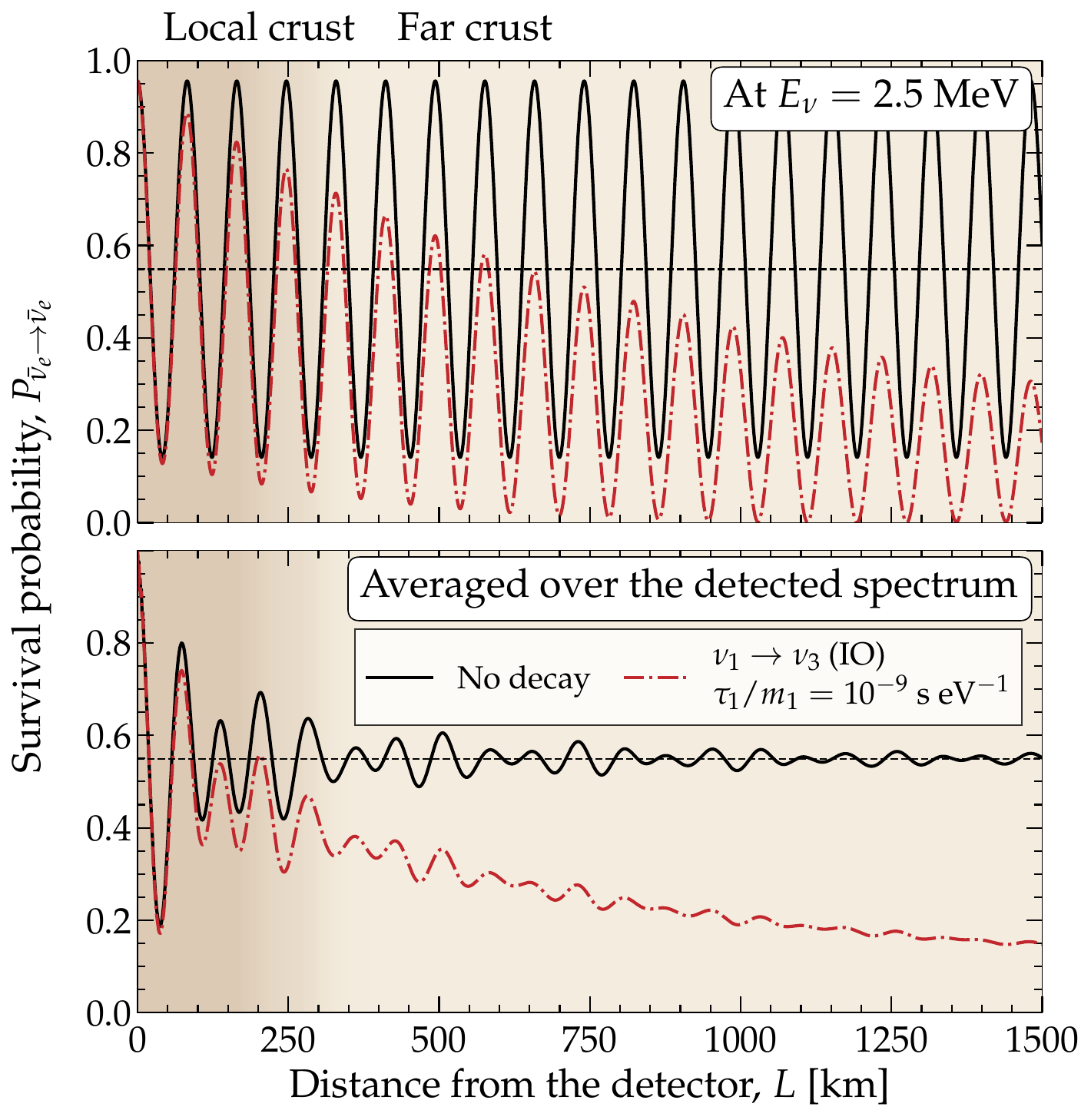}
 \caption{\textbf{Survival probability of geoneutrinos against distance.} Equation~(\ref{equ:psurv}), in vacuum, at the best-fit mixing parameters~\cite{Esteban:2024eli, NuFIT61}. \textit{Top:} The terms of $\Delta m^2_{31}$, whose oscillation length is 2--3~km, are averaged. \textit{Bottom:} Uranium and thorium at the bulk-Earth ratio Th/U = 3.9 (Sec.~\ref{sec:prediction}). The darker shading is the local crust at Borexino: all of the crust out to 172~km, and none beyond 404~km. The dashed line is the fully averaged value, 0.549. Matter, which our analysis includes, changes the bottom curves by at most 2.3\% and shifts the top ones in phase, by up to 0.13 near 1500~km. See Secs.~\ref{sec:survival} and \ref{sec:flux} for details.}
 \label{fig:survival}
\end{figure}

Figure~\ref{fig:survival} shows \equ{psurv} against distance for geoneutrinos. At a single energy (top panel), and with the fast oscillations of $\Delta m^2_{31}$ averaged, the probability without decay oscillates with the oscillation length of $\Delta m^2_{21}$, 82~km at 2.5~MeV. Decay of $\nu_1$ into $\nu_3$, for $\tau_1/m_1 = 10^{-9}$~s~eV$^{-1}$, depletes $\nu_1$ with distance. Both the probability and its oscillations shrink, since the oscillations come from the interference between $\nu_1$ and $\nu_2$. A spread of energies averages the oscillations out at large distances (bottom panel). We return to this in Sec.~\ref{sec:flux}.

Decay makes the flavor transitions non-unitary, although the mixing matrix stays unitary: the probabilities of arriving as $\bar\nu_e$, $\bar\nu_\mu$, or $\bar\nu_\tau$ add up to $\sum_i |U_{ei}|^2 \exp[-L m_i/(E_\nu \tau_i)]$, less than one. The difference is the probability that the antineutrino decayed on the way; visible decay returns part of it at lower energies. For a decay $\nu_i \to \nu_j$ and a source at distance $L$ whose antineutrinos would arrive with spectrum $F(E_\nu)$ without oscillations or decay, the arrival spectrum of $\bar\nu_e$ is~\cite{Lindner:2001fx, Abdullahi:2020rge}
\begin{multline}
 \mathcal{T}_L[F](E_\nu)
 =
 \Pee(L, E_\nu)\, F(E_\nu)
 \\
 + |U_{ei}|^2 |U_{ej}|^2
 \int_{E_\nu}^{x_{ij}^2 E_\nu} dE_\nu^\prime\,
 F(E_\nu^\prime)\, W_{ij}(E_\nu \vert E_\nu^\prime)
 \\
 \times
 \left[1 - \exp\left(-\frac{L}{E_\nu^\prime} \frac{m_i}{\tau_i}\right)\right] \;.
 \label{equ:arrival}
\end{multline}
The first term counts the survivors, through \equ{psurv}. The second counts the daughters of parents of higher energy, $E_\nu^\prime$, that decayed on the way, with probability one minus the survival probability of $\nu_i$, weighted by the electron-flavor content of the parent at production and of the daughter at detection. In vacuum, the daughter, a single mass eigenstate, does not oscillate. In matter, it does, since there the eigenstates of propagation differ from the mass eigenstates; our analysis includes this (Sec.~\ref{sec:matter}).

\subsection{Decay channels}
\label{sec:channels}

\begin{table}[t!]
 \caption{\label{tab:channels}\textbf{Decay channels.} In each, only the decays shown occur; every other eigenstate is stable. A chain is two decays in succession; we constrain the lifetimes of both of its parents. The last two columns assume that every parent decays, and give shares of the $\bar\nu_e$ that would arrive without decay, averaged over oscillations in vacuum: the share lost, $|U_{ei}|^4/\sum_k |U_{ek}|^4$, and the most that the daughters regain, $|U_{ei}|^2|U_{ej}|^2/\sum_k |U_{ek}|^4$. The daughters regain that much only if each keeps all of its parent's energy and conserves helicity. They come closest in $\nu_2 \to \nu_1$ in the inverted ordering, where each keeps at least 97\% of it (Sec.~\ref{sec:signal}). In a chain, $i$ runs over both parents and $j$ is the last eigenstate. Values at the best fit of NuFIT~6.1~\cite{Esteban:2024eli, NuFIT61}, with $|U_{ei}|^2$ as marked in \figu{flavor_content} (the same in normal ordering, to within 0.001).}
 \centering
 \begin{ruledtabular}
 \begin{tabular}{llcc}
  & & \multicolumn{2}{c}{Share of $\bar\nu_e$ if every parent decays} \\
  \cline{3-4}
  Ordering & Channel & Lost & Regained, at most \\
  \hline
  Inverted & $\nu_1 \to \nu_3$ & 83\% & 2.8\% \\
  ($\nu_3$ stable) & $\nu_2 \to \nu_3$ & 17\% & 1.2\% \\
   & $\nu_2 \to \nu_1$ & 17\% & 37\% \\
   & $\nu_2 \to \nu_1 \to \nu_3$ & 99.9\% & 4.0\% \\
  \hline
  Normal & $\nu_2 \to \nu_1$ & 17\% & 37\% \\
  ($\nu_1$ stable) & $\nu_3 \to \nu_1$ & 0.1\% & 2.8\% \\
   & $\nu_3 \to \nu_2$ & 0.1\% & 1.2\% \\
   & $\nu_3 \to \nu_2 \to \nu_1$ & 17\% & 40\% \\
 \end{tabular}
 \end{ruledtabular}
\end{table}

\tabl{channels} shows the eight decay channels we consider. In each, only the decays shown occur, each through its own coupling; every other eigenstate is stable, as limits are commonly quoted~\cite{Gago:2017zzy, Porto-Silva:2020gma}. In the inverted mass ordering (IO), $\nu_3$ is the lightest. In the normal ordering (NO), $\nu_1$ is. Therefore, in $\nu_2 \to \nu_1$, the daughter is stable in both orderings, even though $\nu_1$ is not the lightest in the inverted one. Two channels are chains, in which the daughter of the first decay decays in turn: in $\nu_2 \to \nu_1 \to \nu_3$, $\nu_1$ decays too, both as a daughter of $\nu_2$ and as a component of the produced $\bar\nu_e$. Their calculation integrates over the point along the path where the first decay occurs~\cite{Abdullahi:2020rge}. Later, we show limits on the neutrino lifetime for each of these channels.

\begin{figure}[t!]
 \centering
 \includegraphics[width=\columnwidth]{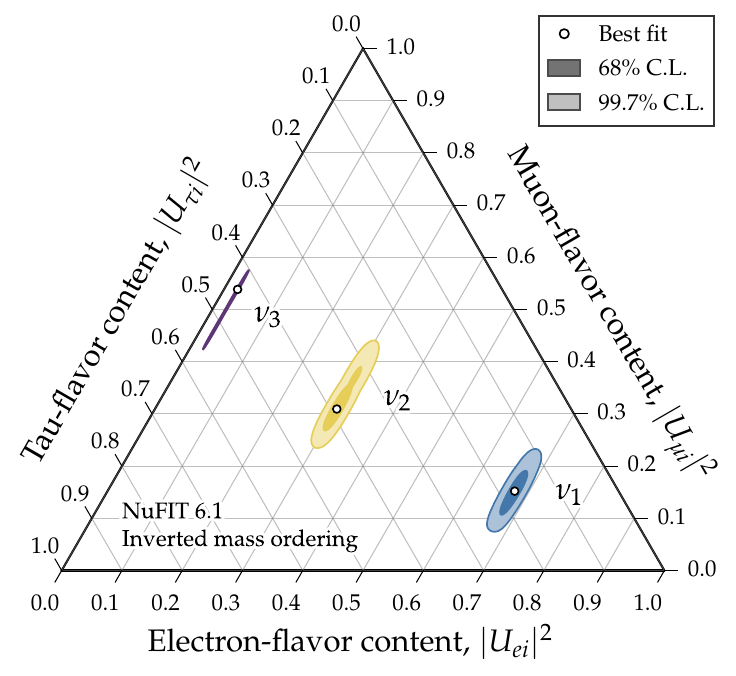}
 \caption{\textbf{Flavor content of the neutrino mass eigenstates.} Regions of the flavor content, $|U_{\alpha i}|^2$ with $\alpha = e, \mu, \tau$, of $\nu_1$, $\nu_2$, and $\nu_3$ allowed by the global fit NuFIT~6.1, with the atmospheric-neutrino data of Super-Kamiokande included, in the inverted mass ordering~\cite{Esteban:2024eli, NuFIT61}. They combine NuFIT's $\chi^2$ tables over $(\theta_{12}, \theta_{13})$ and over $(\theta_{23}, \delta_{\rm CP})$, the third mixing angle and the CP-violation phase (pairing the four parameters differently moves the boundaries by less than 0.01 in any flavor fraction). IBD detects each eigenstate in proportion to its electron-flavor content, $|U_{e i}|^2$. See Sec.~\ref{sec:channels} for details.}
 \label{fig:flavor_content}
\end{figure}

Figure~\ref{fig:flavor_content} shows the flavor content of $\nu_1$, $\nu_2$, and $\nu_3$. Their electron-flavor content, $|U_{e i}|^2$, the only part that IBD sees, is 0.68 for $\nu_1$, 0.30 for $\nu_2$, and 0.02 for $\nu_3$, each known to within 0.025 at 99.7\%~C.L. Their muon- and tau-flavor content, which depend also on $\theta_{13}$ and $\delta_{\rm CP}$, are far less certain, but IBD does not see them.

Without decay, the averaged survival probability, $\sum_i |U_{ei}|^4 = 0.549$ (Sec.~\ref{sec:survival}), splits among the eigenstates in proportion to $|U_{ei}|^4$: about 83\% to $\nu_1$, 17\% to $\nu_2$, and almost none to $\nu_3$. Therefore, decay is most visible when it removes $\nu_1$, which is possible only in the inverted ordering. There, the daughter, $\nu_3$, carries little electron flavor, so the decay $\nu_1 \to \nu_3$ removes antineutrinos nearly as if it were an invisible decay. In the normal ordering, $\nu_1$ is stable; decay can remove at most the 17\% carried by $\nu_2$ and $\nu_3$. The decay $\nu_2 \to \nu_1$ can add events, since the daughter carries more electron flavor than its parent (Sec.~\ref{sec:signal}). Figure~\ref{fig:flavor_content} shows that the uncertainty of the mixing parameters barely changes any of this.

\section{Geoneutrinos}
\label{sec:geoneutrinos}

\subsection{Origin of geoneutrinos}
\label{sec:origin}

\begin{figure}[t!]
 \centering
 \includegraphics[width=\columnwidth]{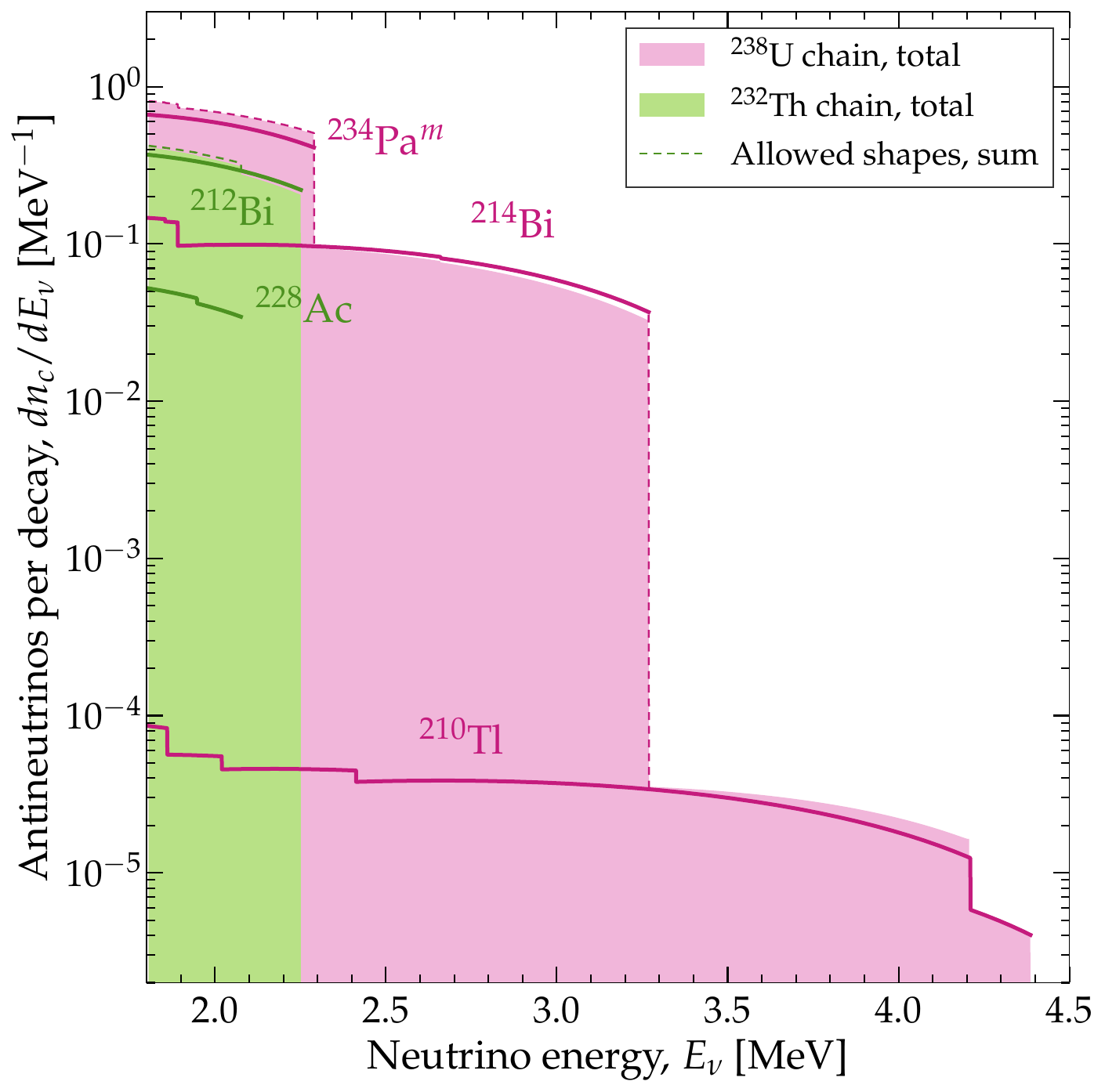}
 \caption{\textbf{Antineutrino spectra of the uranium and thorium chains above the IBD threshold.} Each line is the spectrum of the antineutrinos that one nuclide emits, per decay of the parent of its chain, $\U$ or $\Th$, computed with the allowed shape for every branch (Appendix~\ref{app:spectra}); a step within a line marks the endpoint of one of its weaker branches. The total of each chain is the spectrum $dn_c/dE_\nu$ of \equ{flux} that our fits use, from the tables of \Ref~\cite{Li:2024mqz}, which give only these totals. Near the endpoints, the totals lie below the sum of the allowed shapes, because the first-forbidden branches emit fewer antineutrinos there.}
 \label{fig:chains}
\end{figure}

Geoneutrinos are the $\bar\nu_e$ emitted in the beta decays of radioactive nuclides inside the Earth~\cite{Bellini:2013wsa, Smirnov:2019pnj, Bellini:2021sow, Sramek:2021}. Those of the $\U$ and $\Th$ decay chains are detectable. On its way to stable lead, the uranium chain emits six $\bar\nu_e$, in the beta decays of $^{234}$Th, $^{234}$Pa$^m$, $^{214}$Pb, $^{214}$Bi or, rarely, $^{210}$Tl, $^{210}$Pb, and $^{210}$Bi; the thorium chain emits four, in those of $^{228}$Ra, $^{228}$Ac, $^{212}$Pb, and $^{212}$Bi or $^{208}$Tl. Every other nuclide in the chains has a half-life shorter than 250\,000 years, against 4.5 and 14 billion years for $\U$ and $\Th$. Only five nuclides have beta transitions with endpoints above the 1.806-MeV threshold of IBD (see \figu{chains}): $^{234}$Pa$^m$, $^{214}$Bi, and $^{210}$Tl in the uranium chain, and $^{212}$Bi and $^{228}$Ac in the thorium chain. Nearly all of the detectable flux lies below 3.27~MeV, the endpoint of the main $^{214}$Bi branch. The other emitters, among them the $^{235}$U chain and $^{40}$K, emit only below the threshold: the highest of their endpoints, that of $^{207}$Tl in the $^{235}$U chain, is 1.42~MeV.

The spectrum of each chain is the sum of the spectra of its beta decays. A beta decay can leave the daughter nucleus in its ground state or in one of its excited states; each such transition, or branch, has its own endpoint, the most energy that its antineutrino can carry, and its own share of the decays. The branches that produce most of the detectable antineutrinos, those of $^{234}$Pa$^m$, $^{214}$Bi, and $^{212}$Bi, are first-forbidden: in them, the electron and the antineutrino carry away orbital angular momentum, which changes the shape of the spectrum from the allowed one that phase space alone sets. Therefore, we take the spectrum of each chain from the tables of \Ref~\cite{Li:2024mqz}, which include the shapes of these branches and the Coulomb attraction between the electron and the daughter nucleus (Appendix~\ref{app:spectra}).

Figure~\ref{fig:chains} shows the spectra of the two chains above the IBD threshold, nuclide by nuclide. Most of each chain's detectable antineutrinos come from a single branch that ends near 2.3~MeV: that of $^{234}$Pa$^m$ in the uranium chain and of $^{212}$Bi in the thorium chain. Above 2.3~MeV, only the uranium chain emits, almost entirely through $^{214}$Bi.

\subsection{Earth model}
\label{sec:prediction}

The geoneutrino flux at a detector depends on how much uranium and thorium the Earth holds, and where. Our Earth model is spherically symmetric, with the densities of the Preliminary Reference Earth Model (PREM)~\cite{Dziewonski:1981xy}. It has four reservoirs of uranium and thorium: the crust in a region of $6^\circ \times 4^\circ$ around the detector (the \textit{local crust}), the rest of the crust (the \textit{far crust}), the rigid top of the mantle beneath the crust (the \textit{lithospheric mantle}), and the mantle below it. Uranium and thorium are spread uniformly within each reservoir, in amounts set by a model of the bulk silicate Earth (see below); none are in the core, since both elements bind to silicates rather than to iron~\cite{Bellini:2013wsa}. Each detector sits underground, at the depth listed in \tabl{experiments}, so geoneutrinos reach it from the rock above it as well as from below. Appendix~\ref{app:earth} gives the depths of the reservoirs.

\begin{figure}[t!]
 \centering
 \includegraphics[width=\columnwidth]{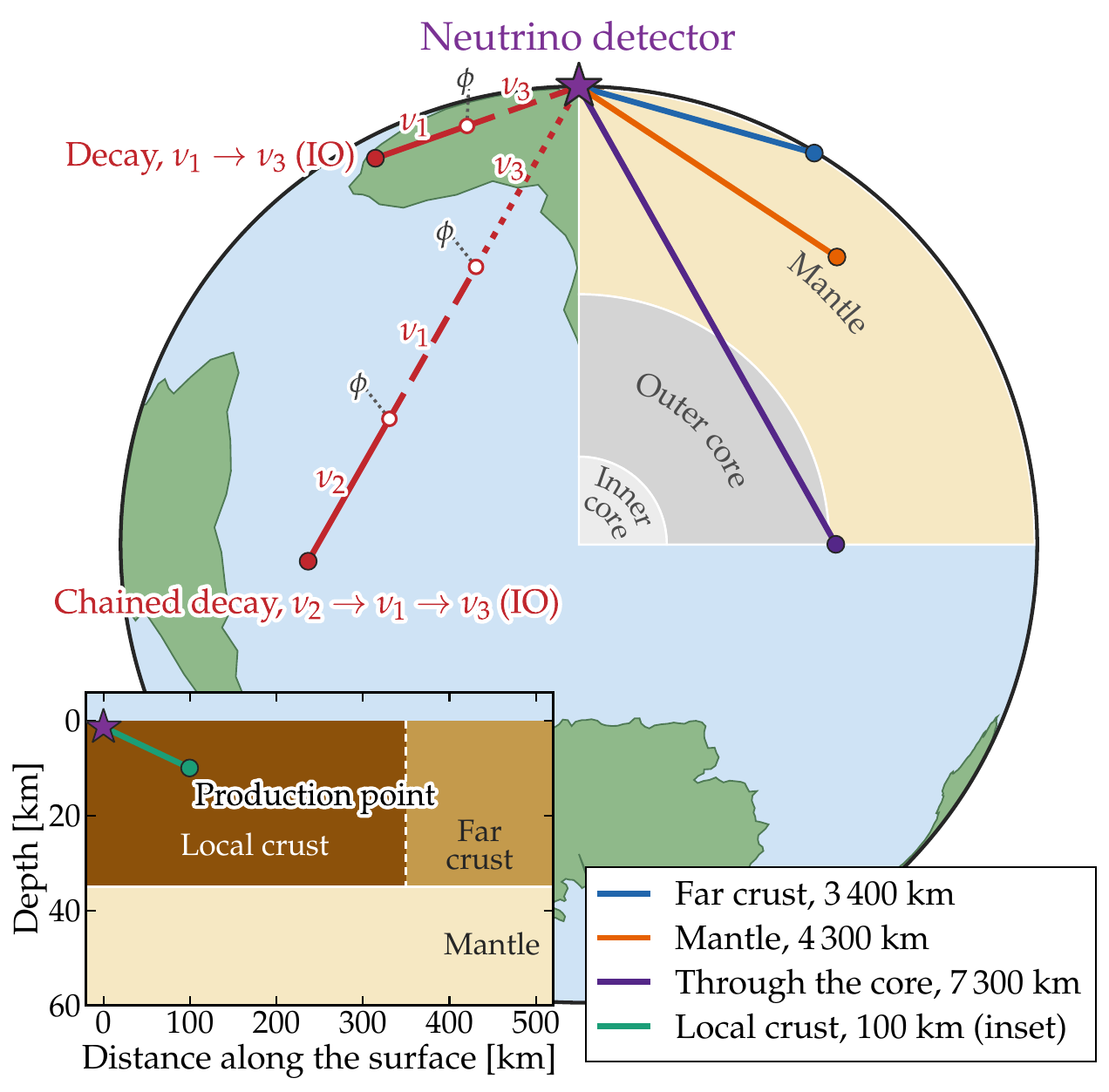}
 \caption{\textbf{Where geoneutrinos come from.} A cross section of the Earth, with the detector at Gran Sasso, the site of Borexino. Geoneutrinos are produced in the crust and the mantle. The chords on the right, and the one in the inset, run from typical production points to the detector; the inset magnifies the crust near the detector. The chords on the left illustrate decay along the path. About three quarters of the signal comes from beyond the local crust, across distances over which decay can act.}
 \label{fig:schematic}
\end{figure}

Figure~\ref{fig:schematic} shows where geoneutrinos come from. They are produced throughout the crust and the mantle. The local crust contributes much of the signal. The far crust and the mantle send geoneutrinos across thousands of kilometers. Along those paths, a heavier neutrino can decay into a lighter one before it arrives at the detector.

\subsubsection{The bulk silicate Earth}

\tabl{bse} lists the Earth models we use. Models of the bulk silicate Earth (BSE), \ie, the Earth without its metallic core, fix the total amounts of uranium and thorium. They fall into three classes~\cite{Sramek:2012nk, Bellini:2013wsa}, which differ in their radiogenic power, the heat released by the uranium and thorium chains. Potassium also heats the Earth, but its antineutrinos lie below the threshold of inverse beta decay. Therefore, the radiogenic powers in \tabl{bse}, from Ref.~\cite{Borexino:2019gps}, count only uranium and thorium. Cosmochemical models build the Earth from enstatite chondrites (non-metallic meteorites), the chondrites closest to the Earth's mantle in isotopic composition~\cite{Bellini:2013wsa}. Geochemical models are based on the composition of chondrites and terrestrial rocks~\cite{McDonough:1995}. Geodynamical models are based on the heat needed to drive mantle convection. In all three, the crust produces 6.0~TW~\cite{Huang:2013ozo, Borexino:2019gps} and the lithospheric mantle 0.7~TW~\cite{Borexino:2019gps}; the mantle produces the rest.

Besides fixing the power to that of each class, in some analyses we leave it free to float between 9.4~TW, the power of the coldest BSE model class, and 39.6~TW, the power of uranium and thorium if all 47~TW of the Earth's surface heat flow were radiogenic~\cite{Bellini:2013wsa, Borexino:2019gps}. A higher power adds geoneutrinos and can make up for those that decay removes, so this treatment yields weaker, more conservative limits. We call it \textit{free radiogenic power}.

\begin{table}[t]
 \caption{\label{tab:bse}\textbf{Earth models.} Radiogenic power of uranium and thorium in the bulk silicate Earth for each class of model~\cite{Borexino:2019gps}, and the range over which we leave it free, with the Th/U of the mantle and its uranium and thorium per gram of rock. The crust and the lithospheric mantle are the same in every case. See Sec.~\ref{sec:prediction} for details.}
 \begin{ruledtabular}
 \begin{tabular}{lcccc}
  & & \multicolumn{3}{c}{Mantle} \\
  \cline{3-5}
  Earth model & Power [TW] & Th/U & U [ng/g] & Th [ng/g] \\
  \hline
  Cosmochemical & $9.4 \pm 1.2$ & 2.8 & 4.0 & 11 \\
  Geochemical & $16.5 \pm 3.0$ & 3.6 & 13.1 & 47 \\
  Geodynamical & $28.8 \pm 3.1$ & 3.7 & 28.7 & 108 \\
  Free radio.~power & 9.4--39.6 & 2.8--3.8 & 4.0--42.5 & 11--161 \\
 \end{tabular}
 \end{ruledtabular}
\end{table}

The three classes span a factor of three in power, and so in how much uranium and thorium the BSE holds, the largest uncertainty in the predicted geoneutrino flux. The ratio of thorium to uranium is Th/U$= 3.9$ for the BSE as a whole, so each power fixes the abundances of both elements (Appendix~\ref{app:earth}). The crust and the lithospheric mantle are the same in all three classes, and have Th/U$>3.9$. The mantle holds the rest of the uranium and thorium, so its Th/U is lower and depends on the power (\tabl{bse}). Geoneutrinos born in the mantle travel the farthest, so decay removes the largest share of them (\figu{distance}); the power sets how many the mantle emits and, through its Th/U, the shape of their spectrum.

\subsubsection{Calibrating the lithosphere}
\label{sec:calibration}

The crust produces a large share of the geoneutrinos that each detector sees. Their number varies from site to site with the thickness and composition of the crust around the detector. As \Ref~\cite{Bellini:2021sow} does, at each site we scale the uranium and thorium of the local crust, the far crust, and the lithospheric mantle so that each reproduces the signal that published geological models give for it (\tabl{experiments}; Appendix~\ref{app:calibration}). These scale factors are the $f_{rc}$ of \equ{rate} later. The published geology then sets how much uranium and thorium the lithosphere around each detector holds; our Earth model sets how its geoneutrinos spread over distance and energy, and so what fraction of them decay removes. We do not scale the mantle, whose signal is what each BSE model predicts.

\subsection{Geoneutrino flux at a detector}
\label{sec:flux}

Above the IBD threshold, where only the $\U$ and $\Th$ chains emit antineutrinos (Sec.~\ref{sec:origin}), the flux of geoneutrinos that reach a detector as $\bar\nu_e$, per unit neutrino energy $E_\nu$, is~\cite{Mantovani:2003yd, Smirnov:2019pnj, Sramek:2021, Bellini:2021sow}
\begin{equation}
 \frac{d\Phi}{dE_\nu}
 =
 \sum_{r,c} A_{rc}
 \int_r dV \, \frac{\rho}{4\pi L^2} \, \mathcal{T}_{L,\psi}\!\left[\frac{dn_c}{dE_\nu}\right]\!(E_\nu) \;,
 \label{equ:flux}
\end{equation}
where $r$ runs over the four reservoirs (Sec.~\ref{sec:prediction}) and $c$ over the $\U$ and $\Th$ decay chains (Sec.~\ref{sec:origin} and \figu{chains}). The integral runs over the volume of reservoir $r$, with $\rho$ the density. We locate each volume element by its distance from the detector, $L$, and by the angle between the downward vertical and the direction from the detector to it, $\psi$; in these coordinates, $dV = 2\pi L^2 \sin\psi\, dL\, d\psi$. 

The operator $\mathcal{T}_{L,\psi}$ turns the spectrum that would arrive from the volume element without oscillations or decay into that of the $\bar\nu_e$ that do arrive, the survivors plus the daughters of decay. In vacuum, it is the operator $\mathcal{T}_L$ of \equ{arrival}, which depends only on the distance. In the matter of the Earth, it depends on the whole path to the detector, which $L$ and $\psi$ fix in our spherically symmetric Earth model. Appendix~\ref{sec:matter} defines it in full. 

The specific activity, $A_{rc}$, is the number of decays of the parent of chain $c$, $\U$ or $\Th$, per unit time and kilogram of rock in reservoir $r$. It follows from the parent's abundance in the reservoir, $a_{rc}$, which the BSE model sets (Sec.~\ref{sec:prediction}): $A_{rc} = \lambda_c a_{rc} N_A / M_c$, with $\lambda_c$ the decay constant of the parent, $M_c$ its molar mass, and $N_A$ the Avogadro constant. Every other nuclide in the chains is far shorter-lived than $\U$ and $\Th$, so it decays as fast as it is produced. Therefore, the spectrum, $dn_c/dE_\nu$ (\figu{chains}), counts the antineutrinos per unit energy that the chain emits per decay of its parent (Sec.~\ref{sec:origin}).

\begin{figure}[t!]
 \centering
 \includegraphics[width=\columnwidth]{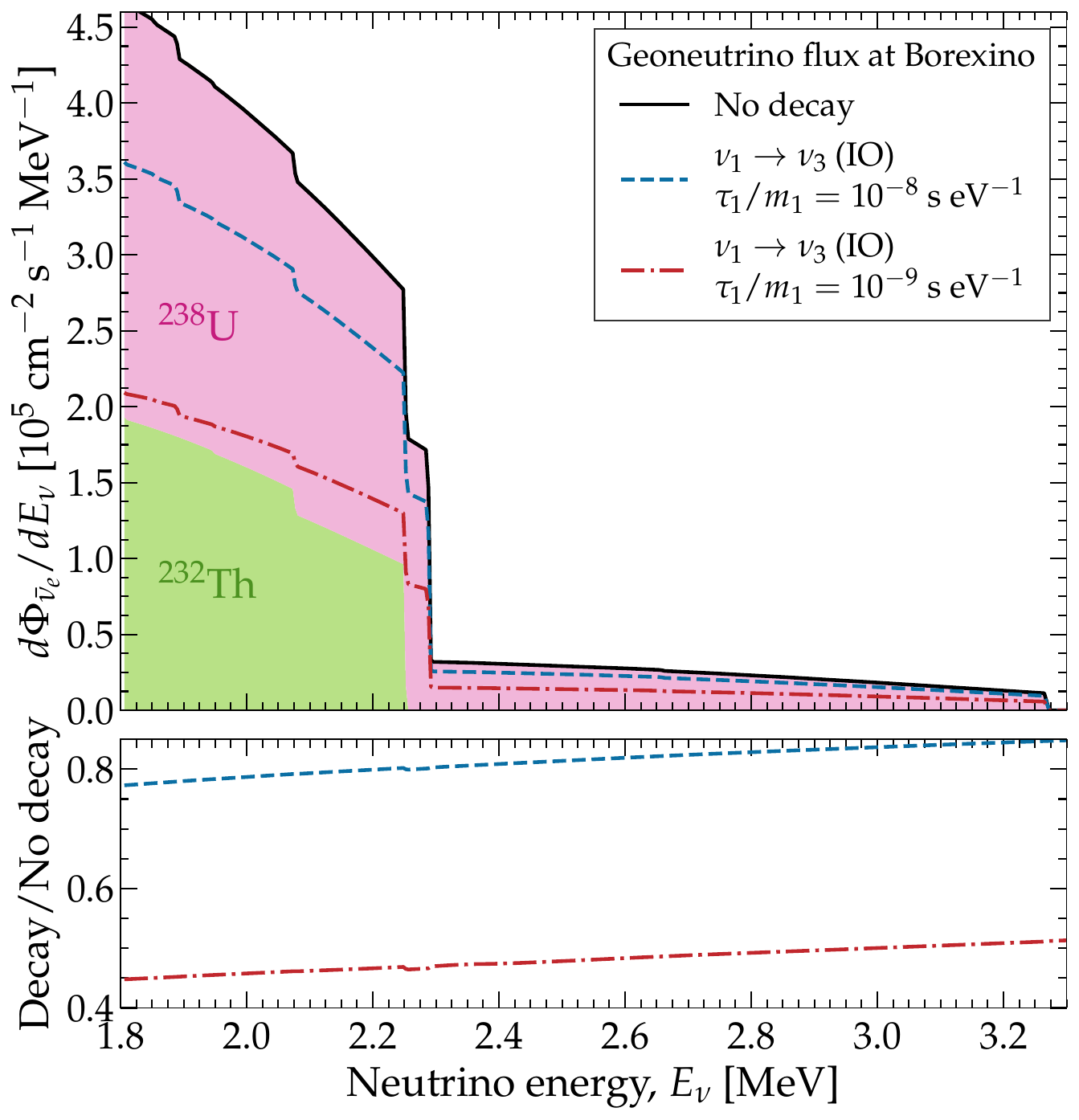}
 \caption{\textbf{Geoneutrino flux at Borexino, with and without decay.} \textit{Top:} Flux above the IBD threshold, for the geochemical Earth model, without decay and with $\nu_1 \to \nu_3$ in the inverted mass ordering, for two lifetimes. The shaded areas show the contributions of the uranium and thorium chains without decay; the sharp drop near 2.3~MeV is where the main branches of $^{212}$Bi and $^{234}$Pa$^m$ end. \textit{Bottom:} Ratio of each flux with decay to the flux without decay. Decay depletes the flux more at low energies, where time dilation is weakest. See Secs.~\ref{sec:flux} and~\ref{sec:signal} for details.}
 \label{fig:spectrum}
\end{figure}

Figure~\ref{fig:spectrum} shows the flux of \equ{flux} at Borexino, for the geochemical Earth. Without decay, it is the sum of the contributions of the uranium and thorium chains, each shaped nearly like its spectrum in \figu{chains}. The other lines show the flux with decay, which we discuss in Sec.~\ref{sec:signal}. 

\begin{figure}[t!]
 \centering
 \includegraphics[width=\columnwidth]{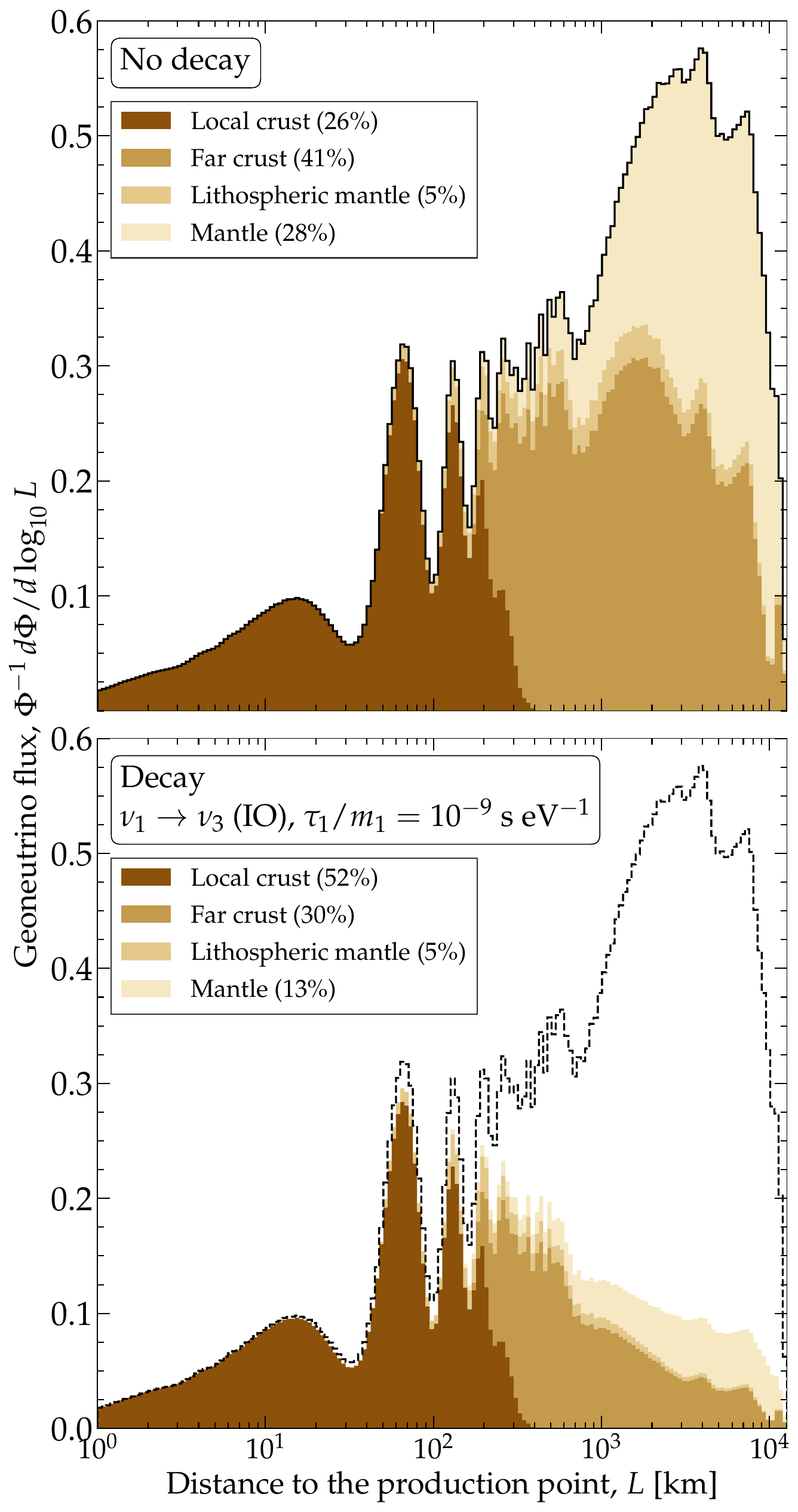}
 \caption{\textbf{Geoneutrino flux at Borexino by distance traveled.} Each panel shows $\Phi^{-1}\, d\Phi/d\log_{10} L$, the distribution of the geoneutrino flux in the distance $L$ from the production point to the detector, for the geochemical Earth model with the lithosphere calibrated (Sec.~\ref{sec:prediction}). Here, $\Phi$ is the flux of \equ{flux} integrated over energy, without decay; $d\Phi$ is the part of the flux from production points between $L$ and $L + dL$. In each panel, the percentages give each reservoir's share of the flux. \textit{Top:} Without decay. \textit{Bottom:} With $\nu_1 \to \nu_3$ in the inverted mass ordering; the dashed line repeats the flux without decay. The far crust and the lithospheric mantle lose most of their flux to decay, as the mantle does, while the local crust keeps nearly all of it. See Secs.~\ref{sec:flux} and~\ref{sec:signal} for details.}
 \label{fig:distance}
\end{figure}

Figure~\ref{fig:distance} shows where the geoneutrino flux at Borexino comes from. Without decay, about a quarter of the flux comes from the local crust, from a few kilometers to a few hundred kilometers away. The other three quarters, from the far crust, the lithospheric mantle, and the mantle, travel hundreds to thousands of kilometers, up to a full Earth diameter. The figure plots the flux per decade of distance; oscillations make the peaks and dips within the local crust (Fig.~30 of \Ref~\cite{Bustamante:2026yce} shows the flux without them). The flux from each unit volume of rock falls as $1/L^2$, but the rock between distances $L$ and $L + dL$ from the detector has a volume proportional to $L\,dL$. Therefore, rock thousands of kilometers away contributes as much per decade of distance as rock nearby, so decay, which acts over hundreds to thousands of kilometers, can remove much of the flux. Figure~\ref{fig:reactor_distance} shows the geoneutrino flux by distance at all four experiments, summed over the reservoirs, with and without decay (Sec.~\ref{sec:reactors}).

In \figu{distance}, the local crust gives way to the far crust between 172 and 404~km, the distances to the nearest edge and to the farthest corner of its region of $6^\circ \times 4^\circ$. Since we calibrate the two separately (Sec.~\ref{sec:prediction}), a kilogram of the far crust at Borexino emits 23\% more geoneutrinos than a kilogram of the local crust. At KamLAND, SNO+, and JUNO, a kilogram of the local crust emits 1.7--3.2 times as many as a kilogram of the far crust, so the local crust supplies more of their signal (\tabl{experiments}). The bottom panel of \figu{distance} shows the same distribution with decay, to which we return in Sec.~\ref{sec:signal}.

\subsection{Oscillations across the Earth}

The bottom panel of \figu{survival} shows the survival probability, \equ{psurv}, averaged over the detected spectrum of uranium and thorium (\figu{chains}). Our analysis computes it in the same way for every source, near or far: in full, interference terms included, and in matter (Appendix~\ref{sec:matter}). How it behaves depends on the distance:
\begin{itemize}
 \item Within about 350~km of the detector, a few oscillation lengths of $\Delta m^2_{21}$ and mostly in the local crust, the interference terms survive the average over the spectrum, so the probability oscillates.
 \item Beyond, the spread of energies averages them out, more so the farther the source. So does the spread of distances within the far crust and mantle.
\end{itemize}
KamLAND and SNO+ use the averaged value at every distance~\cite{KamLAND:2022vbm, SNO:2025chx}; JUNO does not say whether it does~\cite{JUNO:2026tap}. Doing so would change our 90\%~C.L. limits on $\nu_1 \to \nu_3$ by at most 1.3\% (Appendix~\ref{sec:matter}).

\subsection{What decay does to geoneutrinos}
\label{sec:signal}

Decay can lower or raise the geoneutrino signal, depending on the channel and the neutrino masses. We show this below, with Borexino as the example site. The changes are larger in the inverted ordering, where decay can remove nearly all the signal. Decay also changes the flux of reactor $\bar\nu_e$ (Sec.~\ref{sec:reactor_antineutrinos}). It removes as many of them as of geoneutrinos at Borexino, the site farthest from reactors, and fewer at the other sites (\figu{reactor_distance}).

\subsubsection{Inverted ordering}
\label{sec:signal_io}

Decay lowers the signal when its last daughters carry less electron flavor than their parents (Sec.~\ref{sec:channels}). This happens in the three channels that end in $\nu_3$ ($\nu_1 \to \nu_3$, $\nu_2 \to \nu_3$, and $\nu_2 \to \nu_1 \to \nu_3$), since the electron-flavor content of $\nu_3$ is only 2\% (\figu{flavor_content}). Complete decay, in which every parent decays on the way to the detector, lowers the signal at Borexino by about 84\%, 16\%, and 99.5\% in these channels, respectively. (These reductions differ slightly from the shares lost in \tabl{channels}, which are averaged over oscillations in vacuum and not weighted by the IBD cross section.) These values barely change as the lightest mass, $m_3$, varies from 0 to 16~meV, where the sum of the masses reaches its cosmological limit, $\sum_i m_i < 0.12$~eV~\cite{ParticleDataGroup:2024cfk}, which our fits include as a penalty (Sec.~\ref{sec:stats}).

Figure~\ref{fig:spectrum} shows how decay changes the geoneutrino flux at Borexino; at the other sites, the flux changes by less, since the local crust supplies more of the crustal signal there (\tabl{experiments}). For $\nu_1 \to \nu_3$, the decay that lowers the signal the most, a lifetime $\tau_1/m_1 = 10^{-8}$~s~eV$^{-1}$ removes 21\% of the flux above the IBD threshold; $10^{-9}$~s~eV$^{-1}$ removes 54\%. The depletion is largest at the lowest energies, as \equ{psurv} dictates. Decay distorts the spectrum, besides reducing it. This helps distinguish decay from a lower content of uranium and thorium, which would reduce the flux without distorting it.

The bottom panel of \figu{distance} shows where on Earth decay acts. With $\tau_1/m_1 = 10^{-9}$~s~eV$^{-1}$, decay removes 9\% of the flux from the local crust, but 66\% of that from the far crust, 53\% of that from the lithospheric mantle, and 78\% of that from the mantle. In total, 46\% of the flux remains, against 54--58\% at KamLAND, SNO+, and JUNO (\figu{reactor_distance}). At Borexino, the local crust then supplies about half of the flux, twice its share without decay. This makes the limits sensitive to how the crustal signal divides between the local and the far crust (\tabl{experiments}).

For $\nu_2 \to \nu_1$, whose daughter carries more electron flavor than its parent (Sec.~\ref{sec:channels}), complete decay instead raises the signal, by 17\%, since $\nu_1$ and $\nu_2$ have nearly equal masses and each daughter keeps nearly all of its parent's energy. (That falls short of the 20\% that \tabl{channels} allows, because daughters lose $\lesssim 3\%$ of their energy and the IBD cross section falls steeply as energy decreases.)

\subsubsection{Normal ordering}
\label{sec:signal_no}

Here, every channel ends in daughters that carry more electron flavor than their parents ($\nu_2 \to \nu_1$, $\nu_3 \to \nu_1$, $\nu_3 \to \nu_2$, and $\nu_3 \to \nu_2 \to \nu_1$). Still, decay does not always raise the signal. For $\nu_2 \to \nu_1$, whether complete decay lowers or raises it depends on the masses, because a daughter keeps between a fraction $(m_1/m_2)^2$ of its parent's energy and all of it (Sec.~\ref{sec:decays}). If $\nu_1$ is massless, that fraction is zero: the daughters spread over all energies up to their parent's. Many fall below the IBD threshold, so complete decay lowers the signal by 13\%. As $m_1$ grows, the fraction approaches 1, so the daughters keep more of the energy. Above $m_1 = 17$~meV, complete decay raises the signal, by up to 11\% at $m_1 = 30$~meV, where the sum of the masses reaches its cosmological limit.

The chain $\nu_3 \to \nu_2 \to \nu_1$ behaves like its last step, $\nu_2 \to \nu_1$: complete decay lowers the signal by 13\% if $\nu_1$ is massless and raises it above $m_1 = 17$~meV, by up to 11\%. The decays $\nu_3 \to \nu_1$ and $\nu_3 \to \nu_2$ raise it by less than 0.3\%, since $\nu_3$ makes up 2\% of each $\bar\nu_e$ at production.

\section{Reactor antineutrinos}
\label{sec:reactor_antineutrinos}

\subsection{Reactor antineutrinos at the experiments}
\label{sec:reactor_flux}

Nuclear reactors emit $\bar\nu_e$ in the beta decays of the products of fission. Every experiment that detects geoneutrinos detects them too. The spectrum of reactor $\bar\nu_e$ overlaps that of geoneutrinos but extends to higher energies, where the data constrain its normalization. We compute it core by core, from the thermal power of each core, its distance to the detector, and the spectrum of $\bar\nu_e$ that its fissions emit. For this spectrum, we follow each collaboration: each uses spectra measured at reactors, mainly by Daya Bay~\cite{DayaBay:2016ssb, DayaBay:2021dqj, DayaBay:2021owf}, alone or combined with predictions~\cite{Huber:2011wv, Mueller:2011nm}. On their way from each core, reactor $\bar\nu_e$ oscillate and, under each decay hypothesis, decay, as geoneutrinos do. We normalize the spectrum without decay to the number of reactor $\bar\nu_e$ that each collaboration expects to detect in its data.

The reactor flux at each site changes over the time of data taking, as reactors switch on and off. At KamLAND, it falls tenfold between the first and the last period of data, as Japanese reactors stopped after 2007 and 2011~\cite{KamLAND:2022vbm}. We account for these changes by averaging the thermal power of each core over the data taking of each data set (\ie, each period of data taking that an experiment publishes separately; \tabl{experiments}), except at JUNO, whose cores enter at their nominal power. Appendix~\ref{sec:reactors} describes the reactors and the spectra per fission of each experiment. Figures~\ref{fig:reactor_spectra} and~\ref{fig:reactor_distance} show the resulting spectrum of each data set, reactor by reactor, and how far its reactor antineutrinos travel, with and without decay.

\subsection{What decay does to reactor antineutrinos}
\label{sec:reactor_signal}

\begin{figure}[t!]
 \centering
 \includegraphics[width=\columnwidth]{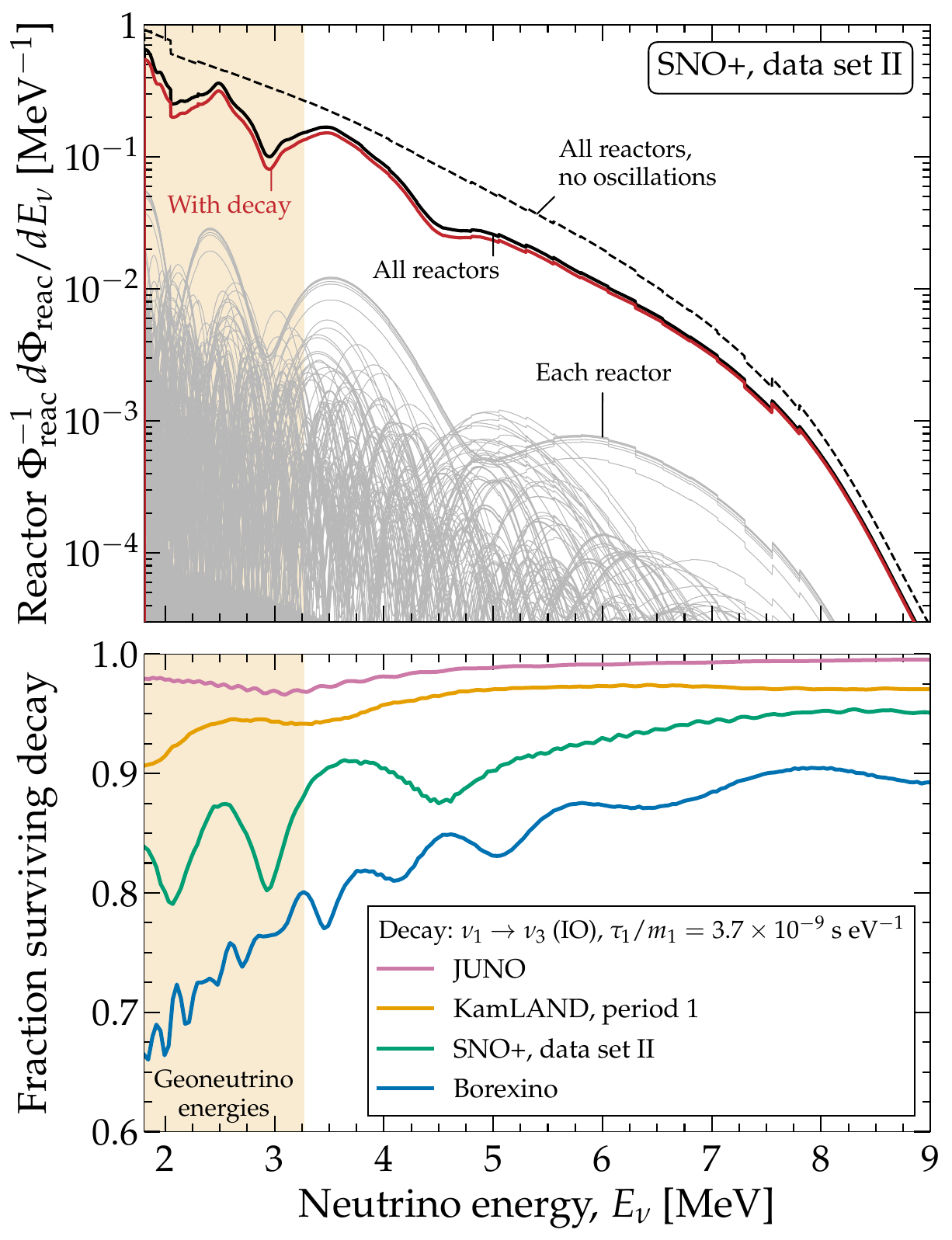}
 \caption{\textbf{Reactor antineutrinos with and without decay.} \textit{Top:} Flux of reactor $\bar\nu_e$ above the IBD threshold, for one data set, divided by $\Phi_{\rm reac}$, its integral without oscillations or decay. Each reactor core oscillates at its own distance. \textit{Bottom:} Fraction of the reactor $\bar\nu_e$ of each experiment that survives decay. Decay removes the most at Borexino, whose reactors are the farthest, and the least at JUNO. The mixing parameters are at the best fit of NuFIT 6.1~\cite{Esteban:2024eli, NuFIT61}. See Sec.~\ref{sec:reactor_signal} and \figu{reactor_spectra} for details.}
 \label{fig:reactor_decay}
\end{figure}

Figure~\ref{fig:reactor_decay} shows the reactor $\bar\nu_e$ with and without decay. In the top panel, each core detected by SNO+, shown as illustration, contributes its own spectrum, which oscillates at the distance of that core. The bottom panel shows the fraction of the reactor $\bar\nu_e$ that survives the decay $\nu_1 \to \nu_3$, at the lifetime of the limit that we find (Sec.~\ref{sec:nu1}). This fraction is smallest at Borexino, the site farthest from reactors, and largest at JUNO, 52.5~km from its own: the longer the $\bar\nu_e$ travel, the more of them decay. The fraction also falls toward low energies, because the laboratory-frame lifetime grows with energy.

\begin{figure}[t!]
 \centering
 \includegraphics[width=\columnwidth]{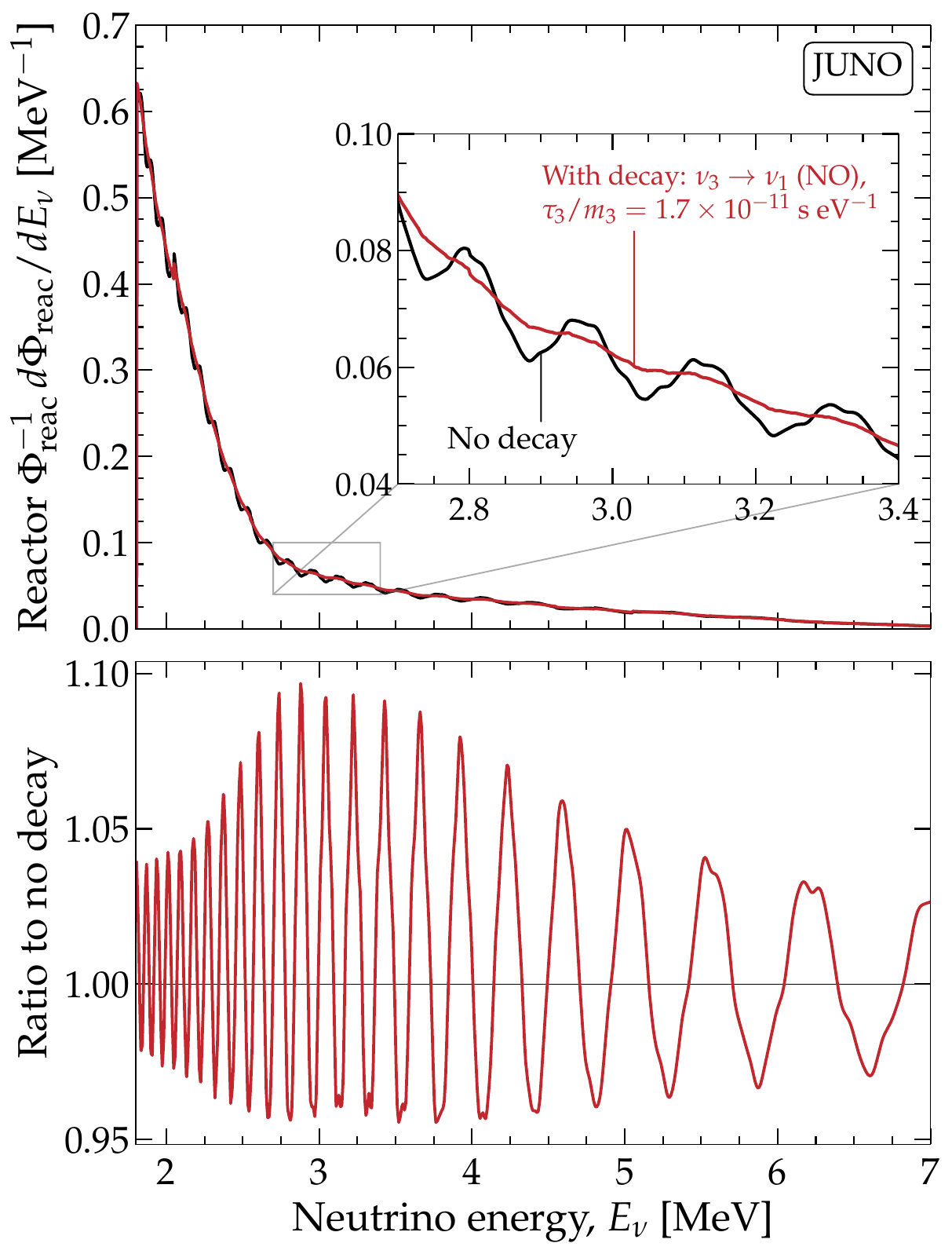}
 \caption{\textbf{JUNO's reactor antineutrinos with and without the decay of $\nu_3$.} \textit{Top:} Flux of reactor $\bar\nu_e$ above the IBD threshold, divided by $\Phi_{\rm reac}$, its integral without oscillations or decay. The inset magnifies the boxed region. \textit{Bottom:} Ratio of the two spectra. The lifetime is that of our limit (Sec.~\ref{sec:allchannels}); the mixing parameters are at the best fit of NuFIT~6.1~\cite{Esteban:2024eli, NuFIT61}. Decay flattens the ripple of the oscillation driven by $\Delta m^2_{31}$ but barely changes the total flux. See Sec.~\ref{sec:reactor_signal} for details.}
 \label{fig:juno_nu3_decay}
\end{figure}

Figure~\ref{fig:juno_nu3_decay} shows a second effect of decay, which only JUNO sees. At 52.5~km from its reactors, JUNO resolves the oscillation driven by $\Delta m^2_{31}$, which ripples its spectrum about every 0.2~MeV near 3~MeV. The ripple comes from the interference of $\nu_3$ with $\nu_1$ and $\nu_2$, so its depth depends on how much $\nu_3$ reaches the detector. In the normal ordering, where $\nu_3$ can decay, its decay flattens the ripple. The total reactor flux changes by under 1\%, since $\nu_3$ carries only 2\% of the electron flavor (\figu{flavor_content}). Therefore, JUNO tests the decay of $\nu_3$~\cite{Abrahao:2015rba, deGouvea:2022cmo}, which geoneutrinos cannot (Sec.~\ref{sec:signal_no}), through the shape of its spectrum, not its normalization.

This test needs the shape of the reactor spectrum at the scale of the ripple, which JUNO constrains in part with data from the Taishan Antineutrino Observatory (TAO) that are not public~\cite{JUNO:2026tap}. With our model of that shape, we reproduce JUNO's measurement of $\Delta m^2_{31}$, which comes from the ripple (Appendix~\ref{app:reproduction}).

Reactor $\bar\nu_e$ test decay wherever the fit constrains their flux: at KamLAND, SNO+, and JUNO, where the power of the reactors fixes it to within a few percent (\tabl{parameters}). At Borexino, whose fit leaves their rate free, they do not (Sec.~\ref{sec:nu1}). Unlike the flux of geoneutrinos, the reactor flux does not depend on the Earth model, so no fit can raise it beyond those few percent to make up for decay. Therefore, reactor $\bar\nu_e$ matter most to our limits when the radiogenic power is left free (Sec.~\ref{sec:results}).

\section{Detection}
\label{sec:detection}

\subsection{Detection via inverse beta decay}
\label{sec:ibd}

All four experiments we consider---Borexino, KamLAND, SNO+, and JUNO---detect geoneutrinos through IBD (Sec.~\ref{sec:decays}) in large volumes of liquid scintillator underground. (The targets are the free protons of the scintillator's hydrogen; its carbon does not take part at these energies.) The positron loses its kinetic energy in the scintillator and annihilates with an electron into two photons. Both steps take a few nanoseconds, so their scintillation light forms one prompt signal. The deposited energy of the positron and the two photons is $E_{\rm dep} = E_\nu - Q - T_n$, where $Q = 0.78$~MeV is the $Q$-value of free-neutron beta decay, the energy that the decay releases, and $T_n$ is the kinetic energy of the recoiling neutron, at most about 15~keV for geoneutrinos~\cite{Vogel:1999zy}. Since $T_n$ is small, $E_{\rm dep} \approx E_\nu - Q$, although our calculation keeps $T_n$ (Sec.~\ref{sec:response}). The detector reconstructs the deposited energy as the prompt energy, $E_{\rm prompt}$, which differs from $E_{\rm dep}$ because of the detector's energy scale, nonlinear in general, and its resolution (Sec.~\ref{sec:response}).

About 0.2~ms after the prompt signal, the neutron is captured on hydrogen; the capture emits a 2.2-MeV photon, the delayed signal. The coincidence of the prompt and the delayed signals tags the event and suppresses backgrounds. Every experiment also detects the $\bar\nu_e$ from reactors, through the same reaction. Geoneutrino signals are quoted in terrestrial neutrino units (TNU): one TNU is one IBD event per $10^{32}$ target protons per year.

\subsection{The experiments}
\label{sec:experiments}

\tabl{experiments} shows the four experiments and the data sets we use from each. Below, we present their main features; Appendix~\ref{app:reproduction} contains further detail.

\begin{table*}[t!]
 \caption{\label{tab:experiments}\textbf{The four experiments and their seven data sets.} Each data set $d$ has a column; the rows that describe an experiment $e$ span the columns of its data sets. KamLAND's three periods differ from one another in the reactor flux, which fell to about a tenth after 2011; in the $(\alpha, n)$ background, which the purification of the scintillator in 2007--2009 reduced; and in the fiducial volume, which is smaller in Period~3~\cite{KamLAND:2022vbm}. Compared with SNO+'s data set~I, its data set~II has a reactor flux per unit exposure about 2\% higher and an energy resolution 10\% narrower~\cite{SNO:2025chx} (Appendix~\ref{app:reproduction}). The depth is the thickness of rock above each detector; the scintillator mass is that of the whole target (KamLAND's from \Ref~\cite{KamLAND:2008dgz}). The prompt-energy window is the one in which we fit each data set. The geoneutrino signals of the local crust, the far crust, and the lithospheric mantle are those to which we calibrate them (Sec.~\ref{sec:prediction}): the local crust from each site's own model, the other two from \Ref~\cite{Wipperfurth:2019idn}, except KamLAND's far crust, from \Ref~\cite{Huang:2013ozo}, and SNO+'s two, as \Ref~\cite{Semenec:2023} computes them around its own local crust.}
 \centering
 \begin{ruledtabular}
 \begin{tabular}{lccccccc}
  Experiment, $e$ & Borexino~\cite{Borexino:2019gps} & \multicolumn{3}{c}{KamLAND~\cite{KamLAND:2022vbm}} & \multicolumn{2}{c}{SNO+~\cite{SNO:2025chx}} & JUNO~\cite{JUNO:2026tap} \\
  Data set, $d$ & & Period 1 & Period 2 & Period 3 & \makebox[1.45cm]{I} & \makebox[1.45cm]{II} & \\
  \hline
  Site & Gran Sasso, Italy & \multicolumn{3}{c}{Kamioka, Japan} & \multicolumn{2}{c}{Sudbury, Canada} & Jiangmen, China \\
  Depth [km] & 1.4 & \multicolumn{3}{c}{1.0} & \multicolumn{2}{c}{2.1} & 0.7 \\
  Liquid scintillator [kton] & 0.28 & \multicolumn{3}{c}{1.0} & \multicolumn{2}{c}{0.78} & 20 \\
  Days of data & 3263 & 1486 & 1151 & 2590 & 246 & 439 & 207.2 \\
  IBD candidates in the window & 154 & 682 & 374 & 167 & 77 & 108 & 8294 \\
  Prompt-energy window [MeV] & 0.8--7.8 & \multicolumn{3}{c}{0.9--2.7} & \multicolumn{2}{c}{0.9--8.0} & 0.7--12.0 \\
  Local crust [TNU] & 9.2 & \multicolumn{3}{c}{17.7} & \multicolumn{2}{c}{18.6} & 16.4 \\
  Far crust [TNU] & 14.1 & \multicolumn{3}{c}{7.3} & \multicolumn{2}{c}{16.8} & 11.8 \\
  Lithospheric mantle [TNU] & 1.8 & \multicolumn{3}{c}{1.3} & \multicolumn{2}{c}{2.1} & 1.7 \\
  Main backgrounds & Accidentals, $^9$Li & \multicolumn{3}{c}{$(\alpha, n)$, accidentals} & \multicolumn{2}{c}{$(\alpha, n)$} & $^9$Li/$^8$He \\
 \end{tabular}
 \end{ruledtabular}
\end{table*}

\subsubsection{Borexino}
\label{sec:borexino}

Borexino, at the Gran Sasso laboratory, is the farthest of the four from nuclear reactors, so geoneutrinos make up a larger share of its events than at the other three sites (Sec.~\ref{sec:reproduction}). Its 2020 analysis measured a geoneutrino signal of 47~TNU, to a precision of 18\%~\cite{Borexino:2019gps}. The fit left the reactor rate free and fixed Th/U to 3.9, the value in chondrites (Sec.~\ref{sec:prediction}). Borexino attributed nearly half of this signal, 21.2~TNU, to the mantle; it excluded a mantle signal of zero at 99\% C.L.

\subsubsection{KamLAND}
\label{sec:kamland}

KamLAND, in the Kamioka mine, is surrounded by Japanese reactors. Most of them stopped after the Fukushima accident of March 2011, which left KamLAND nearly free of reactor $\bar\nu_e$ for eight years~\cite{KamLAND:2022vbm}. KamLAND's 2022 analysis, which includes those years, fitted the uranium and thorium signals separately and found $173.7^{+29.2}_{-27.7}$ geoneutrinos: 116.6 from uranium, 57.5 from thorium~\cite{KamLAND:2022vbm}. With this measurement, KamLAND disfavored the geodynamical models (Sec.~\ref{sec:prediction}) at 99.76\% C.L., for a mantle of uniform composition. Its data cover $E_{\rm prompt} = 0.9$--2.7~MeV, where nearly all geoneutrinos lie.

\subsubsection{SNO+}
\label{sec:snoplus}

SNO+, at SNOLAB, sits on the Canadian Shield, where the continental crust is thick~\cite{SNO:2025chx}. Most of its reactor $\bar\nu_e$ come from three nuclear plants in Ontario, 240--350~km away~\cite{SNO:2025chx}. Its analysis of two data sets measured $\Delta m^2_{21}$ from their spectrum and, with the geoneutrino flux free, a geoneutrino signal of $48^{+14}_{-12}$~TNU~\cite{SNO:2025chx}. It constrained the uranium-to-thorium signal ratio to $3.8 \pm 1.3$, from its geological model of the region. 

\subsubsection{JUNO}
\label{sec:juno}

JUNO, near Jiangmen, is the largest of the four, 52.5~km from the Taishan and Yangjiang nuclear power plants. Its first 59.1 days of data measured $\sin^2\theta_{12}$ and $\Delta m^2_{21}$ with reactor $\bar\nu_e$~\cite{JUNO:2025gmd}. In that analysis, geoneutrinos are a background, constrained to their predicted rate, 1.75 events per day, with an uncertainty of 40\%. From 207 days of data, JUNO has since measured the oscillation parameters again and, for the first time at its site, the geoneutrino signal, $73 \pm 11$~TNU, with the rate of geoneutrinos left free~\cite{JUNO:2026tap}. We use these 207 days of data.

\subsection{Event-rate calculation}
\label{sec:rates}

The coincidence of the prompt and the delayed signals selects IBD events, but it cannot tell geoneutrinos from reactor $\bar\nu_e$, or from the backgrounds that mimic the coincidence (Sec.~\ref{sec:backgrounds}). Their spectra of $E_{\rm prompt}$ differ, though: geoneutrinos lie below about 2.6~MeV; reactor $\bar\nu_e$ extend to higher energies; each background has its own shape. A fit of the spectrum separates them, statistically rather than event by event, and reveals the distortion that decay causes at low energies (Sec.~\ref{sec:signal}). 

The experiments publish their data as numbers of events binned in $E_{\rm prompt}$; uniquely, SNO+ publishes a list of its events, which we bin. Each experiment $e$ contributes one or more data sets, numbered by $d$, one for each period of data taking whose data it publishes separately (\tabl{experiments}). We fit the data bin by bin, comparing the observed and expected numbers of events (Sec.~\ref{sec:stats}). 

The expected number of IBD events in bin $i$ of prompt energy of data set $d$ of experiment $e$ is
\begin{multline}
 N_i^{(e,d)}
 =
 \int_i dE_{\rm prompt} \int dE_\nu\, R^{(e,d)}(E_{\rm prompt}, E_\nu)
 \\
 \times
 \bigg[ (N_p T)^{(e,d)}\, \sigma_{\rm IBD}(E_\nu) \sum_{r,c} \nu_{rc}^{(e,d)} f_{rc}^{(e)} \frac{d\Phi_{rc}^{(e)}}{dE_\nu}
 \\
 + \nu_{\rm reac}^{(e,d)} \frac{dN_{\rm reac}^{(e,d)}}{dE_\nu} \bigg]
 + \sum_b \nu_b^{(e,d)} B_{b,i}^{(e,d)} \;.
 \label{equ:rate}
\end{multline}
The two integrals run over bin $i$ of prompt energy, $E_{\rm prompt}$, and over true neutrino energy, $E_\nu$. The response function, $R^{(e,d)}(E_{\rm prompt}, E_\nu)$, is the probability density that an IBD event of neutrino energy $E_\nu$ is selected and has prompt energy $E_{\rm prompt}$; it includes the energy scale, the energy resolution, and the selection efficiency of the detector. The resolution sets how sharply the detected spectrum shows the endpoints of the uranium and thorium branches, which separate geoneutrinos from the backgrounds and the two chains from each other (\figu{chains}). Decay changes the spectrum smoothly across the whole window, so the resolution matters for it mainly through this separation: widening or narrowing the resolution by 20\% moves our expected limits on $\nu_1 \to \nu_3$ by at most 2\% and those from data by at most 2.4\%, except JUNO's, which moves by 12 to 14\%. Figure~\ref{fig:response} shows the response function of each experiment and data set.
The geoneutrino flux from reservoir $r$ and chain $c$, $d\Phi_{rc}^{(e)}/dE_\nu$, is the term $rc$ of \equ{flux}, with distances measured from the detector of experiment $e$. Through the operator $\mathcal{T}_{L,\psi}$ (Sec.~\ref{sec:flux}), it includes oscillations and decay, the daughters among them. 

The IBD cross section, $\sigma_{\rm IBD}$~\cite{Vogel:1999zy, Strumia:2003zx}, and the exposure, $(N_p T)^{(e,d)}$, the number of target protons times the time of data taking, turn this flux into events. The factor $f_{rc}^{(e)}$ calibrates chain $c$ of the local crust, the far crust, and the lithospheric mantle at the site of experiment $e$ to their signals in \tabl{experiments} (Sec.~\ref{sec:prediction}); for the mantle, $f_{rc}^{(e)} = 1$. 

The reactor term, $dN_{\rm reac}^{(e,d)}/dE_\nu$, is the spectrum of reactor IBD events in true energy (Sec.~\ref{sec:reactor_antineutrinos}); the backgrounds, $B_{b,i}^{(e,d)}$, are numbers of events measured in $E_{\rm prompt}$ (Sec.~\ref{sec:backgrounds}). The normalizations $\nu_{rc}^{(e,d)}$, $\nu_{\rm reac}^{(e,d)}$, and $\nu_b^{(e,d)}$, relative to their expected values, are free parameters of the fits, some shared among data sets (Sec.~\ref{sec:stats}).

\subsection{Backgrounds}
\label{sec:backgrounds}

We compute the spectrum of reactor $\bar\nu_e$ but take the backgrounds from each experiment's own estimates. In our fits, the rates of both are free parameters, constrained as in each experiment's own fit (\tabl{parameters}). Most of their events are of three kinds. In $(\alpha, n)$ reactions, alpha particles from $^{210}$Po strike $^{13}$C and free neutrons, whose scattering on protons mimics the prompt signal and whose capture mimics the delayed one. Cosmic muons break up carbon nuclei of the scintillator into $^9$Li and $^8$He, which decay by emitting an electron and then a neutron. In accidental coincidences, two unrelated events, such as two radioactive decays, occur by chance so close in time and space that they mimic the prompt and delayed signals of an IBD event. In all, the backgrounds make up 5--30\% of the candidates, depending on the depth of each detector and the purity of its scintillator (Appendix~\ref{sec:other_backgrounds}). We include every background that each experiment fits, in each of its periods of data taking, with the normalization, shape, and uncertainty that the experiment quotes.

\subsection{Reproducing the experiments}
\label{sec:reproduction}

Our model of each experiment follows its collaboration's analysis, as far as it is published (Appendix~\ref{app:reproduction}). Without decay, it reproduces each collaboration's best-fit spectrum to 0.5--2.4\% per bin on average (\figu{reproduction}). 

In our fits of Borexino's and KamLAND's data, the collaborations' own spectra of detected uranium, thorium, and reactor events replace ours without decay. Borexino published the simulated spectra that its own fit used; KamLAND's spectra have features that our calculation cannot reproduce. Decay and the mixing parameters then change each bin of these spectra in the same proportion as they change ours (Appendix~\ref{app:reproduction}). At SNO+ and JUNO, we use our own spectra, computed from the fluxes, the IBD cross section, and the detector response.

\begin{figure}[t!]
 \centering
 \includegraphics[width=\columnwidth]{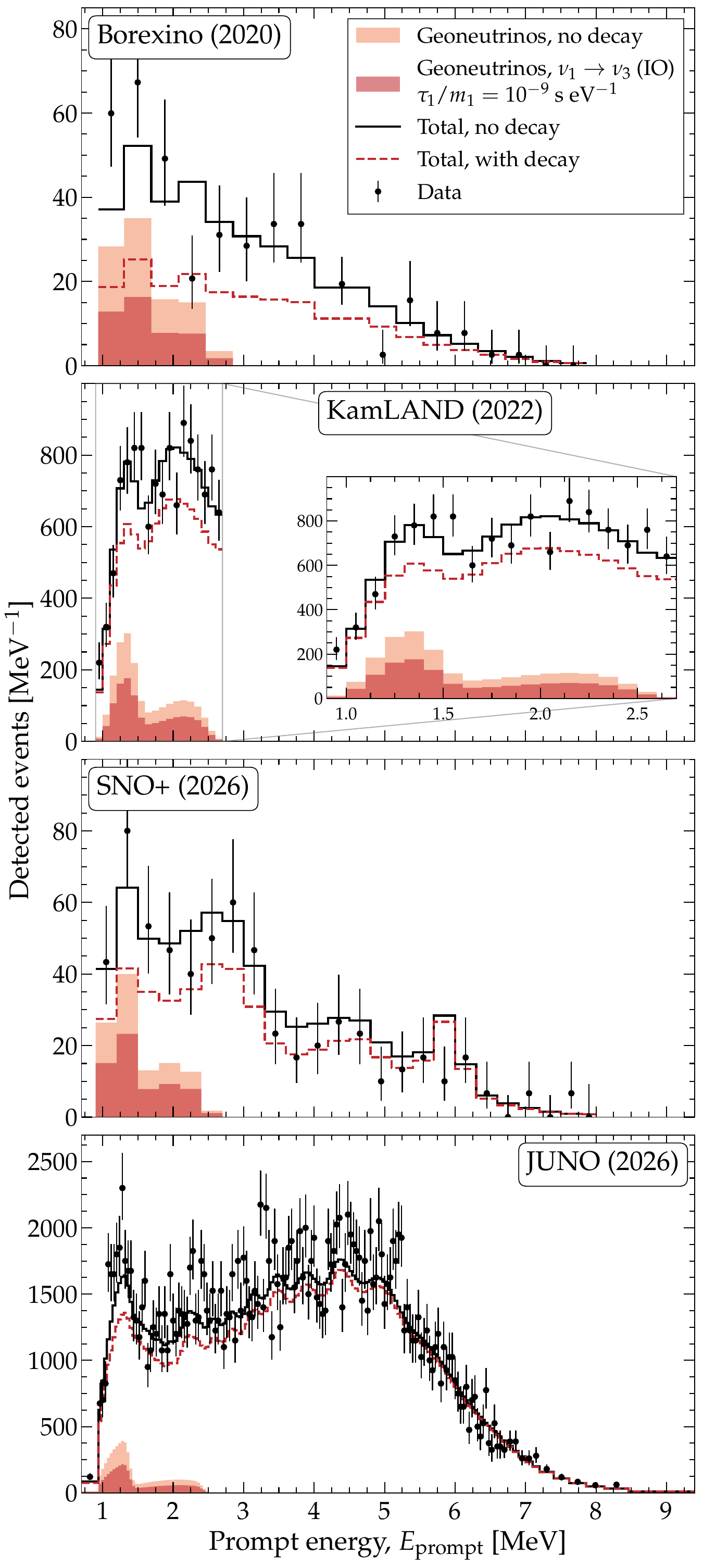}
 \caption{\textbf{Detected IBD spectra at the four experiments, with and without decay.} The expected spectra are for the geochemical Earth model, with every background at its expected value, without decay and with $\nu_1 \to \nu_3$ in the inverted mass ordering. The data are from \Refs~\cite{Borexino:2019gps, KamLAND:2022vbm, SNO:2025chx, JUNO:2026tap}, with error bars at 68\%~C.L. KamLAND's three periods and SNO+'s two data sets are added up. See Sec.~\ref{sec:reproduction} for details.}
 \label{fig:events}
\end{figure}

Fitted to each data set as its collaboration fits it, the model also recovers the collaboration's results (Appendix~\ref{app:reproduction}): the geoneutrino signals of Borexino and SNO+ to within 2\%, and JUNO's oscillation parameters to within $0.3\sigma$. At KamLAND, it finds 156 geoneutrino events, $0.6\sigma$ below KamLAND's 174; on simulated data, our KamLAND fit is unbiased. These differences come from information that the collaborations used but did not publish: KamLAND's spectrum above 2.7~MeV and the times of its events; the full response of each detector, which we infer from what its collaboration published; and the power of each JUNO reactor over time.

Figure~\ref{fig:events} shows the data of the four experiments and their expected spectra, with and without decay. At KamLAND and SNO+, the figure adds up the data sets of \tabl{experiments} (the statistical analysis of Sec.~\ref{sec:stats} uses the spectrum of each data set, not their sum). Figures~\ref{fig:response}--\ref{fig:reactor_distance} show the data sets one by one: their detector responses, their reactor spectra, and how far their reactor antineutrinos travel. For the geochemical Earth model, geoneutrinos make up 26\% of the expected events at Borexino, 17\% at KamLAND, 17\% at SNO+, and 3\% at JUNO. Borexino has the fewest events, but the largest share of geoneutrinos; JUNO has the most events, but reactor $\bar\nu_e$ dilute its geoneutrinos. With $\tau_1/m_1 = 10^{-9}$~s~eV$^{-1}$, decay removes 41 to 52\% of the geoneutrinos. It also removes reactor $\bar\nu_e$, more where the reactors are far: 47\% of them at Borexino, against 5\% at JUNO (Sec.~\ref{sec:reactors}).

\section{Statistical analysis}
\label{sec:stats}

\subsection{Likelihood function}

For each experiment, the likelihood is binned Poisson in prompt energy, in the bins in which its data are published, or, for SNO+, which publishes a list of its events, in bins of 0.1~MeV. Parameters that external measurements constrain enter through penalty terms. The negative logarithm of the likelihood of experiment $e$ is
\begin{align}
 -\ln \mathcal{L}^{(e)}(\boldsymbol\theta)
 ={}&
 \sum_d \sum_i \left[ N_i^{(e,d)}(\boldsymbol\theta) - n_i^{(e,d)} \ln N_i^{(e,d)}(\boldsymbol\theta) \right]
 \nonumber \\
 &+ \frac{1}{2} \sum_k \left( \frac{\theta_k - \bar\theta_k}{\sigma_k} \right)^2
 + \frac{1}{2} \left( \frac{\sum_i m_i}{\sigma_\Sigma} \right)^2
 \nonumber \\
 &+ \frac{1}{2} \left( \Delta\chi^2_{12} + \Delta\chi^2_{13} \right) \;,
 \label{equ:likelihood}
\end{align}
where $\boldsymbol\theta$ holds all the free parameters of the fit, listed in \tabl{parameters}: the lifetimes, the mass of the lightest neutrino, the mixing parameters, the normalizations of the Earth's reservoirs, the radiogenic power when it is free, and the background normalizations and detector parameters of the experiment. The sums run over the data sets of the experiment, $d$, and over the bins of prompt energy, $i$. In bin $i$ of data set $d$, $n_i^{(e,d)}$ is the number of observed events and $N_i^{(e,d)}(\boldsymbol\theta)$ is the expected number, \equ{rate}. Equation~(\ref{equ:likelihood}) leaves out the terms that do not depend on the parameters, such as $\ln n_i^{(e,d)}!$ from the Poisson probability; they change neither the best fit nor the test statistic, \equ{test_statistic}, in which they cancel.

\subsubsection{Penalties}

The second term of \equ{likelihood} holds a Gaussian penalty for each parameter, $\theta_k$, that an external measurement constrains, with central value $\bar\theta_k$ and uncertainty $\sigma_k$. For the normalizations of the lithosphere (Sec.~\ref{sec:free_parameters}), $\theta_k$ stands for the logarithm of the normalization, with $\bar\theta_k = 0$, so that their priors are log-normal.

The third term enforces the cosmological upper limit on the sum of neutrino masses, $\sum_i m_i < 0.12$~eV at 95\%~C.L.~\cite{ParticleDataGroup:2024cfk}. It is a half-Gaussian penalty above zero, of width $\sigma_\Sigma = 0.12~\mathrm{eV}/1.960 = 0.061$~eV, so that 95\% of its probability lies below the limit. 

The last term constrains the mixing parameters with NuFIT~6.1~\cite{Esteban:2024eli, NuFIT61}: $\Delta\chi^2_{12}$ and $\Delta\chi^2_{13}$ are its tabulated pairwise $\Delta\chi^2$ over $(\sin^2\theta_{12}, \Delta m^2_{21})$ and over $(\sin^2\theta_{13}, \vert\Delta m^2_{31}\vert)$. Because NuFIT also fits the reactor antineutrinos of KamLAND, SNO+, and JUNO, a fit to any of them uses $\Delta\chi^2_{12}$ less that experiment's own measurement---to prevent double-counting those neutrinos---and the fit of all four uses the fit of the solar data alone instead (Appendix~\ref{app:nufit}).

\subsubsection{Free parameters}
\label{sec:free_parameters}

The parameters of interest are the lifetimes, $\tau_i / m_i$; all the others are nuisance parameters. Each experiment contributes the backgrounds, energy scale, and detector parameters of its own analysis, with the priors that its collaboration uses. A fit to one experiment has 15--44 parameters; the combined fit of all four has 75. Appendix~\ref{app:parameters} lists them with their priors and the value that each fit to data finds for it.

\smallskip
\textit{Earth.}---The Earth contributes four normalizations, the $\nu_{rc}$ of \equ{rate}: one each for the local crust, the far crust, the lithospheric mantle, and the mantle. Each scales its uranium and thorium together and has a prior from the published uncertainty of its signal or, for the mantle, of the radiogenic power of each Earth model (\tabl{parameters}). Without these priors, the normalizations could absorb any deficit that decay produces. The combined fits share the normalizations of the far crust and the lithospheric mantle across sites, as \Ref~\cite{Bellini:2021sow} does for Borexino and KamLAND. The local crusts stay independent. In the fits with the radiogenic power free, the power replaces the mantle's normalization. These fits still set limits, because the Earth's surface heat flow bounds the power from above (Sec.~\ref{sec:prediction} and Appendix~\ref{app:brazil}).

\smallskip
\textit{Neutrinos.}---The mixing angles and mass splittings float in every fit, as one set shared by geoneutrinos and reactor $\bar{\nu}_e$ of every experiment. The mass of the lightest neutrino, which sets the energies of the daughters in visible decay through the ratio of the parent and daughter masses, $x_{ij}$ in \equ{daughter}, is free, apart from the penalty that the cosmological limit puts on the sum of the masses.

\subsubsection{Single-experiment and combined fits}

We fit each experiment alone and in combinations. For a combination, we multiply its experiments' likelihoods,
\begin{equation}
 -\ln \mathcal{L}(\boldsymbol\theta)
 =
 \sum_e \left[ -\ln \mathcal{L}^{(e)}(\boldsymbol\theta) \right] \;,
 \label{equ:likelihood_combined}
\end{equation}
in which each penalty on a shared parameter counts once. All the experiments, and their data sets, share the lifetime, the mass of the lightest neutrino, the mixing parameters, the radiogenic power when it is free, and the mantle normalizations, since the mantle is the same at every site. Each experiment keeps its own local crust, backgrounds, and detector parameters. 

We show two combinations. In the first, Borexino, KamLAND, and SNO+ (``BX+KL+SNO+''), the Earth model sets the rate of geoneutrinos at each site, so the geoneutrinos test decay. The second adds JUNO (``All four''). At JUNO, the measured rate of geoneutrinos, $73 \pm 11$~TNU~\cite{JUNO:2026tap}, exceeds what every Earth model that we use predicts, 32--51~TNU. To match it, a fit must raise the lithospheric mantle to ten times its nominal content, which no geological model holds. Therefore, we leave JUNO's rate free, as JUNO's fit does (Sec.~\ref{sec:juno}), so JUNO tests decay only through its reactor $\bar\nu_e$.

\subsection{Setting limits on the neutrino lifetime}
\label{sec:limits}

We set frequentist limits from the profile likelihood ratio. For each trial value of $\log_{10}(\tau_i/m_i)$, we minimize the likelihood, \equ{likelihood} or \equ{likelihood_combined}, over all other parameters, which in a chain ($\nu_2 \to \nu_1 \to \nu_3$ or $\nu_3 \to \nu_2 \to \nu_1$) include the lifetime of the other parent. The test statistic is
\begin{equation}
 q(\tau_i/m_i)
 \!=\!
 2 \left[ -\ln \mathcal{L}\big(\tau_i/m_i, \hat{\hat{\boldsymbol\theta}}\big)
 + \ln \mathcal{L}\big(\widehat{\tau_i/m_i}, \hat{\boldsymbol\theta}\big) \right] \;,
 \label{equ:test_statistic}
\end{equation}
where $\hat{\hat{\boldsymbol\theta}}$ holds the values of the other parameters that minimize $-\ln\mathcal{L}$ at fixed $\tau_i/m_i$, and $\widehat{\tau_i/m_i}$ and $\hat{\boldsymbol\theta}$ are the values that minimize it globally. 

The one-sided lower limit on $\tau_i/m_i$ is the value at which $q$ reaches a threshold: 1.64 for 90\%~C.L., or 2.71 for 95\%~C.L. (\tabl{limits})~\cite{Wilks:1938dza, Cowan:2010js}. When the data prefer a finite lifetime, $q$ rises on both sides of the best fit, so the limit is where it reaches the threshold on the side of shorter lifetimes. In that case, we also report how strongly the data disfavor no decay, in standard deviations, \ie, $\sqrt{q}$ without decay~\cite{Chernoff:1954eli}.

For the two preferences that Sec.~\ref{sec:results} quotes (KamLAND and all four experiments, both for $\nu_1 \to \nu_3$ on the geodynamical Earth), we instead count how often simulated experiments without decay, fitted as the data are, give a $q$ without decay at least as large as the data's, and express it in standard deviations. 

\subsection{Expected limits and coverage}
\label{sec:validation}

\subsubsection{Expected limits}

To tell whether a limit from data is typical, we compare it with the limits that the experiments would set if neutrinos did not decay. We compute these for the channel of our main result, $\nu_1 \to \nu_3$ in the inverted ordering. We simulate experiments without decay on the geochemical Earth, the middle of the three BSE classes: 200 for each individual experiment and for each of the two combinations (``BX+KL+SNO+'' and ``All four''). We fit each simulated experiment as we fit the data, with its events and the centers of its priors drawn at random (Appendix~\ref{app:brazil}). The \textit{expected limit} is the limit from the Asimov data set~\cite{Cowan:2010js}, a simulated experiment in which every bin holds its expected number of events.

For both combinations, the expected limit is about $10^{-8}$~s~eV$^{-1}$, as the estimate of Sec.~\ref{sec:intro} anticipates. At this lifetime, decay removes a fifth of the geoneutrino flux at Borexino (Sec.~\ref{sec:signal}). The expected limits of Borexino and KamLAND lie within a factor of two of this value, whereas those of SNO+ and JUNO are several times weaker (\figu{brazil}). Therefore, the two combinations, which differ only by JUNO, have the same expected limit. Fluctuations alone spread the limits of the simulated experiments over a wide range.

\subsubsection{Coverage}

We read the limits from $q$ at thresholds such as 1.64 for 90\%~C.L. (Sec.~\ref{sec:limits}). These thresholds come from the asymptotic distribution of the test statistic, given by Wilks' theorem~\cite{Wilks:1938dza, Chernoff:1954eli}, which requires that the data contain many events and that every parameter changes the prediction. Our fits do not always meet these requirements. Borexino and SNO+ record few events. Nor does every parameter always change the prediction: once decay is complete, a shorter lifetime changes nothing. The mass of the lightest neutrino changes it only when there is decay. The \textit{coverage} of a limit is the share of repeated experiments in which it does not exclude the true lifetime; at 90\%~C.L., it should be 90\%. With wrong thresholds, the coverage could fall below 90\%, which would make the limits too strong.

The Feldman-Cousins construction~\cite{Feldman:1997qc} ensures the coverage by calibrating the thresholds with simulated experiments at every lifetime, but is computationally expensive in our case. So, instead, we test whether the asymptotic thresholds already give the right coverage (Appendix~\ref{app:coverage}). For $\nu_1 \to \nu_3$, we simulate 200 experiments with decay for each experiment and combination, at two true lifetimes: the expected limit and a lifetime 3.2 times shorter. We fit them as we fit the data. The share of simulated experiments whose limit excludes the true lifetime never exceeds 10\% by more than its uncertainty. With the radiogenic power free, and at SNO+, the share does not exceed 10\%, so these limits are correct or conservative. Therefore, the asymptotic thresholds are valid for the fits of this decay. For the decay of $\nu_3$, which only JUNO tests, they are not; there, we calibrate the thresholds with simulated experiments (Appendix~\ref{app:coverage}).

\section{Results}
\label{sec:results}

The data show no evidence of decay that holds across Earth models. With the radiogenic power free, the data of each experiment and combination disfavor no decay at $1.4\sigma$ or less, in every channel. On a class of BSE model, a preference for decay can appear; it is strongest on the geodynamical Earth, which predicts the most geoneutrinos. There, KamLAND's data disfavor no decay at $2.7\sigma$ for $\nu_1 \to \nu_3$, because they hold fewer geoneutrinos than predicted (Sec.~\ref{sec:nu1}). On the same Earth, the data of all four disfavor no decay at $1.8\sigma$, for the same channel. On the other two classes, no preference exceeds $1.8\sigma$. 

The preference on the geodynamical Earth measures the tension between the data and that Earth model, not decay: when the radiogenic power is left free, KamLAND's data choose the coldest Earth allowed, 9.4~TW, and their preference for decay disappears; the combined data choose about 15~TW and no decay. KamLAND's deficit of geoneutrinos does not point to decay either. First, the prediction at KamLAND depends on how its crust is modeled, so the deficit may come from that model. Second, decay would also remove geoneutrinos at Borexino, whose data hold more geoneutrinos than predicted. Borexino's data exclude, at 90\%~C.L., the lifetime that KamLAND's data prefer (Appendix~\ref{app:parameters}). Therefore, we set limits on the lifetimes.

Our strongest limits are on the decay of $\nu_1$, for the two reasons of Sec.~\ref{sec:channels}. First, $\nu_1$ carries most of the detected flux. Second, $\nu_1$ can decay only in the inverted ordering, into $\nu_3$, which IBD barely detects, so its decay removes geoneutrinos almost as an invisible decay would.

\subsection{The decay of $\nu_1$}
\label{sec:nu1}

\tabl{limits} shows the limits from each experiment and the two combinations. Our main result is the limit on the decay $\nu_1 \to \nu_3$ in the inverted ordering from the four experiments combined, at 90\%~C.L. To be conservative, we lead with the limit obtained with the radiogenic power free, which assumes the least about the Earth (Sec.~\ref{sec:prediction}):
\begin{equation*}
 \tau_1/m_1 >
 \begin{cases}
  3.7\times10^{-9}~\mathrm{s~eV^{-1}} & \text{power free,} \\
  (0.44 \text{ to } 1.6)\times10^{-8}~\mathrm{s~eV^{-1}} & \text{BSE models.}
 \end{cases}
\end{equation*}
With the power free, the limit from data lies within 2\% of the limit expected without decay. On a class of BSE model, the limit from data follows how far the data lie from that model: it is twice the expected limit on the cosmochemical Earth, 0.83 times on the geochemical, and 0.24 times on the geodynamical. 

\begin{table}[t!]
 \caption{\label{tab:limits}\textbf{Lower limits on the $\nu_1$ lifetime.} Limits on $\tau_1/m_1$ from $\nu_1 \to \nu_3$ in the inverted ordering, at 90\% and 95\%~C.L., in units of $10^{-9}$~s~eV$^{-1}$, for each class of BSE model and for the radiogenic power free. ``BX+KL+SNO+'' combines Borexino, KamLAND, and SNO+; ``All four'' adds JUNO. See Sec.~\ref{sec:nu1} for details and \Ref~\cite{NuDecayGeoReactorRepo} for limits on all $\nu_i$ across all decay channels.}
 \centering
 \begin{ruledtabular}
 \begin{tabular}{lcccc}
  & \multicolumn{3}{c}{Radiogenic power assumed} & Free \\
  \cline{2-4}
  & Cosmochem. & Geochem. & Geodyn. & power \\
  \hline
  \multicolumn{5}{l}{\textit{90\% C.L.}} \\
  Borexino & 21.2 & 25.8 & 14.9 & 8.01 \\
  KamLAND & 7.24 & 4.07 & 2.53 & 3.41 \\
  SNO+ & 0.865 & 0.861 & 0.844 & 0.858 \\
  JUNO & 2.67 & 2.64 & 2.61 & 2.60 \\
  \hline
  BX+KL+SNO+ & 16.1 & 10.1 & 4.42 & 3.60 \\
  All four & 15.8 & 9.40 & 4.38 & 3.66 \\
  \hline
  \multicolumn{5}{l}{\textit{95\% C.L.}} \\
  Borexino & 10.9 & 15.4 & 9.94 & 5.97 \\
  KamLAND & 4.93 & 3.19 & 2.15 & 2.35 \\
  SNO+ & 0.721 & 0.718 & 0.706 & 0.711 \\
  JUNO & 1.76 & 1.75 & 1.74 & 1.74 \\
  \hline
  BX+KL+SNO+ & 8.19 & 7.11 & 3.67 & 3.01 \\
  All four & 8.26 & 6.74 & 3.67 & 3.06 \\
 \end{tabular}
 \end{ruledtabular}
\end{table}


\subsubsection{Each experiment}

Decay removes geoneutrinos and some reactor $\bar\nu_e$ (Appendix~\ref{sec:reactors}) but leaves the backgrounds untouched. Therefore, an excess of geoneutrinos or reactor $\bar\nu_e$ over their prediction strengthens a limit. A deficit weakens it. On the geochemical Earth, Borexino's data hold 52 geoneutrino events, against 37 predicted, so Borexino sets the strongest limit of the four. KamLAND's data hold 156, against 209 predicted, so its limit is 2.3 times weaker than expected. SNO+'s data hold about as many geoneutrinos as predicted, but above 1.5~MeV, where reactor $\bar\nu_e$ dominate, they hold 148 events, against 166 predicted. This deficit makes SNO+'s limit the weakest of the four, 2.5 times weaker than expected; if only the geoneutrinos decayed, its limit would be 3.9 times stronger.

\subsubsection{Combined experiments}

A single lifetime must describe all the data at once. Thus, the combined limits lie between Borexino's and those of the other experiments. For the decays of $\nu_1$, either combination sets a weaker limit than Borexino alone, on every Earth model (\tabl{limits}), because KamLAND's deficit of geoneutrinos favors decay. The two experiments disagree more than chance makes likely: in only 2 of 200 simulated pairs of them do their limits sit as far apart within their simulated ranges. Adding JUNO changes the combined limits on $\tau_1/m_1$ by 7\% at most (\tabl{limits}), because at these lifetimes decay removes less than 4\% of JUNO's reactor antineutrinos (\figu{reactor_decay}).

\subsubsection{Geoneutrinos vs.~reactor $\bar\nu_e$}

To tell whether the geoneutrinos or the reactor $\bar\nu_e$ set the limit on $\nu_1 \to \nu_3$ from the four experiments combined, we let only one of them decay. On the geochemical Earth, the limit is $1.3\times10^{-8}$~s~eV$^{-1}$ if only the geoneutrinos decay and $1.8\times10^{-9}$~s~eV$^{-1}$ if only the reactor $\bar\nu_e$ do. The limit with both decaying lies near the first (\tabl{limits}). Therefore, on that Earth, the geoneutrinos set the limit. With the radiogenic power free, the first limit weakens to $9.3\times10^{-9}$~s~eV$^{-1}$, because a free power can make up for the geoneutrinos that decay removes. The second stays at $1.8\times10^{-9}$~s~eV$^{-1}$, because the power of each reactor fixes its flux of antineutrinos (Sec.~\ref{sec:reactor_signal}). The limit with both decaying now lies between the two, so the reactor $\bar\nu_e$ count about as much as the geoneutrinos.

\subsubsection{Nuisance parameters}

In Appendix~\ref{app:parameters}, we shift each nuisance parameter by one standard deviation, fit the others again, and compute the limit again. For all four experiments with the radiogenic power free, only 11 of the 73 parameters that we can shift move the limit by more than 10\% (\figu{impacts}). Borexino's reactor flux moves it most: Borexino's fit leaves its reactor rate free, so a lower rate leaves more of its events to geoneutrinos. Next come $\sin^2\theta_{12}$, which sets how much electron flavor $\nu_1$ carries, and the radiogenic power, which the fit raises to make up for decay (Appendix~\ref{app:brazil}). Each of the other eight belongs to a single experiment and can make up for decay only in its own data. The parts of the lithosphere move the limit little.

\subsubsection{Parameter correlations}

Figure~\ref{fig:corner_rows} shows the joint regions of the lifetime and the four nuisance parameters that move most with it, in each fit. Some of these parameters can make up for what decay removes, so their regions extend toward short lifetimes. The clearest are the lithospheric mantle in the combined fit, the energy scale of the $(\alpha,n)$ proton recoils at SNO+, and the rate of geoneutrinos at JUNO. Our one-dimensional limits include these correlations, because at each lifetime the fit lets every nuisance parameter take its best-fitting value (Sec.~\ref{sec:limits}). The four parameters of each fit are nearly independent of each other (correlation coefficients below 0.3 in magnitude), so no pair acts together to make up for decay. Our one-dimensional limits are representative.

\subsubsection{Earth models}

Across the three classes of BSE model, the combined limits change by a factor of three to four (\tabl{limits}). The classes share one crust but differ in their mantle, from the cosmochemical Earth, with the lowest radiogenic power, to the geodynamical, with the highest (Sec.~\ref{sec:prediction}). A hotter Earth predicts more geoneutrinos, so it shrinks any excess in the data and deepens any deficit. Therefore, the combined limits are strongest on the cosmochemical Earth and weakest on the geodynamical. With the power free, they are close to those on the geodynamical Earth, because the fit raises the power to its ceiling, 39.6~TW, to offset decay (Appendix~\ref{app:brazil}).

\subsubsection{Validity}

Figure~\ref{fig:brazil} compares each limit with the limits of simulated experiments, made without decay on the geochemical Earth. Each limit lies within the middle 68\% of its simulated limits, except Borexino's, which is stronger than 98.5\% of them because its data hold more geoneutrinos than predicted. With the radiogenic power free, the combined limits remain typical (Appendix~\ref{app:brazil}). The model fits each experiment and both combinations acceptably, except JUNO. Only 3\% of JUNO's simulated experiments fit worse than its data, but JUNO's own best-fit curve fits them no better, with a $\chi^2$ of 178 against 180 for ours. In the fit of all four, no nuisance parameter lies more than 1.7 prior widths from its prior's center (Appendix~\ref{app:parameters}). Our limits have the coverage they claim (Sec.~\ref{sec:validation}). Therefore, the limits from data are sound.

\subsection{All decay channels}
\label{sec:allchannels}

\begin{figure*}[p]
 \centering
 \includegraphics[width=0.95\textwidth]{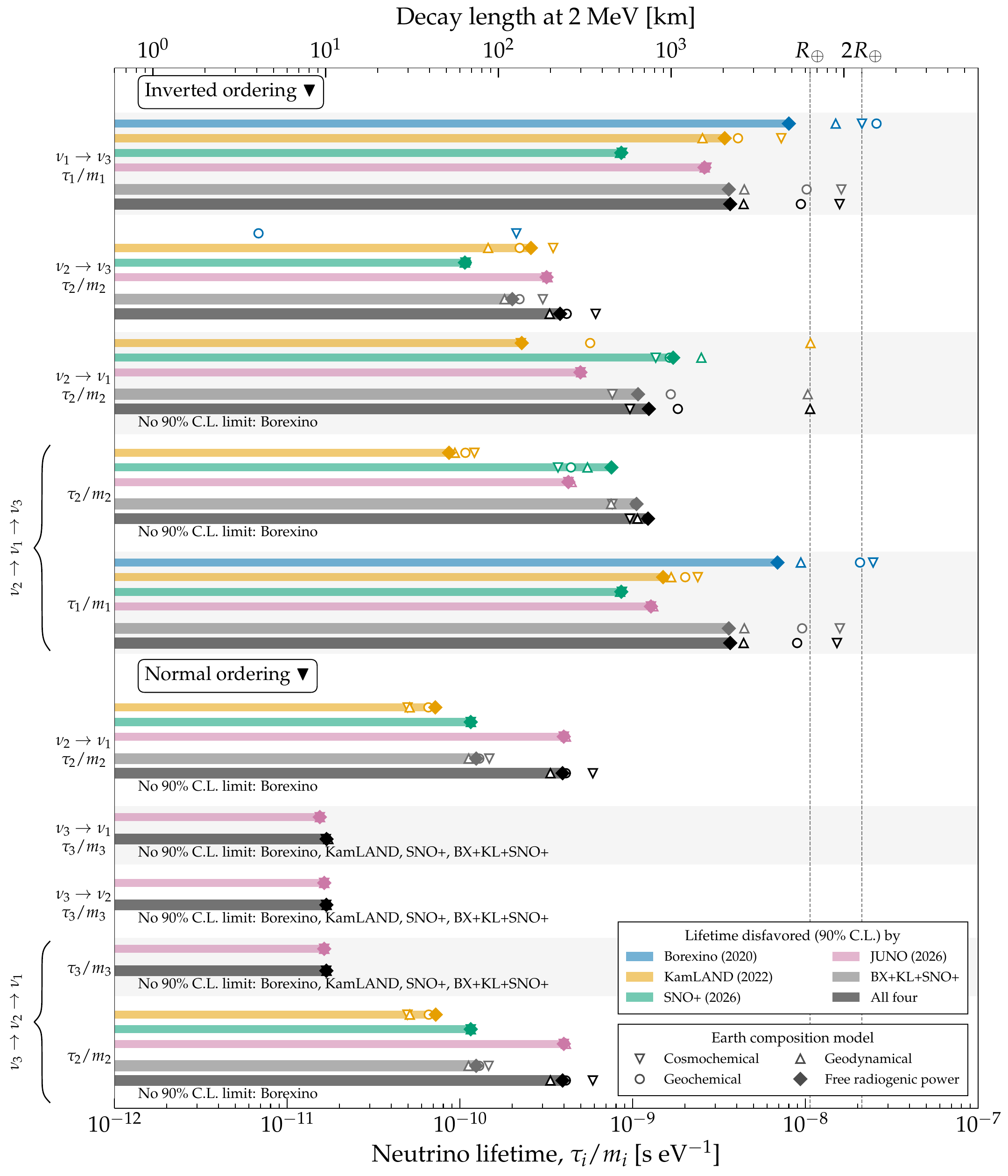}
 \caption{\textbf{Lower limits on the neutrino lifetime-to-mass ratio from every decay channel, experiment, and Earth model.} Each block is one decay channel, or one of the two lifetimes of a chain, inverted mass ordering first. Within a block, each row is one experiment or combination. Its bar runs to the limit with the radiogenic power free. Experiments and combinations that set no limit, or one below the range shown, are named below each block. The top axis shows the decay length at 2~MeV, with the radius and diameter of the Earth marked. Only the decays of $\nu_1$ are limited by every experiment on every Earth model. See Sec.~\ref{sec:allchannels} for details.}
 \label{fig:overview}
\end{figure*}

Figure~\ref{fig:overview} shows the limits for every decay channel, from every experiment and combination, on every Earth model. Only the geoneutrinos depend on the Earth model. Therefore, where the limits on the three classes of BSE model (open markers) lie away from the limit with the radiogenic power free (the end of the bar), the geoneutrinos matter to the limit. In every row of JUNO and of $\nu_3$, the markers lie at the end of the bar. Those limits come from JUNO's reactor antineutrinos. We provide every limit, from data and expected, at 90\% and 95\%~C.L., and the numbers and code that draw every figure of this paper, as supplementary files~\cite{NuDecayGeoReactorRepo}.

\smallskip
\textit{Inverted ordering.}---With the radiogenic power free, the chain $\nu_2 \to \nu_1 \to \nu_3$ limits $\tau_1/m_1$ as strongly as $\nu_1 \to \nu_3$ does. The decays of $\nu_2$ alone are limited more weakly, because $\nu_2$ carries less of the detected flux than $\nu_1$ (Sec.~\ref{sec:channels}). The limit on $\nu_2 \to \nu_3$ comes mainly from JUNO's reactor antineutrinos. For $\nu_2 \to \nu_1$, decay raises the signal of geoneutrinos (Sec.~\ref{sec:signal_io}), so Borexino's excess of them sets no limit; the reactor antineutrinos set it. On the geodynamical Earth, however, this rise deepens KamLAND's deficit of geoneutrinos. Therefore, on that Earth, KamLAND and both combinations limit $\nu_2 \to \nu_1$ more strongly than $\nu_1 \to \nu_3$.

\smallskip
\textit{Normal ordering.}---Here, geoneutrinos test decay only weakly, because $\nu_1$ is stable (Sec.~\ref{sec:signal_no}). With the radiogenic power free, the limit of the four experiments combined equals that of JUNO alone to within 2\%, so it comes from JUNO's reactor antineutrinos.

Geoneutrinos cannot test the decay of $\nu_3$ (Sec.~\ref{sec:signal_no}). JUNO's reactor antineutrinos can, through the ripple in their spectrum (Sec.~\ref{sec:reactor_signal}). The limit is $\tau_3/m_3 > 1.7\times10^{-11}$~s~eV$^{-1}$ at 90\%~C.L., the same in the three channels that decay $\nu_3$, because the ripple depends on how much $\nu_3$ survives and not on its daughters. We read it at a threshold calibrated with simulated experiments, which weakens it by 22\% (Appendix~\ref{app:coverage}).

\subsection{Comparison with existing limits}
\label{sec:comparison}

In \figu{money}, each of our bars is the limit of the four experiments combined, with the radiogenic power free. The published limits rest on other choices of coupling and of detected daughters (Sec.~\ref{sec:decays}) and on other statistics (Appendix~\ref{app:published}). When the daughters carry little electron flavor, as in $\nu_1 \to \nu_3$ and $\nu_2 \to \nu_3$, the limit on $\tau/m$ comes from the disappearance of the parent, so these choices hardly change it~\cite{Porto-Silva:2020gma}.

\smallskip
\textit{Against terrestrial limits.}---Ours are the first terrestrial limits on the visible decay of $\nu_1$. We compare the published terrestrial limits with our 95\%~C.L. limits, which are read at the same threshold (Appendix~\ref{app:published}). An analysis of 207 days of JUNO's data limits the invisible decay of each mass eigenstate~\cite{Beccaria:2026ous}. For $\nu_1$, whose decay into $\nu_3$ is nearly invisible, our limit is 12 times stronger. Besides JUNO's data, ours uses the geoneutrinos, the other three experiments, and a constraint on $\theta_{12}$ from solar neutrinos (Appendix~\ref{app:nufit}), which that analysis leaves out. For $\nu_2$, an analysis of KamLAND's reactor antineutrinos sets a limit 1.7 to 4.5 times stronger than ours~\cite{Porto-Silva:2020gma}. It lets $\nu_1$ decay too, through the same coupling, but quotes no limit for it. Our limit on $\nu_2 \to \nu_3$ agrees to within 7\% with JUNO's limit on the invisible decay of $\nu_2$, since ours comes mainly from the same data. For $\nu_3$, our limit is 0.95 to 1.3 times the limits on its visible decay from accelerator neutrinos~\cite{Gago:2017zzy} and about 1.25 times stronger than JUNO's limit on its invisible decay. Long-baseline accelerator neutrinos limit its invisible decay more strongly, to $\tau_3/m_3 > 2.4\times10^{-11}$~s~eV$^{-1}$ at 90\%~C.L.~\cite{Ternes:2024qui}, and to $2.9\times10^{-10}$ with atmospheric neutrinos~\cite{Gonzalez-Garcia:2008mgl}.

\smallskip
\textit{Against astrophysical limits.}---Astrophysical neutrinos travel astronomical distances, so their limits are $3\times10^5$ to $3\times10^{14}$ times stronger than ours, depending on the channel~\cite{KamLAND:2003gfh, Picoreti:2021yct, deGouvea:2023jxn, Ivanez-Ballesteros:2023lqa, Bustamante:2020niz}. One analysis of high-energy astrophysical neutrinos disfavors the complete decay of the two heavier mass eigenstates at about two standard deviations~\cite{Pagliaroli:2015rca}. Another prefers the invisible decay of $\nu_2$ and $\nu_3$ at $\tau/m \approx 10^2$~s~eV$^{-1}$, at 2.8--$3.4\sigma$, because it removes a tension between IceCube's track and cascade samples under one model of the sources~\cite{Denton:2018aml}.

\smallskip
\textit{Against cosmological limits.}---Cosmology limits the same decays, because decays and inverse decays would keep the relic neutrinos from streaming freely, which the CMB constrains~\cite{Chen:2022idm}. For a parent mass of 0.05~eV, Planck data give $\tau/m \gtrsim 10^7$~s~eV$^{-1}$ for a single decay across the atmospheric mass splitting, such as $\nu_1 \to \nu_3$, and $\tau/m \gtrsim 10^4$~s~eV$^{-1}$ for $\nu_2 \to \nu_1$~\cite{Chen:2022idm}. These limits rest on the standard cosmological model, not on detected neutrinos, so \figu{money} leaves them out. The cosmological literature calls these decays invisible, because they emit no photons; in our sense they are visible.

\section{Summary and outlook}
\label{sec:summary}

Geoneutrinos, the $\bar\nu_e$ from uranium and thorium decays inside the Earth, have low energies and travel long distances, which makes them sensitive to neutrino decay. To our knowledge, this work is the first to test neutrino decay with them. We have computed how decay removes geoneutrinos at the Borexino, KamLAND, SNO+, and JUNO detectors, and how the daughter neutrinos, which are also detected, return part of them. We have done the same for the reactor antineutrinos that these experiments detect. We fitted the data of each experiment, and of all four, after validating that our methods reproduce each experiment's own published analysis, without decay.

The data show no evidence of decay that holds across Earth models, so we set lower limits on the ratio of each neutrino's lifetime to its mass, $\tau/m$. KamLAND's data prefer decay on the geodynamical Earth, at $2.7\sigma$, since this Earth model predicts more geoneutrinos than KamLAND detects. However, when the radiogenic power is left free, the preference disappears (Sec.~\ref{sec:results}).

Geoneutrinos and reactor $\bar\nu_e$ are detected only through their electron flavor, of which $\nu_1$ carries the most. Therefore, our strongest limits are on the decay of $\nu_1$. It can decay only into $\nu_3$, which the detectors barely see, and only if $\nu_3$ is the lightest neutrino. The four experiments together give, at 90\%~C.L., $\tau_1/m_1 > 3.7\times10^{-9}$~s~eV$^{-1}$ conservatively, \ie, with the Earth's radiogenic power free, and up to $1.6\times10^{-8}$~s~eV$^{-1}$ on an Earth model. \textbf{These are the first terrestrial limits on the visible decay of $\nu_1$.} On the geochemical Earth, the limit comes from the geoneutrinos. With the power free, the geoneutrinos and the reactor $\bar\nu_e$ count about equally. The most similar existing limit, from JUNO's reactor neutrinos on the invisible decay of $\nu_1$, is 12 times weaker than ours at 95\%~C.L.~(Sec.~\ref{sec:comparison}).

Our limits on the decays of $\nu_2$ and $\nu_3$ are weaker. With the radiogenic power free, they come from reactor antineutrinos, mainly JUNO's; only in the chain $\nu_2 \to \nu_1 \to \nu_3$ do the geoneutrinos strengthen the limit on $\nu_2$, threefold. The limit on $\nu_3$ rests on the fast oscillations in JUNO's spectrum, which the decay of $\nu_3$ would damp. These limits are close to the limits on invisible decay already published from the same JUNO data.

More data will strengthen the limits. JUNO and SNO+ continue to take data. In its first 207 days, JUNO alone detected more geoneutrinos than Borexino, KamLAND, and SNO+ combined~\cite{JUNO:2026tap}. Its reactor neutrinos will also measure precisely the oscillation parameters that now move its own limit by more than a factor of two~\cite{JUNO:2022mxj}. The Jinping Neutrino Experiment, far from nuclear reactors, is expected to detect twice as many geoneutrinos as reactor antineutrinos~\cite{Meng:2026avo}. Better models of the crust near each detector~\cite{Meng:2026avo} will sharpen the prediction against which decay is tested. Combining geoneutrino data from several experiments also constrains how much heat uranium and thorium release inside the Earth, the largest uncertainty in our limits~\cite{Fogli:2010vx, Fiorentini:2012yk, Leyton:2026gev, Bustamante:inprep}. Beyond geoneutrinos, neutrinos from supernovae~\cite{Lindner:2001th, deGouvea:2019goq, Roux:2024zsv} and high-energy astrophysical neutrinos~\cite{Beacom:2002vi, Baerwald:2012kc, Bustamante:2016ciw, Song:2020nfh} could test values of $\tau/m$ many orders of magnitude larger.

\begin{acknowledgments}
MB is supported by {\sc Villum Fonden} under project no.~29388. This work used the Tycho supercomputer hosted at the SCIENCE High Performance Computing Center at the University of Copenhagen. During the preparation of this work, the author used Claude for research assistance in code development, numerical auditing, figure preparation, and editorial revision. All scientific judgments, validation decisions, and responsibility for the manuscript remain with the author.
\end{acknowledgments}

\FloatBarrier
\appendix
\counterwithin*{figure}{section}
\counterwithin*{table}{section}
\renewcommand{\thefigure}{\thesection.\arabic{figure}}
\renewcommand{\thetable}{\thesection.\arabic{table}}

\section{Calculating the geoneutrino signal}
\label{app:signal}

\subsection{Earth reservoirs and their abundances}
\label{app:earth}

The crust is a shell 35~km thick. The lithospheric mantle is a shell beneath it, down to a depth of 91~km, which leaves the mantle below it with its mass of $3.91\times10^{24}$~kg~\cite{Bellini:2021sow}. The mantle extends from there down to the core, at a radius of 3\,480~km. The real lithospheric mantle lies only under the continents, where it is about 140~km thick~\cite{Bellini:2021sow}; our shell has nearly its mass, $9.4\times10^{22}$~kg against $9.7\times10^{22}$~kg, spread over the whole Earth. This shell changes only where the geoneutrinos of the lithospheric mantle come from, since we calibrate its signal at each site (Sec.~\ref{sec:calibration}); that signal is at most 7\% of the total, at any site and for any class of BSE model.

We derive the abundances of uranium and thorium from the power of each class of BSE model (Sec.~\ref{sec:prediction}). The power is the heat that the uranium and thorium of the BSE release per unit time, to which each element contributes in proportion to its abundance, in kilograms of the element per kilogram of rock. Each kilogram of uranium releases $9.8\times10^{-5}$~W, and each kilogram of thorium $2.6\times10^{-5}$~W, counting the decays of every nuclide in its chain~\cite{Bellini:2021sow}. The power alone does not fix the abundances, since more uranium can make up for less thorium.

However, their ratio, Th/U, is known better than the abundances. Both elements condense only at high temperatures and stay out of the core, so neither the loss of volatile elements nor core formation changed Th/U as the Earth formed: the BSE as a whole, crust and mantle together, keeps the value of chondrites, 3.9~\cite{McDonough:1995, Bellini:2013wsa}. With this ratio, each power fixes both abundances. The abundances set how many geoneutrinos the BSE emits; their ratio sets the shape of the spectrum, since only the uranium chain emits above 2.3~MeV (\figu{chains}).

The crust and the lithospheric mantle hold more thorium per uranium: Th/U = 4.4 in the crust~\cite{Huang:2013ozo, Borexino:2019gps} and 5.0 in the lithospheric mantle~\cite{Borexino:2019gps}. The mantle holds the rest of the BSE's uranium and thorium, so its Th/U is lower and depends on the power (\tabl{bse}).

\subsection{Spectra of the uranium and thorium chains}
\label{app:spectra}

We take the endpoint and the share of the decays of each branch from the nuclear data compiled in \Ref~\cite{Li:2024mqz}. The tables of the same reference, from which we take the spectrum of each chain (Sec.~\ref{sec:origin}), include the shape factors of the first-forbidden branches, the Coulomb attraction between the electron and the daughter nucleus, and radiative, finite-size, and weak-magnetism corrections. These tables give only the total spectrum of each chain. For the single nuclides in \figu{chains}, we give every branch the allowed shape and correct it for the Coulomb attraction with the relativistic Fermi function~\cite{Hayen:2017pwg}. In these shapes, the branches of $^{234}$Pa$^m$ and $^{212}$Bi give 69\% and 92\%, respectively, of the antineutrino yield of the uranium and thorium chains above the IBD threshold.

Each contribution ends abruptly at the endpoint of its highest branch, where the $\bar\nu_e$ takes nearly all of the energy and the electron is slow. There, the Coulomb attraction of the daughter nucleus, which the Fermi function describes, enhances the decay rate and keeps the spectrum from falling to zero. The detectable $\bar\nu_e$ come from this high-energy end: only about a quarter of those of the branches of $^{234}$Pa$^m$ and $^{212}$Bi lie above the threshold. For daughter nuclei this heavy, with atomic numbers from 82 to 92, the Coulomb field is too strong for the nonrelativistic approximation to the Fermi function; the relativistic form raises the yield above threshold by about a third.

\subsection{Oscillations in matter}
\label{sec:matter}

Inside the Earth, the electrons of the rock add a potential energy, $-\sqrt{2}\, G_F n_e$, to the $\bar\nu_e$ component of an antineutrino, where $G_F$ is the Fermi constant and $n_e$ is the number density of electrons~\cite{Wolfenstein:1977ue, Mikheyev:1985zog}. The potential comes from charged-current forward scattering off the electrons, which the other flavors lack. Along a path, the amplitudes of the mass eigenstates, $a_i$, evolve as
\begin{multline}
 i\,\frac{da_i}{dx}
 =
 \left( \frac{m_i^2}{2E_\nu} - i\,\frac{m_i}{2E_\nu \tau_i} \right) a_i
 \\
 - \sqrt{2}\, G_F\, n_e(x)\, U_{ei} \sum_{j=1}^{3} U_{ej}^* a_j \;,
 \label{equ:evolution}
\end{multline}
where $x$ is the distance traveled from the production point, at which $a_i = U_{ei}$. The first term gives each $\nu_i$ its phase and its decay, as in \equ{amplitude}. The second term acts on the $\bar\nu_e$ component, $\sum_j U_{ej}^* a_j$, so it couples the mass eigenstates to one another. At the detector, a distance $L$ from the production point, the amplitude to detect a $\bar\nu_e$ is $\mathcal{A} = \sum_i U_{ei}^* a_i(L)$. Without matter, each $a_i$ evolves on its own, which gives \equ{amplitude}.

Along a path through the Earth, the density varies, so \equ{evolution} has no closed form. In PREM, the density varies smoothly within each layer and changes abruptly from one layer to the next, most at the core-mantle boundary, from 5.6 to 9.9~g~cm$^{-3}$~\cite{Dziewonski:1981xy}. The number density of electrons is $n_e = Y_e \rho / m_u$, where $\rho$ is the density, $m_u$ is the atomic mass unit, and $Y_e$ is the number of electrons per nucleon, which depends on the composition of the rock. The value of $Y_e$ jumps at the core-mantle boundary too, from 0.4656 in the core, taken as iron, to 0.4957 in the mantle, to 0.4952 in the crust~\cite{Bustamante:2026yce}. PREM's outermost 3~km are ocean, over 12~km of upper crust. Since the detectors are on land, we give both layers the density of PREM's lower crust below them, 2.9~g~cm$^{-3}$. Keeping PREM's upper crust, at 2.6~g~cm$^{-3}$, changes the expected number of events by less than 0.1\%.

At geoneutrino energies, the matter term of \equ{evolution} is 0.5--2\% of $\Delta m^2_{21}/(2E_\nu)$ and changes slowly along each path. Therefore, the propagation is adiabatic: each eigenstate of \equ{evolution} at the production point turns into the corresponding eigenstate at the detector, without transitions to the others. Averaged over the fast oscillations, the survival probability then depends only on the mixing in matter at the production point and at the detector. We check this against the code \texttt{Magnus}, which solves \equ{evolution} without its decay term along each path~\cite{Bustamante:2026yce}: along paths through the crust and the mantle, the two agree to $1.4\times10^{-5}$.

Near the detector, within a few oscillation lengths of $\Delta m^2_{21}$, the interference terms do not average out, so the survival probability also depends on the phases accumulated along the path. Most of these geoneutrinos are produced in the crust around the detector and cross only that crust. Therefore, we compute their interference terms from \equ{psurv}, with the mixing and the mass-squared splittings replaced by their values in the matter of the crust at the detector. Decay acts on the mass eigenstates in vacuum; against a direct numerical solution of \equ{evolution} with its decay term, this changes the survival probability by less than $10^{-3}$.

In \equ{flux}, matter changes only the arrival operator. Along the path fixed by $L$ and $\psi$, the operator becomes
\begin{multline}
 \mathcal{T}_{L,\psi}[F](E_\nu)
 =
 |\mathcal{A}(E_\nu)|^2\, F(E_\nu)
 \\
 + \int_{E_\nu}^{x_{ij}^2 E_\nu} dE_\nu^\prime\,
 F(E_\nu^\prime)\, W_{ij}(E_\nu \vert E_\nu^\prime)
 \\
 \times
 \int_0^L dx\, \frac{m_i}{E_\nu^\prime \tau_i}\, |a_i(x, E_\nu^\prime)|^2\, |\mathcal{A}_j(x, E_\nu)|^2 \;.
 \label{equ:arrival_matter}
\end{multline}
The first term counts the survivors. The second counts the daughters: a parent $\bar\nu_i$ of energy $E_\nu^\prime$ decays between $x$ and $x + dx$ with probability $|a_i(x, E_\nu^\prime)|^2\, m_i\, dx/(E_\nu^\prime \tau_i)$. Its daughter, a $\bar\nu_j$ of energy $E_\nu$, is detected as a $\bar\nu_e$ with amplitude $\mathcal{A}_j(x, E_\nu)$, which we compute like $\mathcal{A}$, but starting at $x$ with $a_k = \delta_{jk}$. In vacuum, where $|\mathcal{A}|^2 = \Pee$, $|a_i(x, E_\nu^\prime)|^2 = |U_{ei}|^2 \exp[-x\, m_i/(E_\nu^\prime \tau_i)]$, and $|\mathcal{A}_j|^2 = |U_{ej}|^2$, \equ{arrival_matter} becomes \equ{arrival}. (The daughters of the two decays of a chain can also interfere with each other~\cite{Abdullahi:2020rge}, but the phase between them depends on how far the antineutrino travels before it decays, oscillating with a period of 2--3~km, so their interference averages out for the typical baselines of decaying geoneutrinos. We leave it out, which changes the geoneutrino signal at each detector by less than $5\times10^{-5}$.)

In their own analyses, Borexino and SNO+ compute the survival probability of geoneutrinos in vacuum~\cite{Borexino:2019gps, SNO:2025chx}, as KamLAND did in 2013~\cite{KamLAND:2013rgu}. The latest analyses of KamLAND and JUNO do not say whether they include matter~\cite{KamLAND:2022vbm, JUNO:2026tap}. Borexino and KamLAND estimate that matter changes the predicted signal far less than their other uncertainties~\cite{Borexino:2019gps, KamLAND:2013rgu}. We include matter nonetheless, to make our calculation as accurate as we can. Relative to assuming propagation in vacuum, matter changes our limits on $\nu_1 \to \nu_3$ by at most 2.3\%.

Within the local crust, the size of the interference terms depends on how uranium and thorium are distributed within about 100~km of the detector. Our calibration fixes the total signal of the local crust at each site (Sec.~\ref{sec:prediction}), but not where within the local crust that signal comes from. We spread uranium and thorium uniformly through it. In their own analyses of geoneutrinos, KamLAND and SNO+ average the oscillations~\cite{KamLAND:2022vbm, SNO:2025chx}; Borexino keeps them~\cite{Borexino:2019gps}. JUNO does not say which it does. Its 5\% uncertainty on the shape of the geoneutrino spectrum covers, among other effects, the distortions by oscillations that depend on where uranium and thorium lie~\cite{JUNO:2026tap}. In our analysis of neutrino decay, we use the full oscillatory probability for all four experiments. Averaging the probability everywhere would change our 90\%~C.L.~limits on $\nu_1 \to \nu_3$, from each experiment and from both combinations, by at most 1.3\%; it would make all of them slightly stronger except JUNO's, which it would weaken by 0.1\%.

\subsection{Calibrating the lithosphere}
\label{app:calibration}

We calibrate the lithosphere at each site in three independent parts, as \Ref~\cite{Bellini:2021sow} does (Sec.~\ref{sec:calibration}). The local crust comes from a model of the $6^\circ \times 4^\circ$ region around the site~\cite{Borexino:2019gps, Bellini:2013wsa, Bellini:2021sow, Semenec:2023, JUNO:2025sfc}, which our local crust fills. That region is centered, or nearly so, on the detector, except at SNO+, whose model stops at $47^\circ$N, 58~km north of it~\cite{Semenec:2023}. The far crust and the lithospheric mantle come from one global model, which covers all four sites~\cite{Wipperfurth:2019idn}; at SNO+, they come from that model computed around SNO+'s region~\cite{Semenec:2023}. At KamLAND, the far crust comes from \Ref~\cite{Huang:2013ozo}, whose region fits that of KamLAND's local model~\cite{Bellini:2021sow}. \tabl{experiments} lists the three signals by site.

We compute the scale factors with the spectra of the uranium and thorium chains that the published models used, so that the factors fix how much uranium and thorium each part holds. The exception is KamLAND, whose models of the local and the far crust used other spectra~\cite{Bellini:2013wsa, Huang:2013ozo}. We calibrate these two models with the spectra of the other models as well, which gives KamLAND's crust about 3\% less signal than their own spectra would. Our own spectra (Sec.~\ref{sec:origin}) differ from both sets. Our survival probability is also 1.7\% lower than the one that \Ref~\cite{Wipperfurth:2019idn} assumed for the far crust and the lithospheric mantle. Therefore, our lithosphere gives about 5\% fewer geoneutrinos than the published models.

\subsection{Binned event rates}
\label{app:binned}

We evaluate \equ{rate} on a grid of true energies, $E_k$, and distances from the detector, $L_l$. On this grid, it reads
\begin{multline}
 N_i^{(e,d)}
 =
 \sum_k R_{ik}^{(e,d)}
 \bigg[
 \sum_{r,c} \nu_{rc}^{(e,d)} f_{rc}^{(e)}
 \\
 \times \sum_l \langle \Pee^{(e)} \rangle_{rc}(L_l, E_k)\, K_{rc}^{(e,d)}(L_l, E_k)
 \\
 + \nu_{\rm reac}^{(e,d)} B_{{\rm reac},k}^{(e,d)}
 \bigg]
 + \sum_b \nu_b^{(e,d)} B_{b,i}^{(e,d)} \;.
 \label{equ:rate_binned}
\end{multline}
Here, $R_{ik}^{(e,d)}$ is the probability that an IBD event in bin $k$ of true energy is selected in bin $i$ of prompt energy, \equ{response} averaged over bin $k$ and integrated over bin $i$. The number of reactor events in bin $k$ of true energy, $B_{{\rm reac},k}^{(e,d)}$, is \equ{reactor} integrated over the bin. The backgrounds, $B_{b,i}^{(e,d)}$, stay as in \equ{rate}, since they are already binned in prompt energy. The number of IBD events that reservoir $r$ and chain $c$ would produce in data set $(e,d)$, in bin $l$ of distance and bin $k$ of true energy, without oscillations or decay, is
\begin{multline}
 K_{rc}^{(e,d)}(L_l, E_k)
 =
 (N_p T)^{(e,d)} A_{rc}
 \int_k dE_\nu\, \sigma_{\rm IBD}(E_\nu)\, \frac{dn_c}{dE_\nu}
 \\
 \times
 \int_{V_{rl}^{(e)}} dV\, \frac{\rho}{4\pi L^2} \;,
 \label{equ:kernel}
\end{multline}
where the first integral runs over energy bin $k$; the second, over $V_{rl}^{(e)}$, the part of reservoir $r$ in distance bin $l$ from the detector of experiment $e$. We compute $K_{rc}^{(e,d)}$ once per data set. Each decay hypothesis then weights it with $\langle \Pee^{(e)} \rangle_{rc}(L_l, E_k)$, the survival probability averaged over the paths to the detector and the energies within the two bins, each weighted by its contribution to $K_{rc}^{(e,d)}$. With visible decay, the bracket also includes the daughters, binned and averaged in the same way, with $\sigma_{\rm IBD}$ taken at the daughter's energy.

If $\langle \Pee^{(e)} \rangle_{rc}$ is the exact average within each pair of bins, \equ{rate_binned} equals \equ{rate}, so the grids matter only through how well we compute this average. The grid of true energy has 220 points, evenly spaced from the IBD threshold to 10~MeV; at JUNO, whose resolution brings out the endpoints of the beta branches, it has 880. The grid of distance has bins of 5~km out to 400~km from the detector, where the local crust lies, and 160 bins of 77~km beyond, out to the diameter of the Earth. Doubling the grid of true energy changes our limits by at most 0.7\% at 90\%~C.L.; refining that of distance, by at most 0.03\%.
\section{Modeling the experiments}
\label{app:reproduction}

\begin{figure}[tp]
 \centering
 \includegraphics[width=\columnwidth]{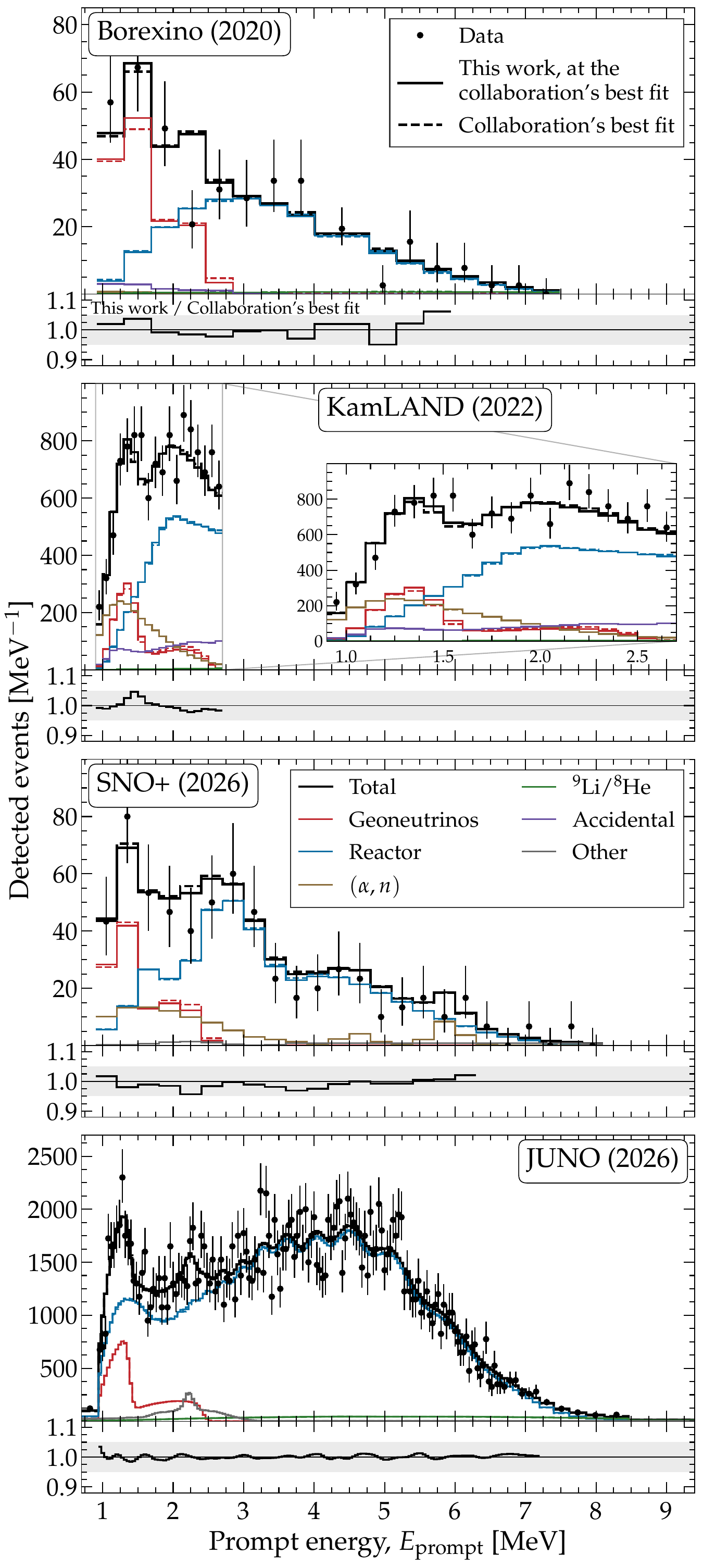}
 \caption{\textbf{Our calculation of the detected spectra, compared with each collaboration's.} Each component of our calculation is at the normalization that the collaboration's best fit gives it, and uses our own spectral shapes throughout. The data are from \Refs~\cite{Borexino:2019gps, KamLAND:2022vbm, SNO:2025chx, JUNO:2026tap}. Below each panel is the ratio of our total to the collaboration's, where the latter holds at least a tenth of its peak; the shaded band is $\pm 5\%$. Our calculation reproduces the four collaborations' spectra to 0.5--2.4\% per bin on average.}
 \label{fig:reproduction}
\end{figure}

\begin{table*}[t!]
 \caption{\label{tab:response}\textbf{Detector response of each data set.} The parameters of \equ{response}: the nonlinearity of the energy scale, $g$; the offset of the energy axis, $\delta$; the coefficients of the resolution, $a$ and $b$; and the selection efficiency, $\varepsilon$. Values marked with an asterisk we fit to the published spectra in the last column, which the fits reproduce to 0.6--1.9\% (Borexino), 0.3--1.6\% (KamLAND), and 0.6--2.3\% (SNO+) of their peaks per bin; unmarked values are as the collaborations report them. Borexino, KamLAND, and SNO+ quote resolutions of the form $a/\sqrt{E}$, so we take $b = 0$ for them. At Borexino, $\varepsilon$ is 0.870 for geoneutrinos and 0.90 for reactor antineutrinos. At KamLAND, it depends on prompt energy; the table gives its maximum within the range of prompt energy that we fit. At SNO+, a classifier against $(\alpha, n)$ events multiplies it by 0.86 to 0.95 below 3.5~MeV, values that we also infer from SNO+'s spectra. See Sec.~\ref{sec:response} for details.}
 \centering
 \begin{ruledtabular}
 \begin{tabular}{llccccll}
  Experiment & Data set & $g$ & $\delta$ [MeV] & $a$ [\%~MeV$^{1/2}$] & $b$ [\%] & $\varepsilon$ & Spectra fitted \\
  \hline
  Borexino & & 1* & $-0.144$* & 8.8* & 0 & 0.870, 0.90~\cite{Borexino:2019gps} & $\U$, $\Th$, reactor templates~\cite{Borexino:2019gps} \\
  KamLAND & Period~1 & 1* & 0* & 6.9* & 0 & 0.88~\cite{KamLAND:2013rgu} & $\U$ and $\Th$ spectra~\cite{KamLAND:2022vbm} \\
   & Period~2 & 1* & 0* & 7.1* & 0 & 0.84~\cite{KamLAND:2013rgu} & Same \\
   & Period~3 & 1* & 0* & 7.9* & 0 & 0.64~\cite{KamLAND:2013rgu} & Same \\
  SNO+ & I & Slope 0.991* & $-0.087$* & 7.4* & 0 & 0.75* & IBD spectra~\cite{SNO:2025chx} \\
   & II & Slope 1.000* & $-0.080$* & 6.7* & 0 & 0.76* & Same \\
  JUNO & & Measured~\cite{JUNO:2026tap} & 0 & 2.9~\cite{JUNO:2026tap} & 1.0~\cite{JUNO:2026tap} & 0.740~\cite{JUNO:2026tap} & None \\
 \end{tabular}
 \end{ruledtabular}
\end{table*}

For each experiment, we reproduce its collaboration's analysis: the data, the exposure and selection efficiency in each period of data taking, the detector response, the reactor flux over time, every background that the collaboration fits, and the free parameters of its fit, with their priors (\tabl{parameters}). Below, we describe each experiment and show that, without decay, our model recovers the collaboration's best fit.

Figure~\ref{fig:reproduction} shows our calculation of the detected spectra without decay, compared with the best fit of each collaboration, component by component. Since each component holds as many events as in the collaboration's fit, the comparison tests its shape, set by the fluxes, the cross section, and the detector response. The totals agree with the collaborations' to 2.4\% per bin on average at Borexino, 1.4\% at KamLAND, 1.5\% at SNO+, and 0.5\% at JUNO. For Borexino and KamLAND, the limits use the collaborations' own spectral templates (Sec.~\ref{sec:reproduction}), so these differences do not affect their limits.

KamLAND's spectra have features that our calculation cannot reproduce: its uranium spectrum near 1.45~MeV, for example, changes from one period of data taking to the next. For the geoneutrinos, we take the proportion by which decay and the mixing parameters change each bin relative to our spectrum with a survival probability that is the same at every energy. Without decay, however, our survival probability varies slightly with energy, through the oscillations near the detector and the electrons of the rock. Therefore, the uranium and thorium spectra that we use for KamLAND differ from its own by up to 0.9\% per bin. We have not measured this difference for Borexino, whose spectra we use the same way.

\subsection{Detector response}
\label{sec:response}

The response function of \equ{rate} follows from the kinematics of IBD and from the energy scale and the resolution of each detector. An antineutrino of energy $E_\nu$ yields a positron at an angle $\theta$ to its direction, distributed in $\cos\theta$ as the IBD cross section, $\sigma_{\rm IBD}$, is~\cite{Vogel:1999zy}. The deposited energy, $E_{\rm dep}(E_\nu, \theta)$, depends on $\theta$ only through the kinetic energy of the recoiling neutron, $T_n$ (Sec.~\ref{sec:ibd}). The mean prompt energy is $\bar{E}_{\rm prompt} = s\,[\,g(E_{\rm dep})\,E_{\rm dep} + \delta\,]$, where $g$ describes the nonlinearity of the energy scale, $\delta$ is an offset of the energy axis, and $s \approx 1$ is the energy scale, which the fits vary within its uncertainty. About this mean, $E_{\rm prompt}$ follows a normal distribution, $\mathcal{N}$, of width $\sigma_E = (a^2 \bar{E}_{\rm prompt} + b^2 \bar{E}_{\rm prompt}^2)^{1/2}$. Each collaboration describes its response with these same elements, an energy scale, nonlinear in general, and a resolution~\cite{Borexino:2019gps, KamLAND:2013rgu, SNO:2025chx, JUNO:2024jaw}. Altogether,
\begin{align}
 & R^{(e,d)}(E_{\rm prompt}, E_\nu)
 =
 \varepsilon^{(e,d)}(E_{\rm prompt})
 \label{equ:response}
 \\
 & \quad \times
 \int_{-1}^{1} d\cos\theta\, \frac{1}{\sigma_{\rm IBD}} \frac{d\sigma_{\rm IBD}}{d\cos\theta}\,
 \mathcal{N}\!\left[E_{\rm prompt};\, \bar{E}_{\rm prompt}, \sigma_E\right] ,
 \notag
\end{align}
where $\varepsilon^{(e,d)}$ is the selection efficiency. 

\tabl{response} lists the values of $g$, $\delta$, $a$, $b$, and $\varepsilon^{(e,d)}$ for each data set, with their sources. JUNO publishes them. Borexino, KamLAND, and SNO+ quote only a resolution, not the response that their analyses use. We infer that response by folding our spectra in true energy through it and adjusting its resolution and energy scale until they reproduce the published spectra. For example, the thorium spectrum ends sharply at 2.25~MeV of neutrino energy. In KamLAND's published spectrum, the width of this smeared edge gives the resolution; its position gives the energy scale.

\begin{figure*}[t!]
 \centering
 \includegraphics[width=\textwidth]{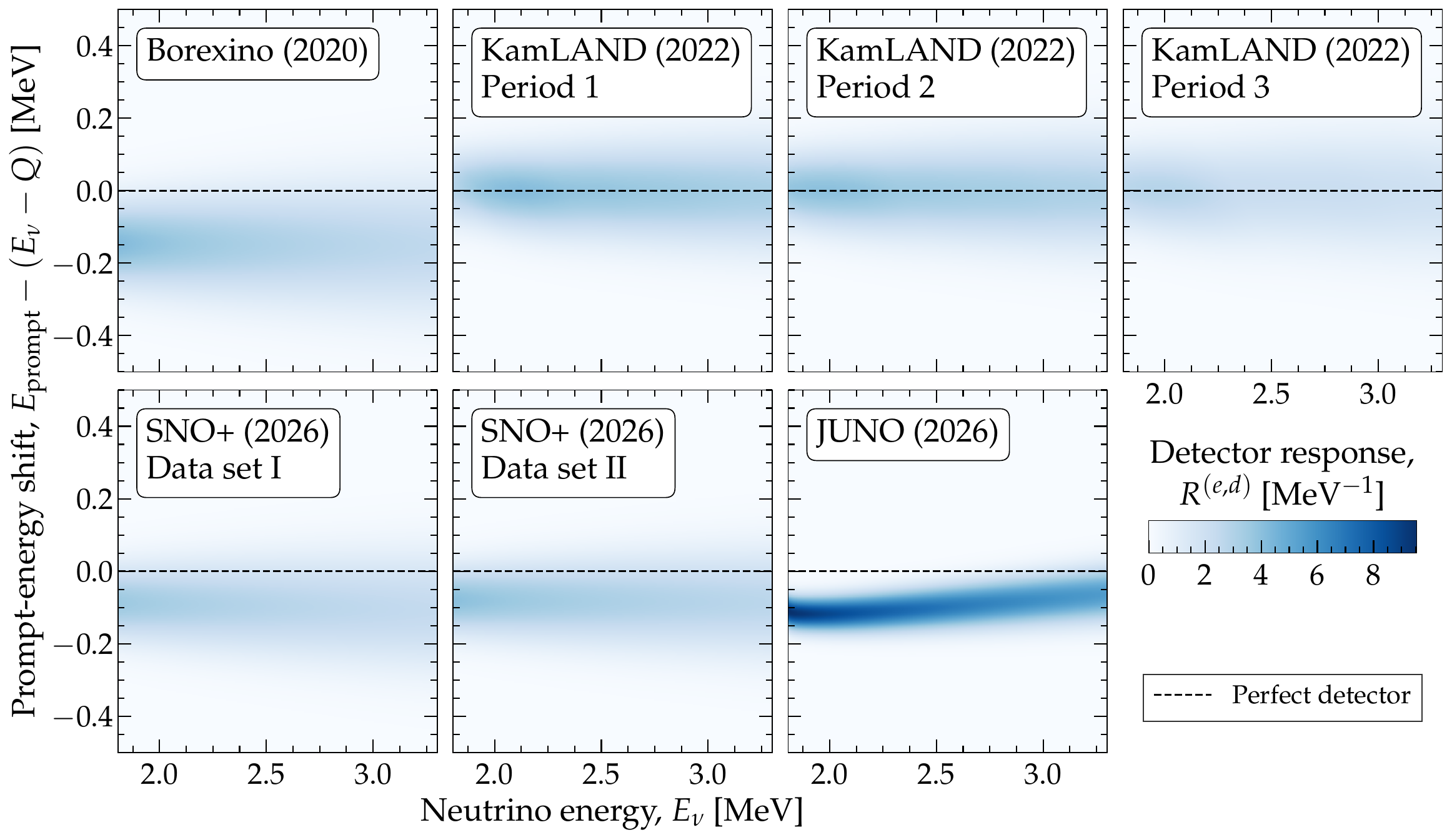}
 \caption{\textbf{Response function of each data set.} Each panel shows the response function of \equ{response}, at the nominal energy scale, against the neutrino energy and the shift of the prompt energy from $E_\nu - Q$. A perfect detector would place every event on the dashed line, up to the recoil energy of the neutron, at most about 15~keV. The color includes the selection efficiency, which at Borexino is that of geoneutrinos. See Sec.~\ref{sec:response} for details.}
 \label{fig:response}
\end{figure*}

Figure~\ref{fig:response} shows the response function of each data set, \equ{response}, at the nominal energy scale, $s = 1$. Its vertical axis is the shift of the prompt energy from $E_\nu - Q$. For a perfect detector, whose prompt energy equals the deposited energy, the shift is $-T_n$ (Sec.~\ref{sec:ibd}). In each panel, the response forms a band centered at the shift
\begin{equation}
 \bar{E}_{\rm prompt} - (E_\nu - Q) = \left[ g(E_{\rm dep}) - 1 \right] E_{\rm dep} + \delta - T_n ,
 \label{equ:response_shift}
\end{equation}
so the band's position shows the detector's energy scale. The width of each band is the resolution, $\sigma_E$, which grows with energy. Since the color includes the selection efficiency, a band is darker where it is narrower or where the efficiency is higher. KamLAND's bands fade at low energy, where its efficiency falls.

\subsection{Reactor antineutrinos}
\label{sec:reactors}

\begin{figure*}[t!]
 \centering
 \includegraphics[width=\textwidth]{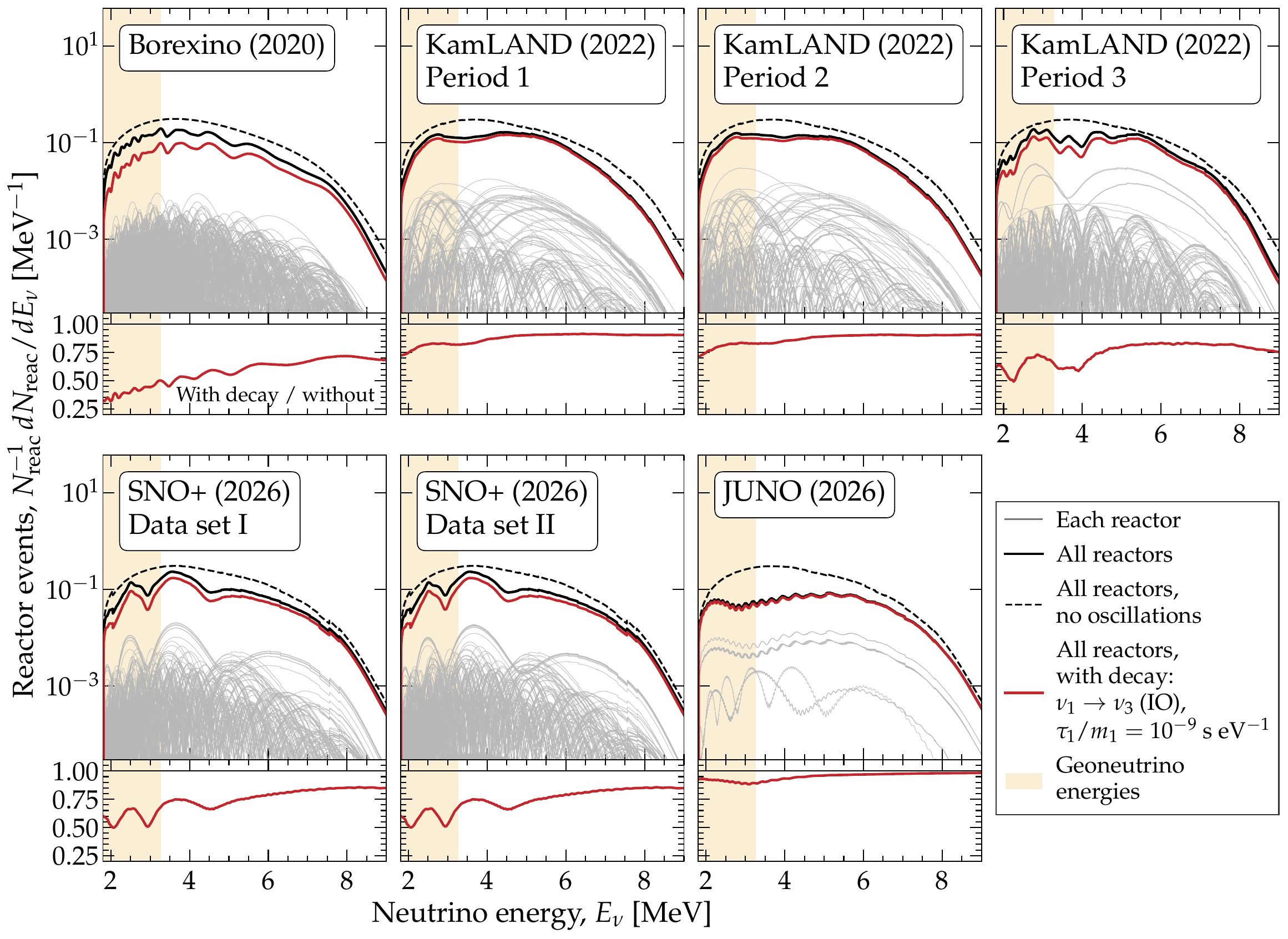}
 \caption{\textbf{Reactor antineutrinos in each data set, reactor by reactor, with and without decay.} Each panel shows, for one data set, the spectrum of reactor IBD events in neutrino energy, \equ{reactor}, before the detector response, divided by the number of reactor events without oscillations or decay, $N_{\rm reac}$. The mixing parameters are at the best fit of NuFIT~6.1 for the inverted ordering~\cite{Esteban:2024eli, NuFIT61}. Each reactor core oscillates at its own distance; where a few distances dominate, the oscillations produce dips in the spectrum, some of them below 3.27~MeV, where the geoneutrinos lie. The strip below each panel is the ratio of the spectra with and without the decay $\nu_1 \to \nu_3$, at $\tau_1/m_1 = 10^{-9}$~s~eV$^{-1}$. It is lowest where the reactors are far and at the dips of the oscillations, which decay deepens. See Sec.~\ref{sec:reactors} for details.}
 \label{fig:reactor_spectra}
\end{figure*}

The spectrum of reactor IBD events in \equ{rate} is a sum over the reactor cores that contribute at the site of experiment $e$, labeled $\kappa$:
\begin{equation}
 \frac{dN_{\rm reac}^{(e,d)}}{dE_\nu}
 \propto
 \sigma_{\rm IBD}(E_\nu) \sum_\kappa
 \frac{\bar{P}_\kappa^{(e,d)}}{4\pi L_\kappa^2\, \bar{E}_{{\rm f},\kappa}}\,
 \mathcal{T}_{L_\kappa}\!\left[ S_\kappa \right]\!(E_\nu) \;.
 \label{equ:reactor}
\end{equation}
Here, $\bar{P}_\kappa^{(e,d)}$ is the thermal power of core $\kappa$, averaged over the time of data taking of data set $d$; $L_\kappa$ is the distance from the core to the detector. The core releases a mean energy $\bar{E}_{{\rm f},\kappa}$ per fission, so it undergoes $\bar{P}_\kappa^{(e,d)}/\bar{E}_{{\rm f},\kappa}$ fissions per unit time. Each fission emits $\bar\nu_e$ with the spectrum $S_\kappa(E_\nu)$. The operator $\mathcal{T}_{L_\kappa}$ is that of \equ{arrival_matter} along the path from the core: it multiplies $S_\kappa$ by the survival probability, $\Pee(L_\kappa, E_\nu)$, which includes the oscillations and, under each decay hypothesis, decay; it also adds the daughters of decay, at lower energies. A chord from a core 1000~km away dips at most 20~km below the surface, within the crust. Along each path, we take the density of the crust as constant, as KamLAND and JUNO do~\cite{KamLAND:2010fvi, JUNO:2025gmd}. We build these ingredients as each collaboration does, within what it has published, and normalize \equ{reactor}, without decay, to the number of reactor events that the collaboration expects.

The spectrum per fission of core $\kappa$ in \equ{reactor}, and its mean energy per fission, are averages over the four isotopes that undergo nearly all fissions:
\begin{equation}
 S_\kappa(E_\nu) = \sum_\xi f_{\kappa\xi}\, s_\xi(E_\nu) , \quad
 \bar{E}_{{\rm f},\kappa} = \sum_\xi f_{\kappa\xi}\, E_{{\rm f},\xi} ,
 \label{equ:per_fission}
\end{equation}
where $\xi$ runs over $^{235}$U, $^{238}$U, $^{239}$Pu, and $^{241}$Pu; $f_{\kappa\xi}$ is the share of all the fissions in core $\kappa$ during the data-taking that are of isotope $\xi$; $s_\xi(E_\nu)$ is the spectrum of $\bar\nu_e$ that one fission of isotope $\xi$ emits; and $E_{{\rm f},\xi}$ is the energy that it releases, from 202 to 214~MeV~\cite{Ma:2012bm}. The shares change during the data taking, as uranium burns and plutonium builds up. The data, however, sum the events over the whole time of data taking. At every moment, each isotope emits the same spectrum per fission. Therefore, the summed spectrum depends only on how many fissions of each isotope occurred in the core during that time; we use each isotope's share of them, as each collaboration estimates it. The spectra $s_\xi$ come in two forms. Predictions derive them from the measured spectra of the electrons that the products of fission emit, for $^{235}$U, $^{239}$Pu, and $^{241}$Pu~\cite{Huber:2011wv}, and from nuclear data, for $^{238}$U~\cite{Mueller:2011nm}. These references give them as
\begin{equation}
 s_\xi(E_\nu) = \exp\!\left( \sum_{p=0}^{5} a_{\xi p}\, E_\nu^p \right) ,
 \label{equ:huber_mueller}
\end{equation}
with $E_\nu$ in MeV and the coefficients $a_{\xi p}$ fitted between 2 and 8~MeV. Measurements give instead the spectrum per fission times the cross section, in bins of neutrino energy: Daya Bay's for the fuel of its own reactors~\cite{DayaBay:2016ssb, DayaBay:2021dqj}, and the joint fit of Daya Bay and PROSPECT for $^{235}$U and $^{239}$Pu alone~\cite{DayaBay:2021owf}. Within each bin, we take the shape of \equ{huber_mueller}. For a core whose fuel differs from Daya Bay's, the predictions correct the measured spectrum~\cite{KamLAND:2022vbm}:
\begin{equation}
 S_\kappa = S_{\rm DB} + \sum_\xi \left( f_{\kappa\xi} - f_{{\rm DB},\xi} \right) s_\xi \;,
 \label{equ:dayabay_correction}
\end{equation}
where $S_{\rm DB}$ is Daya Bay's measured spectrum per fission and $f_{{\rm DB},\xi}$ are its shares of fissions. The measurements lie 5\% below the predictions in total, with an excess near 5~MeV of prompt energy~\cite{DayaBay:2015lja}. The sections below give the form that each experiment uses.

Figure~\ref{fig:reactor_spectra} shows the spectrum of reactor events of each data set, \equ{reactor}, core by core. In every data set, 21--28\% of reactor events lie below 3.27~MeV, among the geoneutrinos. Within this region, the shape of the reactor spectrum depends on the distances to the reactors, through the oscillations. The rest lie above, where they constrain the normalization of the reactor spectrum. 

\begin{figure*}[p]
 \centering
 \includegraphics[width=\textwidth]{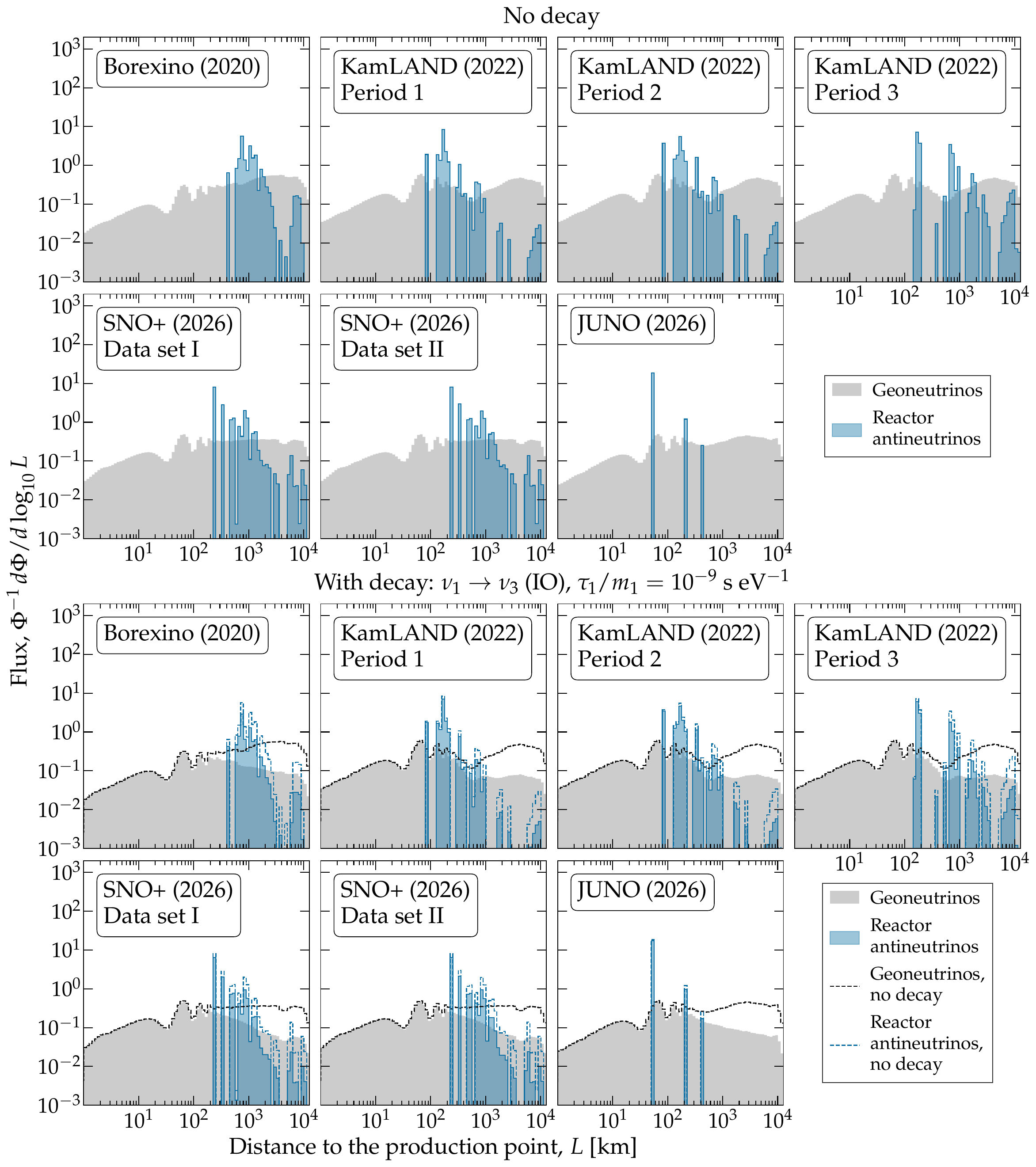}
 \caption{\textbf{How far reactor antineutrinos and geoneutrinos travel to each data set.} Each panel shows the flux of reactor antineutrinos and of geoneutrinos per decade of the distance $L$ that they travel, integrated over energies above the IBD threshold, as in \figu{distance}. Each flux is divided by its value without decay, $\Phi$. Geoneutrinos come from every distance; reactor antineutrinos come from the few distances at which the reactors lie. Decay removes the most from the longest distances, so it removes as many reactor antineutrinos as geoneutrinos at Borexino, and fewer elsewhere. See Sec.~\ref{sec:reactors} for details.}
 \label{fig:reactor_distance}
\end{figure*}

At Borexino, the reactors lie at hundreds to thousands of kilometers, over which the oscillations average out. Therefore, they change the rate of reactor events but not the shape of their spectrum. At SNO+ and JUNO, a few cores dominate. At SNO+, the Bruce plant, at 240~km, produces 39\% of the reactor events; at JUNO, eight cores at 52~km produce 94\%. There, the oscillation driven by $\Delta m^2_{21}$ is resolved, with a dip below 3.27~MeV, among the geoneutrinos. At KamLAND, the spectrum changes between periods of data taking. In Period~3, after most Japanese reactors stopped, the two cores of the Ohi plant, at 179~km, produce a third of the events, so dips appear. Therefore, at KamLAND, SNO+, and JUNO, the reactor events below 3.27~MeV, where the geoneutrinos lie, depend on $\sin^2\theta_{12}$ and $\Delta m^2_{21}$. Our fits vary both, which also set the survival of geoneutrinos (Sec.~\ref{sec:stats}).

Under each decay hypothesis, reactor antineutrinos decay too, over the distance from each core (\equ{reactor}). The strips of \figu{reactor_spectra} show this for $\nu_1 \to \nu_3$ in the inverted ordering at $\tau_1/m_1 = 10^{-9}$~s~eV$^{-1}$. The loss grows toward low energies and with the distance to the reactors, from 5\% of the reactor events at JUNO to 47\% at Borexino. It is largest at the dips of the oscillations, where the contributions of $\nu_1$ and $\nu_2$ to the survival amplitude nearly cancel; removing part of $\nu_1$ deepens the cancellation. A fit that leaves the reactor rate free, as Borexino's does, absorbs the overall loss but not its dependence on energy.

Figure~\ref{fig:reactor_distance} shows how far the reactor antineutrinos of each data set travel, beside the geoneutrinos at the same site, which come from every distance (\figu{distance}). At JUNO, 95\% of the reactor antineutrinos come from within 100~km; at Borexino, 46\% come from beyond 1000~km. Decay grows with $L/E$, so it removes reactor antineutrinos according to these distances. At $\tau_1/m_1 = 10^{-9}$~s~eV$^{-1}$, JUNO keeps 93\% of its reactor flux, against 54\% of its geoneutrinos. Borexino keeps 43\% of its reactor flux and 46\% of its geoneutrinos, since its reactors, like most of its geoneutrinos, lie hundreds to thousands of kilometers away.

KamLAND's panels rest on two approximations. The distances there are those of our own calculation, from the yearly records of the IAEA~\cite{PRIS}: within a year, each reactor runs at its average power, except where the earthquakes of 2007 and 2011 stopped it. The reactors outside Japan, South Korea, Taiwan, and China, which produce 0.4--3\% of the flux, enter at their availability averaged over 2014--2024~\cite{IAEA:RDS2}. KamLAND's own records, day by day, are not public; its expected reactor events, which our fits use, follow them (Sec.~\ref{sec:kamland_model}).

\subsection{Other backgrounds}
\label{sec:other_backgrounds}

\begin{table*}[t!]
 \caption{\label{tab:backgrounds}\textbf{Backgrounds other than reactor antineutrinos.} For each experiment, the number of events of each background that the collaboration expects before its fit, with its uncertainty, and the number that its fit finds. Both count the selected events within the range of prompt energy that we fit (\tabl{experiments}), summed over periods or data sets. SNO+ quotes its expected numbers before its classifier cut against $(\alpha, n)$ events; we give them after it, at the classifier's nominal efficiency. The fitted numbers are those that KamLAND and JUNO publish and those of the best-fit spectra that Borexino and SNO+ draw. A dash marks a background that the collaboration does not fit. JUNO's fit also includes 191 events from reactors beyond 450~km, which are reactor antineutrinos and so are not listed. See Sec.~\ref{sec:other_backgrounds} for details.}
 \centering
 \begin{ruledtabular}
 \begin{tabular}{lcccccccc}
  & \multicolumn{2}{c}{Borexino~\cite{Borexino:2019gps}} & \multicolumn{2}{c}{KamLAND~\cite{KamLAND:2013rgu, KamLAND:2022vbm}} & \multicolumn{2}{c}{SNO+~\cite{SNO:2025chx}} & \multicolumn{2}{c}{JUNO~\cite{JUNO:2026tap}} \\
  Background & Expected & Fitted & Expected & Fitted & Expected & Fitted & Expected & Fitted \\
  \hline
  $(\alpha, n)$ reactions & $0.81 \pm 0.13$ & 0.8 & $194 \pm 28$ & 221 & $21.9 \pm 10.1$ & 30.5 & $10.4 \pm 4.1$ & 10.4 \\
  $^9$Li and $^8$He & $3.6 \pm 1.0$ & 3.6 & $4.3 \pm 0.4$ & 4.7 & --- & --- & $218 \pm 56$ & 292 \\
  Accidental coincidences & $3.85 \pm 0.02$ & 3.8 & $135.2 \pm 0.1$ & 135.2 & --- & --- & $14.5 \pm 2.1$ & 14.5 \\
  $^{214}$Bi--$^{214}$Po coincidences & --- & --- & --- & --- & $6.9 \pm 5.7$ & 1.5 & $87 \pm 64$ & 118 \\
  Atmospheric neutrinos & --- & --- & --- & --- & $3.0 \pm 2.0$ & 3.8 & $18.6 \pm 8.3$ & 22.8 \\
  Neutrons from cosmic muons & --- & --- & --- & --- & --- & --- & $25 \pm 21$ & 31 \\
  \hline
  Total & 8.3 & 8.2 & 333 & 361 & 31.8 & 35.9 & 373 & 489 \\
  IBD candidates & \multicolumn{2}{c}{154} & \multicolumn{2}{c}{1223} & \multicolumn{2}{c}{185} & \multicolumn{2}{c}{8294} \\
  Fitted share of candidates & \multicolumn{2}{c}{5\%} & \multicolumn{2}{c}{30\%} & \multicolumn{2}{c}{19\%} & \multicolumn{2}{c}{6\%} \\
 \end{tabular}
 \end{ruledtabular}
\end{table*}

Each experiment estimates its backgrounds other than reactor antineutrinos from its own measurements. For example, the rate of $(\alpha, n)$ reactions follows from the measured rate of $^{210}$Po decays; that of the spallation products $^9$Li and $^8$He, from the IBD-like events that occur shortly after a cosmic muon crosses the detector; and that of accidental coincidences, from pairs of events seconds apart, which can only have formed by chance, scaled to the time window in which a coincidence is accepted~\cite{Borexino:2019gps, SNO:2025chx}.

\tabl{backgrounds} lists the backgrounds other than reactor $\bar\nu_e$ that each collaboration fits, with the number of events of each that the collaboration expects before its fit and the number that its fit finds; \figu{reproduction} shows their spectra. Their sizes differ widely between the experiments, because the detectors lie at different depths, which set how many muons cross them, and because their scintillators contain different amounts of radioactive contaminants. Below, we describe each background and give its share of the candidates in the collaborations' fits.
\begin{itemize}
 \item \textit{$(\alpha, n)$ reactions.} Alpha particles from $^{210}$Po strike $^{13}$C and free neutrons, whose scattering on protons mimics the prompt signal and whose capture mimics the delayed one. These events make up 18\% of the candidates at KamLAND, most of them from before its scintillator was purified~\cite{KamLAND:2008dgz}, and 17\% at SNO+, but less than 1\% at Borexino and JUNO.
 \item \textit{Spallation products, $^9$Li and $^8$He.} Cosmic muons produce these nuclei by spallation, breaking up carbon nuclei of the scintillator; they decay by emitting an electron and then a neutron. Their number follows the rate of muons, which falls with depth (\tabl{experiments}). They make up 10\% of the candidates at JUNO, the shallowest detector, and are negligible at SNO+, the deepest.
 \item \textit{Accidental coincidences.} Two unrelated events, such as two radioactive decays, can occur by chance so close in time and space that they mimic the prompt and delayed signals of an IBD event. The more radioactive decays a detector records, the more such pairs it finds. These pairs make up 11\% of the candidates at KamLAND and less than 3\% at the other experiments.
 \item \textit{Minor backgrounds.} Unlike accidental coincidences, the decays of $^{214}$Bi and of its daughter, $^{214}$Po, form correlated pairs: $^{214}$Po decays at the same place, with a half-life of 164~$\mu$s, similar to the time that a neutron takes to be captured. The alpha particle of $^{214}$Po usually produces too little light to pass for a neutron capture, but it can produce enough when it gives part of its energy to a proton, which turns more of it into light~\cite{SNO:2025chx, JUNO:2025gmd}. These pairs, atmospheric neutrinos, and neutrons from cosmic muons make up at most 3\% of the candidates.
\end{itemize}
Our fits take the expected numbers of \tabl{backgrounds} and their uncertainties as nominal values and priors (\tabl{parameters}). At each experiment, one or two of these backgrounds dominate: accidental coincidences and $^9$Li at Borexino, $(\alpha, n)$ reactions and accidental coincidences at KamLAND, $(\alpha, n)$ reactions at SNO+, and $^9$Li and $^8$He at JUNO. The collaborations' fits change most of them by less than their uncertainties. The largest relative changes are at SNO+, whose fit raises the $(\alpha, n)$ events by 39\% and lowers the $^{214}$Bi--$^{214}$Po coincidences from 6.9 to 1.5 events, each within its uncertainty.

\subsection{Borexino}

\textit{Data and detector.}---Borexino's 2020 analysis comprises 3\,262.74 days of data, from December 2007 to April 2019, with an exposure of $(1.29 \pm 0.05)\times10^{32}$ target protons~$\times$~year and 154 IBD candidates~\cite{Borexino:2019gps}. We fit them in the bins of prompt charge in which Borexino draws its fit, from 408 to 4000 photoelectrons, \ie, about 0.8 to 7.8~MeV. The detection efficiency is 0.87 for geoneutrinos and 0.90 for reactor antineutrinos. The spectra of uranium, thorium, and reactor events are Borexino's own simulated templates. We calibrate the crust to Borexino's own model of it~\cite{Borexino:2019gps}, not to the lithosphere of our limits (\tabl{experiments}). Borexino's fit works in prompt charge, $N_{\rm pe}$, the number of photoelectrons. We map the deposited energy onto it as $N_{\rm pe} = N_0 + K (E_{\rm dep} - 1.022~{\rm MeV})$ and take $N_{\rm pe}/K$ as the prompt energy, so that $g = 1$ and $\delta = N_0/K - 1.022$~MeV. We fit $N_0 = 454.2$ photoelectrons, $K = 517.6$ photoelectrons per MeV, and $a$ to Borexino's spectral templates of uranium, thorium, and reactor events~\cite{Borexino:2019gps}. The resolution that we find (\tabl{response}) is wider than the 5\% at 1~MeV that Borexino quotes, as expected from the loss of 39\% of its photomultipliers over the data taking~\cite{Borexino:2019gps}.

\smallskip
\textit{Backgrounds.}---Borexino's reactor antineutrinos come from the world's roughly 440 power reactors, at distances of hundreds to thousands of kilometers, over which their oscillations average out~\cite{Borexino:2019gps}. We build \equ{reactor} for them as Borexino does~\cite{Borexino:2019gps}: each core at its nominal thermal power times its load factor averaged over the data taking~\cite{PRIS}, with the fuel composition of its type of reactor, the energies per fission of \Ref~\cite{Ma:2012bm}, and the spectra per fission of \Ref~\cite{Mueller:2011nm}, corrected to Daya Bay's measurement~\cite{DayaBay:2015lja}. Borexino expects 97.6 reactor events in its window and leaves their rate free in its fit; so do we.
The other backgrounds are those of Borexino's fit, $^9$Li, accidental coincidences, and $(\alpha, n)$ reactions, each at its expected rate and with its uncertainty (\tabl{backgrounds}).

\smallskip
\textit{Best fit.}---Without decay, with the geoneutrinos and the reactor antineutrinos free, as in Borexino's fit, the fit gives $52.0^{+9.3}_{-8.6}$ geoneutrino events and 93.8 reactor events, against Borexino's $51.9^{+10.1}_{-7.9}$ and 92.5. With the uranium and thorium signals free separately, it gives 18.3 and 28.0 events, against 27.8 and 21.1, within their uncertainties.

\subsection{KamLAND}
\label{sec:kamland_model}

\textit{Data and detector.}---KamLAND's 2022 data release comprises 5227 days in three periods~\cite{KamLAND:2022vbm}: Period~1, before the scintillator was purified~\cite{KamLAND:2008dgz}; Period~2, after it; and Period~3, with a smaller fiducial volume around the inner balloon of KamLAND-Zen. For each period, the release gives the observed spectrum from 0.9 to 2.7~MeV in 18 bins and every component of KamLAND's fit, including the selection efficiency. Between 0.9 and 2.6~MeV, the region of KamLAND's geoneutrino analysis, the release holds 1159 candidates, 19 fewer than KamLAND's paper reports~\cite{KamLAND:2022vbm}. Scaling each period's spectrum up to that count would strengthen our 90\%~C.L. limits on $\tau_1/m_1$ on the geochemical Earth by 36\% for KamLAND and by about 30\% for both combinations; we use the release as published, which gives the weaker limits.

Each period enters our likelihood separately, with its own exposure, efficiency, resolution, reactor flux, and backgrounds; the normalizations are shared across periods, as in KamLAND's analysis~\cite{KamLAND:2013rgu}. We fit the resolution of each period to the uranium and thorium spectra of the release, leaving out the bins near the uranium edge at 1.45~MeV, which changes between periods. The resulting resolutions (\tabl{response}) are 8 to 23\% wider than the 6.4\% at 1~MeV that KamLAND quotes~\cite{KamLAND:2013rgu}. The offset and the slope of the energy axis, left free, stay within 0.006~MeV of zero and 0.004 of one, so we fix them there. The selection efficiency of each period is the curve of \Ref~\cite{KamLAND:2013rgu}; in Period~3, it includes the smaller fiducial volume.

\smallskip
\textit{Reactors over time.}---KamLAND sees mostly Japanese reactors, at a flux-weighted mean distance of 180~km~\cite{KamLAND:2013rgu}, where the oscillation driven by $\Delta m^2_{21}$ is resolved. The Kashiwazaki-Kariwa plant, which supplied 28\% of the flux in Period~1, stopped after the earthquake of July 2007; most Japanese reactors, after March 2011. We take KamLAND's expected reactor events in each period, 327, 232, and 49, which follow the reactors' daily operating records, and its reactor spectrum in each period.

Our own calculation of \equ{reactor} sets how decay and the mixing parameters, which every experiment shares (Sec.~\ref{sec:stats}), change these spectra. It includes the 137 reactors of Japan, South Korea, Taiwan, and China, each at its thermal power averaged over the period from the yearly records of the IAEA~\cite{PRIS}, and the rest of the world on average. Their spectrum per fission is the one measured by Daya Bay~\cite{DayaBay:2016ssb}, corrected to the average fuel of KamLAND's reactors with predicted spectra~\cite{Huber:2011wv, Mueller:2011nm}, as KamLAND builds it~\cite{KamLAND:2022vbm}. The oscillations include matter of density 2.7~g~cm$^{-3}$, as in KamLAND's analysis~\cite{KamLAND:2010fvi}. Above 1.1~MeV, this calculation reproduces the shape of KamLAND's reactor spectra to 1.1--3.5\% per bin on average, depending on the period. Unlike KamLAND's, it keeps the contributions of spent fuel and of fission products out of equilibrium that Daya Bay's measured spectrum contains, in place of those of KamLAND's reactors; the two differ by about 0.2\% in total~\cite{DayaBay:2016ssb, KamLAND:2022vbm}.

\smallskip
\textit{Other backgrounds.}---KamLAND's data release gives the $(\alpha, n)$ and accidental spectra of each period. The rates are KamLAND's expectations before its fit (\tabl{backgrounds}): for $(\alpha, n)$ reactions and for $^9$Li and $^8$He, those of its 2013 analysis~\cite{KamLAND:2013rgu}, scaled to the live time of each period; for accidental coincidences, the measured rate.

\smallskip
\textit{Best fit.}---Without decay, with the uranium and thorium signals free, as in KamLAND's fit, our fit gives $89^{+42}_{-39}$ uranium and $67.4 \pm 25$ thorium events, $156^{+32}_{-30}$ in all, against KamLAND's 116.6, 57.5, and 173.7. The totals differ by $0.6\sigma$. Our fit uses only the released spectrum below 2.7~MeV; KamLAND's own fit also used the spectrum above it and the times of the events. On 400 simulated data sets made at KamLAND's result, our fit is unbiased: it returns $117.6 \pm 2.0$ uranium and $57.0 \pm 1.2$ thorium events on average. Its results scatter about as widely as KamLAND's quoted uncertainties.

\subsection{SNO+}

\textit{Data and detector.}---SNO+ published 685 days of data in two data sets, with 77 and 108 IBD candidates~\cite{SNO:2025chx}. It also released every candidate with its prompt energy and the output of its classifier, which separates IBD events from $(\alpha, n)$ events. We fit the two data sets jointly with SNO+'s fit model of 21 parameters (\tabl{parameters}). In the fits with decay, our Earth model replaces SNO+'s free geoneutrino signal. SNO+'s energy axis is linear about the point where the positron is at rest, with slope $\lambda$: $g(E_{\rm dep})\,E_{\rm dep} = 1.022~{\rm MeV} + \lambda\,(E_{\rm dep} - 1.022~{\rm MeV})$. We fit $\lambda$, $\delta$, and $a$, data set by data set, to the IBD spectra in SNO+'s supplemental material~\cite{SNO:2025chx}. Across the fiducial volume, the resolutions that result (\tabl{response}) are 1.13 and 1.33 times those that SNO+ quotes for 1-MeV electrons at the center of the detector. The part of the selection efficiency that does not depend on energy follows from the numbers of events that SNO+ expects. Below $E_{\rm prompt} = 3.5$~MeV, SNO+ also applies a classifier against $(\alpha, n)$ events, whose efficiency we read from SNO+'s spectra with and without it: 0.86 at 0.9~MeV, rising to 0.95 at 3.5~MeV.

\smallskip
\textit{Reactors over time.}---SNO+ sees three Canadian complexes, Bruce at 240~km and Pickering and Darlington at 340 and 349~km, and more distant reactors, nearly all in North America~\cite{SNO:2025chx}. As SNO+ does, we follow the thermal power of each core day by day, from the public records of the Ontario grid operator~\cite{IESO} and of the U.S. Nuclear Regulatory Commission~\cite{NRC}; cores without daily records, among them those outside North America, enter at their capacity times their availability averaged over 2014 to 2024~\cite{IAEA:RDS2}. Bruce~3 stopped in March 2023, before data set~II began; Pickering~1 and~4 stopped in late 2024, during it. Each data set carries its own reactor flux. The spectrum per fission combines the $^{235}$U and $^{239}$Pu spectra measured by Daya Bay and PROSPECT~\cite{DayaBay:2021owf} with the predicted ones for $^{238}$U and $^{241}$Pu~\cite{Huber:2011wv, Mueller:2011nm}, at the average fission fractions that SNO+ adopts for the heavy-water reactors of Ontario and for the light-water reactors elsewhere~\cite{SNO:2025chx}, and with the energies per fission of \Ref~\cite{Ma:2012bm}. The oscillations include matter, at SNO+'s constant electron density~\cite{SNO:2025chx}. We normalize the spectrum to the $140 \pm 6$ events that SNO+ expects.

\smallskip
\textit{Other backgrounds.}---The other backgrounds are read from SNO+'s supplemental material~\cite{SNO:2025chx}: $^{13}$C$(\alpha, n)^{16}$O reactions, whose proton recoils lie below 3.5~MeV and whose excited states of $^{16}$O lie near 6~MeV; $\alpha$-$p$ coincidences from decays of $^{214}$Bi and $^{214}$Po; and atmospheric neutrinos. Each enters at SNO+'s expected rate, with its shape and uncertainty (\tabl{backgrounds}). As in SNO+'s fit, the energy scale of the proton recoils and the classifier factor of the $(\alpha, n)$ events are free within broad ranges.

\smallskip
\textit{Best fit.}---Without decay, with the geoneutrino signal free, as in SNO+'s fit, the fit gives $48.7^{+13.4}_{-12.1}$~TNU and a uranium-to-thorium event ratio of 3.36, against SNO+'s $48^{+14}_{-12}$~TNU and 3.23. With the classifier cut, it expects 197.1 events in all, against SNO+'s 198. Between 0.9 and 1.5~MeV of prompt energy, where decay removes the most geoneutrinos, the data hold 37 events, against 32 predicted on the geochemical Earth.

\subsection{JUNO}

\textit{Data and detector.}---JUNO published the spectrum of its first 207.2 days of data, 8294 IBD candidates in 156 bins of prompt energy from 0.7 to 12~MeV~\cite{JUNO:2026tap}. We fit it with the fit model of JUNO's first analysis~\cite{JUNO:2025gmd}, of 39 parameters (\tabl{parameters}). Among them, $\sin^2\theta_{12}$ and $\Delta m^2_{21}$ are free; Daya Bay constrains $\sin^2 2\theta_{13}$ and $\Delta m^2_{32}$~\cite{DayaBay:2022orm}. The geoneutrino rate is free, as in JUNO's fit of these data. Our detector response follows JUNO's, with the values of \tabl{response}. JUNO measured its resolution for positrons in radial shells of its detector~\cite{JUNO:2026tap}; we average it over the fiducial volume, taking the shells to be equally wide in radius, which JUNO does not state. The averaged resolution has a third term, $c/E_{\rm prompt}$ with $c = 0.96\%$~MeV, added in quadrature to the terms in $a$ and $b$ of \tabl{response}. JUNO's fit of these data also carries a covariance for the detector response, which includes a 5\% pile-up of $^{14}$C; we keep a single uncertainty on the energy scale instead. A pile-up of that size moves our fitted parameters by less than 0.1 of JUNO's uncertainties.

\smallskip
\textit{Reactors.}---JUNO sees two Taishan and six Yangjiang cores, at 52.1 to 52.8~km, the Daya Bay complex, at 215~km~\cite{JUNO:2024jaw}, and the Fangchenggang plant, at 412~km~\cite{Beccaria:2026ous}. At these distances, both oscillation frequencies are resolved, so we compute the survival probability exactly at each core's distance, in three flavors and in matter of density 2.55~g~cm$^{-3}$, as JUNO does~\cite{JUNO:2025gmd}. The spectrum per fission is the one unfolded from Daya Bay's data~\cite{DayaBay:2021dqj}, rescaled in each of its 25 intervals of energy, all but the last 0.25~MeV wide, to the spectrum that JUNO's fit returns. JUNO also constrains the shape of this spectrum with its satellite detector, TAO, whose data are not public~\cite{JUNO:2026tap}; our model holds no structure finer than these intervals. JUNO's own model corrects this spectrum for the fuel of its cores, which changes over time, and for spent fuel and fission products out of equilibrium, each under 1\% of the flux~\cite{JUNO:2025gmd}; we leave these corrections out.

JUNO has not published each core's power over these 207 days, so the eight near cores enter in proportion to their nominal thermal powers~\cite{JUNO:2024jaw}. Daya Bay and Fangchenggang supply 4.3\% and 0.9\% of the flux without oscillations, the shares that reproduce the ratio of JUNO's spectra with and without oscillations~\cite{JUNO:2026tap}. Since JUNO gives no expectation before its fit, we normalize the spectrum to the 21,121 IBD events that its fit ends with, after selection and without oscillations~\cite{JUNO:2026tap}. The other backgrounds, the reactors beyond 450~km and those of \tabl{backgrounds}, each enter with JUNO's expected rate, shape, and uncertainty.

\smallskip
\textit{Best fit.}---Without decay, the fit gives $\sin^2\theta_{12} = 0.3020^{+0.0061}_{-0.0060}$ and $\Delta m^2_{21} = (7.40 \pm 0.08)\times10^{-5}$~eV$^2$, against JUNO's $0.3036 \pm 0.0064$ and $(7.39 \pm 0.08)\times10^{-5}$~eV$^2$~\cite{JUNO:2026tap}. Its backgrounds lie within 0.02 events per day of JUNO's fitted ones. Its reactor antineutrinos number 7174, against 7162 under JUNO's best-fit curve. Left free, as in JUNO's fit, its geoneutrinos come out at $2.12 \pm 0.31$ events per day, against JUNO's $2.12 \pm 0.32$. Fitting $\Delta m^2_{31}$ to JUNO's data as JUNO does, we find $(2.506^{+0.026}_{-0.024})\times10^{-3}$~eV$^2$ in the normal ordering, against JUNO's $(2.509^{+0.027}_{-0.025})\times10^{-3}$~eV$^2$~\cite{JUNO:2026tap}.

\FloatBarrier
\section{Limits from data vs.~simulation}
\label{app:brazil}

\begin{figure*}[t!]
 \centering
 \includegraphics[width=\textwidth]{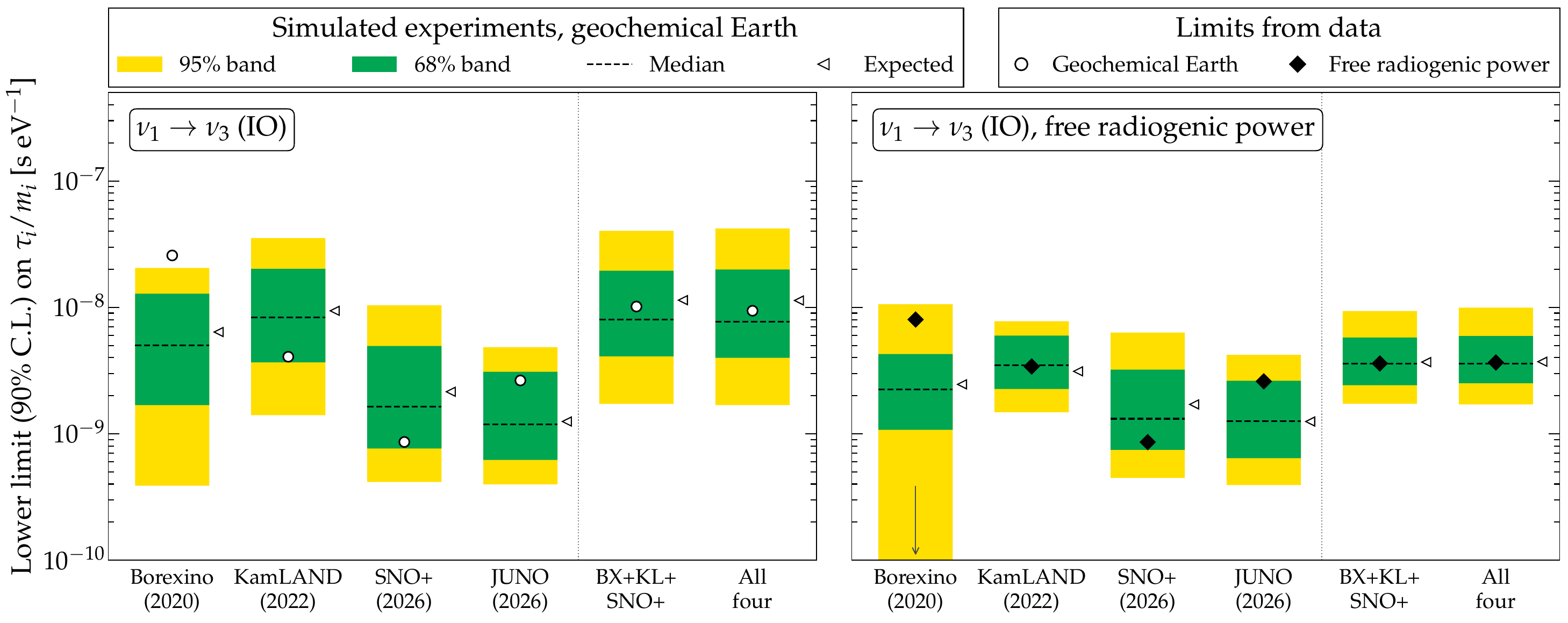}
 \caption{\textbf{Lower limits on the neutrino lifetime-to-mass ratio from data, compared with those from simulated experiments.} The simulated experiments are made without decay, on the geochemical Earth, and fitted as the data are; there are 200 for each experiment and for each combination of experiments. An arrow marks a band that extends to simulated experiments that set no limit. Borexino's limits lie at the top of their bands, because its data hold more geoneutrinos than the geochemical Earth predicts. KamLAND's data hold fewer, which puts its limit with the radiogenic power assumed in the lower part of its band. See Sec.~\ref{sec:validation} for details.}
 \label{fig:brazil}
\end{figure*}

We make the simulated experiments of Sec.~\ref{sec:validation} as follows. For each simulated experiment, we draw the number of events in every bin from a Poisson distribution whose mean is the number expected on the geochemical Earth without decay. Each prior stands for an external measurement, which would also fluctuate in a repeated experiment. Therefore, each simulated experiment also draws the center of every prior, including the constraints on the mixing parameters, at random from that prior. The cosmological limit on the sum of the masses stays fixed, since it is an upper limit rather than a measured value. We then fit each simulated experiment with its priors centered at the drawn values.

Figure~\ref{fig:brazil} compares each limit from data with the limits from simulated experiments. The simulated experiments are made without decay, so their limits show the range expected from each experiment and combination if neutrinos do not decay. For $\nu_1 \to \nu_3$, the limit from data is stronger than 98.5\% of the simulated ones at Borexino, 18.5\% at KamLAND, and 19--79\% elsewhere.

Freeing the power weakens the limits, because a hotter Earth makes up for the geoneutrinos that decay removes. At the limit from data, the fit of all four experiments raises the power from 15~TW to its ceiling, 39.6~TW, as it does at the expected limit. In the actual data, the hotter Earth worsens the fit to Borexino by 1.04 in $-\ln\mathcal{L}$, more than it improves the fit to SNO+, by 0.75; in data without fluctuations, it worsens the fits to KamLAND, SNO+, and Borexino by similar amounts. The combined limits from data remain typical, stronger than 51.5\% (all four) and 50\% (BX+KL+SNO+) of their simulated ones. KamLAND's limit moves most, from 18.5 to 46.5\%, because freeing the power weakens it only 1.2 times, against a median of 2.4 for its simulated limits.

\section{Coverage of the limits}
\label{app:coverage}

\begin{figure}[t!]
 \centering
 \includegraphics[width=\columnwidth]{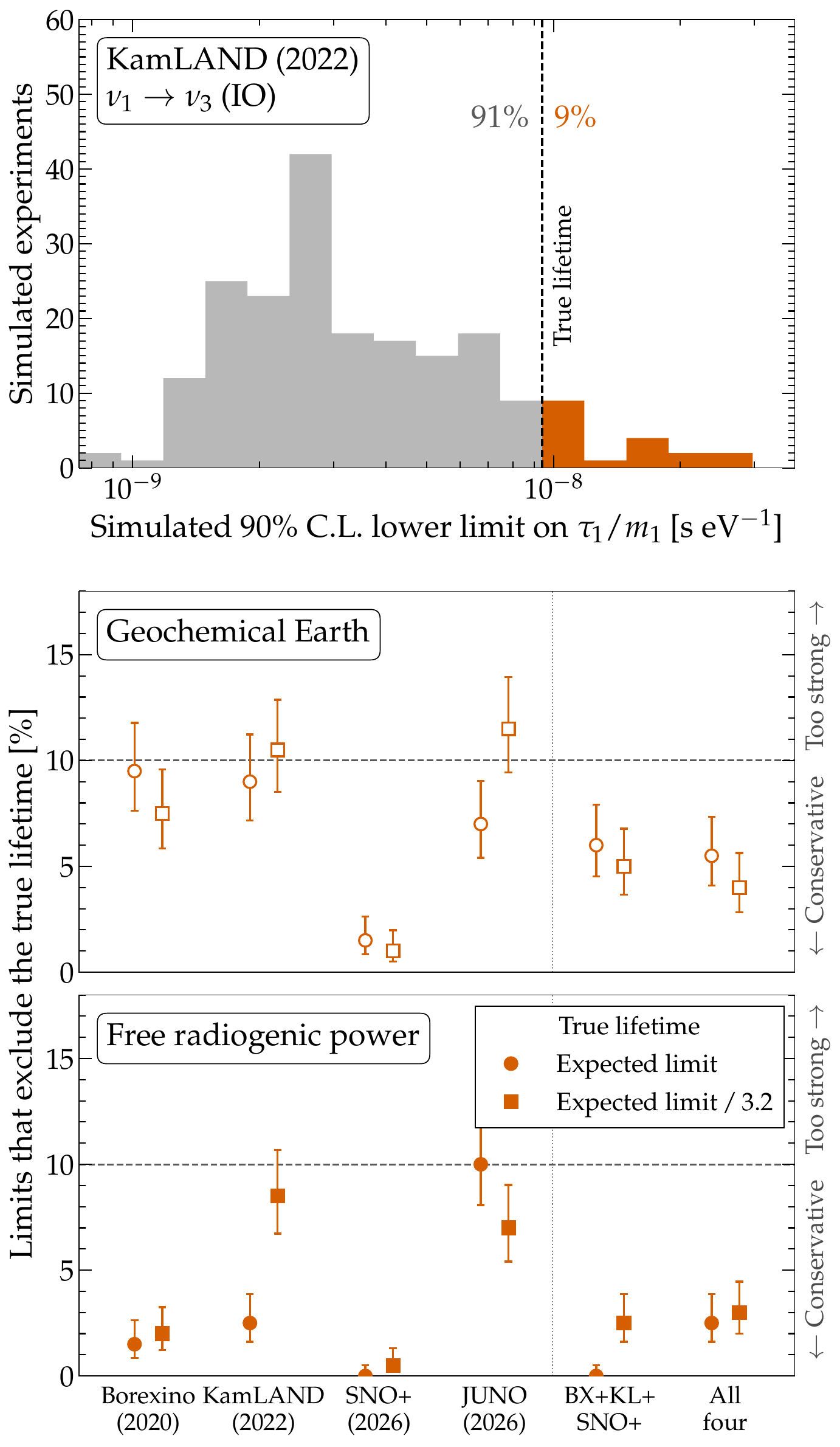}
 \caption{\textbf{Coverage of our 90\%~C.L. limits.} \textit{Top:} Limits on $\tau_1/m_1$ from 200 simulated KamLAND experiments made with $\nu_1 \to \nu_3$ decay in the inverted mass ordering, with the true lifetime equal to KamLAND's expected limit, on the geochemical Earth. The percentages are the shares of the limits below and above the true lifetime; those above exclude it. \textit{Middle and bottom:} Share of the simulated experiments whose limit excludes the true lifetime, for each experiment and combination, with the radiogenic power assumed and free; the error bars are binomial 68\% intervals~\cite{Wilson:1927xyh}. Correct 90\%~C.L. limits exclude the true lifetime in 10\% of the experiments. Our limits are correct or conservative.}
 \label{fig:coverage}
\end{figure}

A procedure for 90\%~C.L. limits has correct coverage if its limit lies above the true lifetime, and so excludes it, in 10\% of repeated experiments. We test this directly. For each experiment and both experiment combinations, we make 200 simulated experiments with $\nu_1 \to \nu_3$ decay at a known lifetime, fit each as the data are fitted, with the centers of its priors drawn at random as in Appendix~\ref{app:brazil}, and count how often the limit excludes the true lifetime. The test uses two true lifetimes: one equal to the expected limit, and one 3.2 times shorter. Coverage must hold for any true lifetime. The two probe decay at the edge of the sensitivity and well within it.

Figure~\ref{fig:coverage} shows the result. In the top panel, 9\% of 200 simulated KamLAND experiments set a limit above the true lifetime and so exclude it. With the radiogenic power assumed, this share is 1 to 11.5\% across the fits, against the 10\% of correct coverage, which 200 experiments measure to $\pm 2.1\%$. With the radiogenic power free, it is 0--10\%. A higher power makes up for part of what decay removes, so every fit but JUNO's excludes the true lifetime less often than with the power assumed; a third to a half of Borexino's simulated experiments set no limit at all. JUNO's fit leaves its rate of geoneutrinos free either way, so its two shares differ by less than twice their uncertainty. No limit on $\tau_1/m_1$ excludes the true lifetime too often. Therefore, these limits are correct or conservative; their thresholds need no calibration.

We repeat the test for JUNO's limit on $\nu_3 \to \nu_1$ in the normal ordering, with 1,000 simulated experiments at each of two true lifetimes: the expected limit and the limit from data. At the asymptotic thresholds, the 90\%~C.L. limit excludes the true lifetime in 14.5\% of them and the 95\%~C.L. limit in 6.7\%, more often than the 10\% and 5\% that their confidence levels allow. Therefore, for every limit on $\tau_3/m_3$, we replace the asymptotic threshold by the value that $q$ exceeds at the true lifetime in 10\% (or 5\%) of the simulated experiments: 2.13 at 90\%~C.L., in place of 1.64, and 3.07 at 95\%~C.L., in place of 2.71. An experiment counts toward these shares only if its best-fit lifetime is longer than the true one, since only then can its limit exclude it. We measure these thresholds for JUNO alone and apply them to the three channels that decay $\nu_3$ and to the four experiments combined, whose limits on $\tau_3/m_3$ rest on JUNO's spectrum.

For the decays of $\nu_2$, whose limits rest mainly on JUNO's reactor antineutrinos, we test JUNO's limits on $\nu_2 \to \nu_3$ and $\nu_2 \to \nu_1$ in the same way, at their expected limits: they exclude the true lifetime in $12.2 \pm 1.6\%$ and $10.8 \pm 1.7\%$ of 400 and 324 simulated experiments at 90\%~C.L., and in 5.2\% at 95\%~C.L., as they should.

\section{Parameters of the fits}
\label{app:parameters}

\smallskip
\textit{Parameters and priors.}---\tabl{parameters} lists the free parameters of the fits, with their priors and the source of each. Every fit varies the lifetime, the mass of the lightest neutrino, the mixing parameters, and the normalizations of the lithosphere and the mantle; the fits with the radiogenic power free vary the power too. Each experiment adds the parameters of its own analysis.

\begin{figure*}[p]
 \centering
 \includegraphics[width=\textwidth]{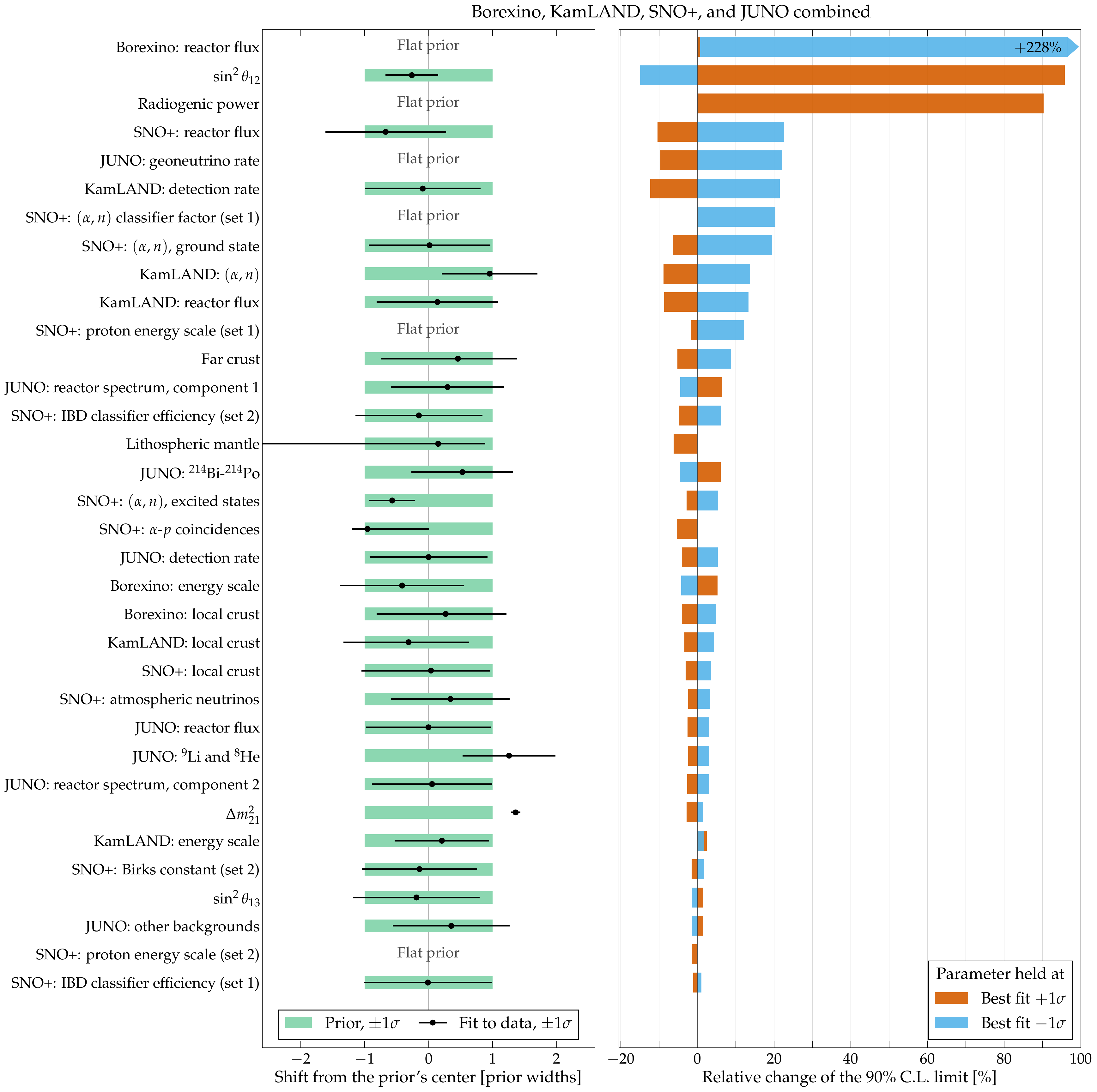}
 \caption{\textbf{Fitted values of the parameters of the four experiments combined, and how much each moves the limit.} The fit is for $\nu_1 \to \nu_3$ in the inverted ordering with free radiogenic power; each row is one parameter. \textit{Left:} the prior and the fit to data, as shifts from the center of the prior, in prior widths; a parameter with a flat prior has no width, and so no shift. \textit{Right:} the relative change of the 90\%~C.L. lower limit on $\tau_1/m_1$ from data when the parameter is held at its best fit plus or minus one standard deviation, with the others fitted; a missing bar would take the parameter outside its allowed range. Rows are ranked by the larger change; the 39 parameters that move the limit by less than 1\% are not shown. A bar beyond 100\% ends in a triangle that gives its value. Borexino's reactor flux moves the limit most.}
 \label{fig:impacts}
\end{figure*}

\smallskip
\textit{Fitted values.}---Tables~\ref{tab:fitted} and \ref{tab:fitted_continued} give the value that each fit to data finds for every parameter, with its uncertainty. Figure~\ref{fig:impacts} shows, in its left panel, how far the data shift each parameter from its prior and how much they narrow it, for the fit of all four with the radiogenic power free. For most parameters, the fit returns the center and width of the prior. The largest shift from the prior is that of $|\Delta m^2_{31}|$, 1.7 prior widths, in JUNO's fit and in the fit of all four; \figu{impacts} omits it, since it moves the limit by less than 1\%. Next come $\Delta m^2_{21}$ in the fit of all four, 1.4, JUNO's background of $^9$Li and $^8$He, 1.3, and the $(\alpha,n)$ background in KamLAND's own fit, 1.2.

The data narrow few priors by much. In the fit of all four, whose prior on $\Delta m^2_{21}$ and $\sin^2\theta_{12}$ comes from the solar data alone, the reactor antineutrinos of KamLAND, SNO+, and JUNO, which measure both directly, narrow these priors to under a tenth and about two-fifths of their widths. Of the other priors, the data narrow most that of SNO+'s $(\alpha,n)$ events from excited states of $^{16}$O, to about a third of its width.

Of the parameters with flat priors, Borexino's reactor flux and the energy scales of the $(\alpha,n)$ proton recoils at SNO+ lie within about one standard deviation of their nominal values. JUNO's rate of geoneutrinos lies 2.4 and 2.5 standard deviations above the rate that the geochemical Earth predicts, in JUNO's fit and in the fit of all four. The fitted rate reproduces the excess that JUNO measures (Sec.~\ref{sec:nu1}).

Combining the experiments moves the parameters of KamLAND: its local crust, detection rate, and $(\alpha,n)$ background fall. The mantle, which all sites share, hardly moves. KamLAND's own fit accounts for its deficit of events with decay, at $\tau_1/m_1 \approx 10^{-8}$~s~eV$^{-1}$ (Sec.~\ref{sec:nu1}). Borexino's data exclude that lifetime at 90\%~C.L. Therefore, the combined fits account for the deficit with KamLAND's own parameters instead.

\smallskip
\textit{Impacts.}---To see how much each nuisance parameter matters, we hold it at its best fit plus or minus its uncertainty, fit all the other parameters, and compute the limit from data again. The uncertainty of each parameter is its standard deviation at the best fit, from the curvature of the likelihood there. One parameter, SNO+'s classifier factor for $(\alpha,n)$ events in its second data set, cannot be shifted: its uncertainty is wider than its whole allowed range (\tabl{fitted}). A parameter that moves the limit when it is shifted is correlated with the lifetime. The right panel of \figu{impacts} shows how much each parameter moves the limit of all four (Sec.~\ref{sec:nu1}).

\begin{figure*}[t!]
 \centering
 \includegraphics[width=\textwidth]{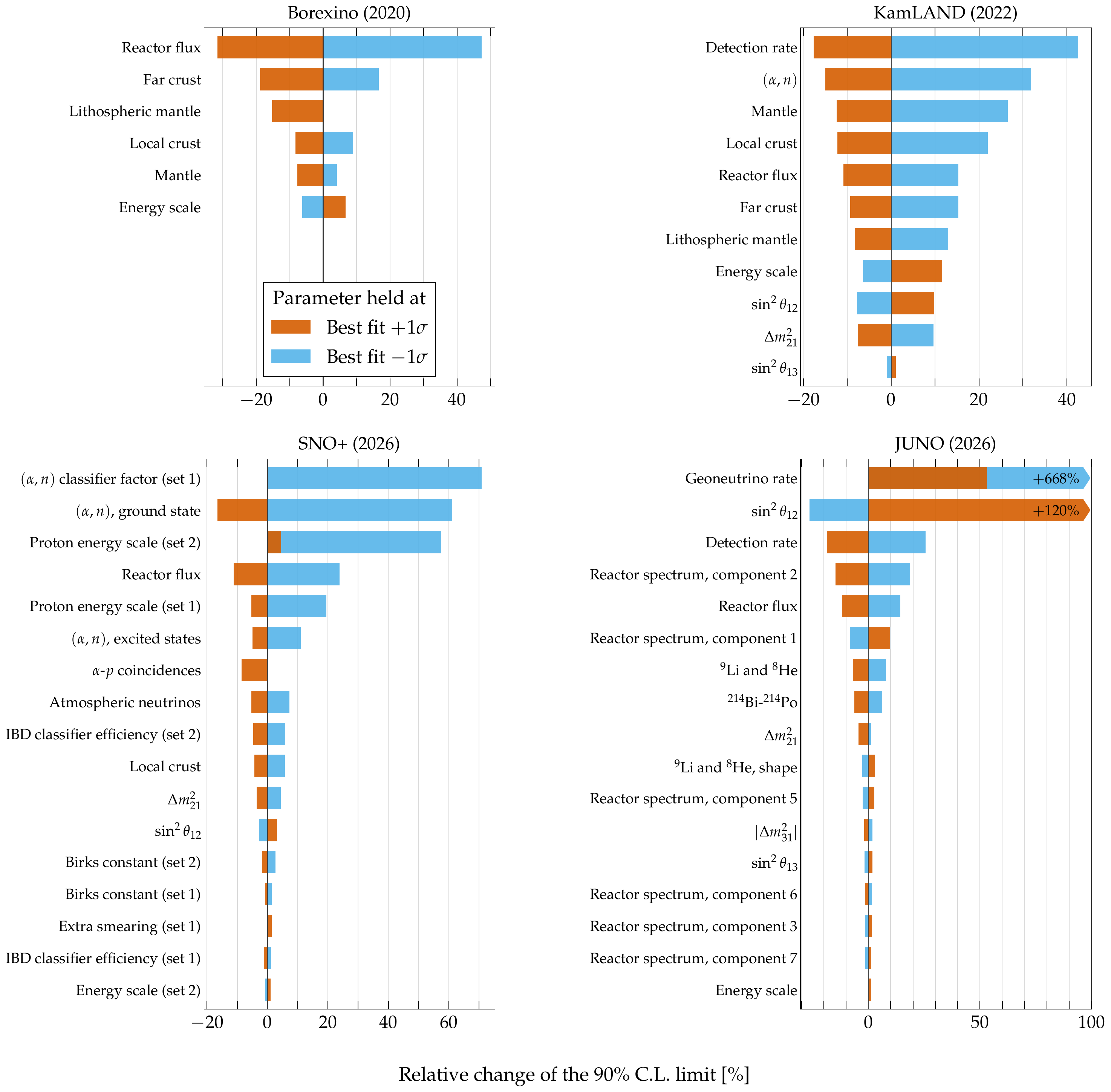}
 \caption{\textbf{How much each parameter moves the limit of each experiment.} Same as the right panel of \figu{impacts}, but for each experiment alone, on the geochemical Earth with the radiogenic power assumed. The 8, 4, 8, and 26 parameters that move these limits by less than 1\% are not shown. At JUNO, the rate of geoneutrinos moves the limit most.}
 \label{fig:impacts_experiments}
\end{figure*}

Figure~\ref{fig:impacts_experiments} shows how much each parameter moves the limit of each experiment alone, as the right panel of \figu{impacts} does for the four experiments combined. Each experiment depends most on parameters of its own analysis. At SNO+, these describe its $(\alpha,n)$ events: the classifier factor in its first data set, the normalization of their ground-state branch, and the energy scale of their proton recoils. At KamLAND, they are the detection rate, the $(\alpha,n)$ background, and the uranium and thorium of the mantle and of the local crust. At Borexino, the reactor flux moves the limit most, since Borexino's fit leaves the reactor rate free.

At JUNO, the rate of geoneutrinos and $\sin^2\theta_{12}$ move the limit most; the detection rate and the second component of the reference reactor spectrum follow. JUNO's fit leaves the rate of geoneutrinos free (Sec.~\ref{sec:stats}), so a higher rate can offset the events that decay removes; the other three change how many reactor antineutrinos JUNO expects. The combination of Borexino, KamLAND, and SNO+, not shown, behaves as that of all four: Borexino's reactor flux moves it most.

\begin{table*}[t!]
 \caption{\label{tab:parameters}\textbf{Free parameters of the fits and their priors.} Normalizations are relative to their expected values. A Gaussian prior is written as its central value $\pm$ its width; a width in percent applies to a scale factor centered on one. A log-normal prior is centered on one; its width is that of the logarithm. See Sec.~\ref{sec:stats} for details.}
 \centering
 \begin{ruledtabular}
 \begin{tabular}{lll}
  Parameter & Prior & Source \\
  \hline
  \multicolumn{3}{l}{\textit{Parameters of interest, neutrino mass, and mixing (all experiments)}} \\
  Lifetimes, $\log_{10}[(\tau_i/m_i)/(\mathrm{s~eV}^{-1})]$ & Flat in $[-14, 4]$ & --- \\
  Mass of the lightest neutrino, $m_{\rm lightest}$ & Flat in $[0, 2]$~eV; cosmology, \equ{likelihood} & \cite{ParticleDataGroup:2024cfk} \\
  $\sin^2\theta_{12}$ and $\Delta m^2_{21}$ & NuFIT~6.1 $\Delta\chi^2$ over the pair, as in Appendix~\ref{app:nufit} & \cite{Esteban:2024eli, NuFIT61, Super-Kamiokande:2023jbt} \\
  $\sin^2\theta_{13}$ and $\vert\Delta m^2_{31}\vert$ & NuFIT~6.1 $\Delta\chi^2$ over the pair & \cite{Esteban:2024eli, NuFIT61} \\
  \hline
  \multicolumn{3}{l}{\textit{Earth (all experiments)}} \\
  Local crust, $\U$ and $\Th$ together & Log-normal, width 0.08 to 0.17, by site & \cite{Borexino:2019gps, Bellini:2013wsa, Bellini:2021sow, Semenec:2023, JUNO:2025sfc} \\
  Far crust, $\U$ and $\Th$ together & Log-normal, width 0.24; shared by sites & \cite{Wipperfurth:2019idn, Semenec:2023} \\
  Lithospheric mantle, $\U$ and $\Th$ together & Log-normal, width 0.97; shared by sites & \cite{Wipperfurth:2019idn, Semenec:2023} \\
  Mantle, $\U$ and $\Th$ (power assumed) & $1 \pm 0.44$, $0.31$, or $0.14$, by model & \cite{Borexino:2019gps} \\
  Radiogenic power (free-power fits only) & Flat in $[9.4, 39.6]$~TW & \cite{Borexino:2019gps} \\
  \hline
  \multicolumn{3}{l}{\textit{Borexino}} \\
  Reactor antineutrinos & Free & \cite{Borexino:2019gps} \\
  $^9$Li; accidental coincidences; $(\alpha, n)$ & $1 \pm 0.28$; $1 \pm 0.004$; $1 \pm 0.16$ & \cite{Borexino:2019gps} \\
  Energy scale & $\pm 1\%$ & \cite{Borexino:2015ucj} \\
  \hline
  \multicolumn{3}{l}{\textit{KamLAND (shared by the three periods)}} \\
  Reactor antineutrinos & $1 \pm 0.021$ & \cite{KamLAND:2022vbm} \\
  $(\alpha, n)$; accidental coincidences; $^9$Li and $^8$He & $1 \pm 0.142$; $1 \pm 0.001$; $1 \pm 0.086$ & \cite{KamLAND:2013rgu, KamLAND:2022vbm} \\
  Energy scale & $\pm 1.2\%$ & \cite{KamLAND:2008dgz} \\
  Detection rate of IBD events & $\pm 2.3\%$; $\pm 3.0\%$ after Period~1 & \cite{KamLAND:2013rgu} \\
  \hline
  \multicolumn{3}{l}{\textit{SNO+ (shared by the two data sets unless marked)}} \\
  Reactor antineutrinos & $1 \pm 0.043$ & \cite{SNO:2025chx} \\
  $(\alpha, n)$, ground and excited states of $^{16}$O & $1 \pm 0.30$; $1 \pm 1.0$ & \cite{SNO:2025chx} \\
  $\alpha$-$p$ coincidences; atmospheric neutrinos & $1 \pm 0.83$; $1 \pm 0.68$ & \cite{SNO:2025chx} \\
  Energy scale (per data set) & $\pm 1.1\%$; $\pm 1.0\%$ & \cite{SNO:2025chx} \\
  Birks quenching constant (per data set) & $0.074 \pm 0.004$ & \cite{SNO:2025chx} \\
  Extra energy smearing (per data set) & $0 \pm 0.044$; $0 \pm 0.049$~MeV$^{1/2}$ & \cite{SNO:2025chx} \\
  Classifier efficiency for IBD events (per data set) & $\pm 2.9\%$ & \cite{SNO:2025chx} \\
  Energy scale of $(\alpha, n)$ proton recoils (per data set) & Flat in $[0.5, 1.5]$ & \cite{SNO:2025chx} \\
  Classifier factor for $(\alpha, n)$ events, $a^2$ (per data set) & Flat in $[0, 0.449]$ & \cite{SNO:2025chx} \\
  \hline
  \multicolumn{3}{l}{\textit{JUNO}} \\
  Geoneutrino rate & Flat in $[0, 10]$ & \cite{JUNO:2026tap} \\
  Reactor flux; detection efficiency; energy scale & $\pm 0.89\%$; $\pm 1.4\%$; $\pm 0.6\%$ & \cite{JUNO:2026tap} \\
  Reference reactor spectrum, 25 coefficients & Daya Bay's covariance & \cite{JUNO:2025gmd, DayaBay:2021dqj} \\
  $^9$Li and $^8$He: rate; shape & $1 \pm 0.26$; $\pm 20\%$ tilt & \cite{JUNO:2026tap, JUNO:2025gmd} \\
  Reactors beyond 450~km; $^{214}$Bi--$^{214}$Po; others & $1 \pm 0.10$; $1 \pm 0.74$; $1 \pm 0.34$ & \cite{JUNO:2026tap} \\
 \end{tabular}
 \end{ruledtabular}
\end{table*}

\begin{table*}[tp]
 \caption{\label{tab:fitted}Fitted values of the parameters in each fit to data, for $\nu_1 \to \nu_3$ in the inverted ordering on the geochemical Earth, with the radiogenic power assumed: the best fit and its uncertainty, one standard deviation from the curvature of the likelihood. A dash means that the fit does not vary the parameter. Uncertainties stop at the edges of each parameter's range. Normalizations are relative to their expected values; energy scales, detection rates, Birks constants, classifier efficiencies, components of the reactor spectrum, and the shape of $^9$Li and $^8$He are shifts in units of their prior widths. SNO+'s energy scale of proton recoils and its classifier factor for $(\alpha,n)$ events are their values. \tabl{fitted_continued} continues this table for SNO+ and JUNO. The lifetime is in \tabl{limits}; the mass of the lightest neutrino fits at zero in every fit.}
 \centering
 \footnotesize
 \begin{ruledtabular}
 \begin{tabular}{lcccccc}
  Parameter & Borexino & KamLAND & SNO+ & JUNO & BX+KL+SNO+ & All four \\
  \hline
  \multicolumn{7}{l}{\textit{Shared by the experiments}} \\
  $\sin^2\theta_{12}$ & $0.3084 \pm 0.0066$ & $0.3076 \pm 0.0066$ & $0.3097 \pm 0.0066$ & $0.3039 \pm 0.0051$ & $0.3097 \pm 0.0065$ & $0.3025 \pm 0.0055$ \\
  $\sin^2\theta_{13}$ [$10^{-2}$] & $2.263 \pm 0.056$ & $2.262 \pm 0.056$ & $2.263 \pm 0.056$ & $2.251 \pm 0.056$ & $2.263 \pm 0.056$ & $2.252 \pm 0.056$ \\
  $\Delta m^2_{21}$ [$10^{-5}$~eV$^2$] & $7.54 \pm 0.10$ & $7.53 \pm 0.10$ & $7.470 \pm 0.099$ & $7.416 \pm 0.068$ & $7.44 \pm 0.11$ & $7.390 \pm 0.071$ \\
  $\vert\Delta m^2_{31}\vert$ [$10^{-3}$~eV$^2$] & $2.408 \pm 0.019$ & $2.408 \pm 0.019$ & $2.408 \pm 0.019$ & $2.443 \pm 0.014$ & $2.408 \pm 0.019$ & $2.443 \pm 0.014$ \\
  Mantle & $1.13 \pm 0.29$ & $0.97 \pm 0.30$ & $1.02 \pm 0.31$ & $1.00 \pm 0.31$ & $0.97 \pm 0.27$ & $0.98 \pm 0.30$ \\
  Far crust & $1.16 \pm 0.28$ & $0.99 \pm 0.24$ & $1.05 \pm 0.26$ & $1.00 \pm 0.24$ & $1.10 \pm 0.25$ & $1.10 \pm 0.28$ \\
  Lithospheric mantle & $1.4^{+1.7}_{-1.4}$ & $0.96 \pm 0.92$ & $1.1^{+1.2}_{-1.1}$ & $1.00 \pm 0.97$ & $1.1 \pm 1.1$ & $1.1 \pm 1.1$ \\
  \hline
  \multicolumn{7}{l}{\textit{Borexino}} \\
  Local crust & $1.02 \pm 0.13$ & -- & -- & -- & $1.03 \pm 0.14$ & $1.03 \pm 0.14$ \\
  Reactor flux & $1.00 \pm 0.11$ & -- & -- & -- & $1.01 \pm 0.11$ & $1.02 \pm 0.13$ \\
  $^9$Li & $0.99 \pm 0.27$ & -- & -- & -- & $0.99 \pm 0.27$ & $0.99 \pm 0.27$ \\
  Accidental coincidences & $1.0000 \pm 0.0044$ & -- & -- & -- & $1.0000 \pm 0.0044$ & $1.0000 \pm 0.0044$ \\
  $(\alpha,n)$ & $1.00 \pm 0.16$ & -- & -- & -- & $1.01 \pm 0.16$ & $1.01 \pm 0.16$ \\
  Energy scale & $-0.37 \pm 0.96$ & -- & -- & -- & $-0.41 \pm 0.96$ & $-0.41 \pm 0.97$ \\
  \hline
  \multicolumn{7}{l}{\textit{KamLAND}} \\
  Local crust & -- & $0.996 \pm 0.079$ & -- & -- & $0.974 \pm 0.074$ & $0.974 \pm 0.075$ \\
  Reactor flux & -- & $1.005 \pm 0.020$ & -- & -- & $1.003 \pm 0.020$ & $1.003 \pm 0.020$ \\
  $(\alpha,n)$ & -- & $1.17 \pm 0.11$ & -- & -- & $1.14 \pm 0.11$ & $1.13 \pm 0.11$ \\
  Accidental coincidences & -- & $1.0000 \pm 0.0010$ & -- & -- & $1.0000 \pm 0.0010$ & $1.0000 \pm 0.0010$ \\
  $^9$Li and $^8$He & -- & $1.000 \pm 0.086$ & -- & -- & $0.999 \pm 0.086$ & $0.999 \pm 0.086$ \\
  Energy scale & -- & $0.04 \pm 0.75$ & -- & -- & $0.23 \pm 0.73$ & $0.22 \pm 0.74$ \\
  Detection rate & -- & $0.23 \pm 0.95$ & -- & -- & $-0.11 \pm 0.88$ & $-0.11 \pm 0.91$ \\
 \end{tabular}
 \end{ruledtabular}
\end{table*}

\begin{table*}[tp]
 \caption{\label{tab:fitted_continued}Fitted values of the parameters of SNO+ and JUNO in each fit to data, as in \tabl{fitted}. The other 23 components of JUNO's reference reactor spectrum fit within 0.22 of zero, with uncertainties of 0.98 to 1.00.}
 \centering
 \footnotesize
 \begin{ruledtabular}
 \begin{tabular}{lcccc}
  Parameter & SNO+ & JUNO & BX+KL+SNO+ & All four \\
  \hline
  \multicolumn{5}{l}{\textit{SNO+}} \\
  Local crust & $1.04 \pm 0.17$ & -- & $1.01 \pm 0.16$ & $1.01 \pm 0.16$ \\
  Reactor flux & $0.994 \pm 0.043$ & -- & $0.972 \pm 0.041$ & $0.972 \pm 0.041$ \\
  $(\alpha,n)$, ground state & $1.17 \pm 0.29$ & -- & $1.01 \pm 0.28$ & $1.00 \pm 0.28$ \\
  $(\alpha,n)$, excited states & $0.49 \pm 0.36$ & -- & $0.44 \pm 0.35$ & $0.43 \pm 0.36$ \\
  $\alpha$-$p$ coincidences & $0.53^{+0.79}_{-0.53}$ & -- & $0.22^{+0.80}_{-0.22}$ & $0.20^{+0.79}_{-0.20}$ \\
  Atmospheric neutrinos & $1.30 \pm 0.63$ & -- & $1.24 \pm 0.63$ & $1.23 \pm 0.63$ \\
  Energy scale (set 1) & $0.03 \pm 0.97$ & -- & $0.03 \pm 0.97$ & $0.03 \pm 0.97$ \\
  Energy scale (set 2) & $0.10 \pm 0.95$ & -- & $0.14 \pm 0.95$ & $0.15 \pm 0.95$ \\
  Birks constant (set 1) & $-0.18 \pm 0.93$ & -- & $-0.20 \pm 0.94$ & $-0.20 \pm 0.94$ \\
  Birks constant (set 2) & $-0.05 \pm 0.88$ & -- & $-0.13 \pm 0.89$ & $-0.14 \pm 0.90$ \\
  Proton energy scale (set 1) & $1.09 \pm 0.15$ & -- & $1.08 \pm 0.18$ & $1.08 \pm 0.18$ \\
  Proton energy scale (set 2) & $0.75 \pm 0.25$ & -- & $0.90^{+0.58}_{-0.40}$ & $0.88^{+0.54}_{-0.38}$ \\
  Extra smearing (set 1) [MeV$^{1/2}$] & $0.000^{+0.052}$ & -- & $0.000^{+0.053}$ & $0.000^{+0.053}$ \\
  Extra smearing (set 2) [MeV$^{1/2}$] & $0.003^{+0.030}_{-0.003}$ & -- & $0.000^{+0.041}$ & $0.003^{+0.030}_{-0.003}$ \\
  IBD classifier efficiency (set 1) & $0.0 \pm 1.0$ & -- & $0.0 \pm 1.0$ & $0.0 \pm 1.0$ \\
  IBD classifier efficiency (set 2) & $0.04 \pm 0.99$ & -- & $-0.15 \pm 0.99$ & $-0.15 \pm 0.99$ \\
  $(\alpha,n)$ classifier factor (set 1) & $0.45_{-0.28}$ & -- & $0.43^{+0.02}_{-0.32}$ & $0.43^{+0.02}_{-0.33}$ \\
  $(\alpha,n)$ classifier factor (set 2) & $0.45_{-0.45}$ & -- & $0.00^{+0.45}$ & $0.00^{+0.45}$ \\
  \hline
  \multicolumn{5}{l}{\textit{JUNO}} \\
  Local crust & -- & $1.00 \pm 0.17$ & -- & $1.00 \pm 0.17$ \\
  Reactor flux & -- & $1.0010 \pm 0.0087$ & -- & $1.0000 \pm 0.0087$ \\
  $^9$Li and $^8$He & -- & $1.33 \pm 0.19$ & -- & $1.32 \pm 0.19$ \\
  Reactors beyond 450 km & -- & $1.007 \pm 0.098$ & -- & $1.003 \pm 0.098$ \\
  $^{214}$Bi-$^{214}$Po & -- & $1.43 \pm 0.58$ & -- & $1.39 \pm 0.59$ \\
  Other backgrounds & -- & $1.12 \pm 0.31$ & -- & $1.12 \pm 0.31$ \\
  Energy scale & -- & $0.09 \pm 0.58$ & -- & $0.16 \pm 0.58$ \\
  Detection rate & -- & $0.20 \pm 0.92$ & -- & $0.00 \pm 0.92$ \\
  Reactor spectrum, component 1 & -- & $0.27 \pm 0.88$ & -- & $0.29 \pm 0.88$ \\
  Reactor spectrum, component 2 & -- & $0.21 \pm 0.94$ & -- & $0.06 \pm 0.94$ \\
  $^9$Li and $^8$He, shape & -- & $0.76 \pm 0.88$ & -- & $0.79 \pm 0.88$ \\
  Geoneutrino rate & -- & $1.88 \pm 0.37$ & -- & $1.93 \pm 0.37$ \\
 \end{tabular}
 \end{ruledtabular}
\end{table*}


\begin{figure*}[t!]
 \centering
 \includegraphics[width=\textwidth]{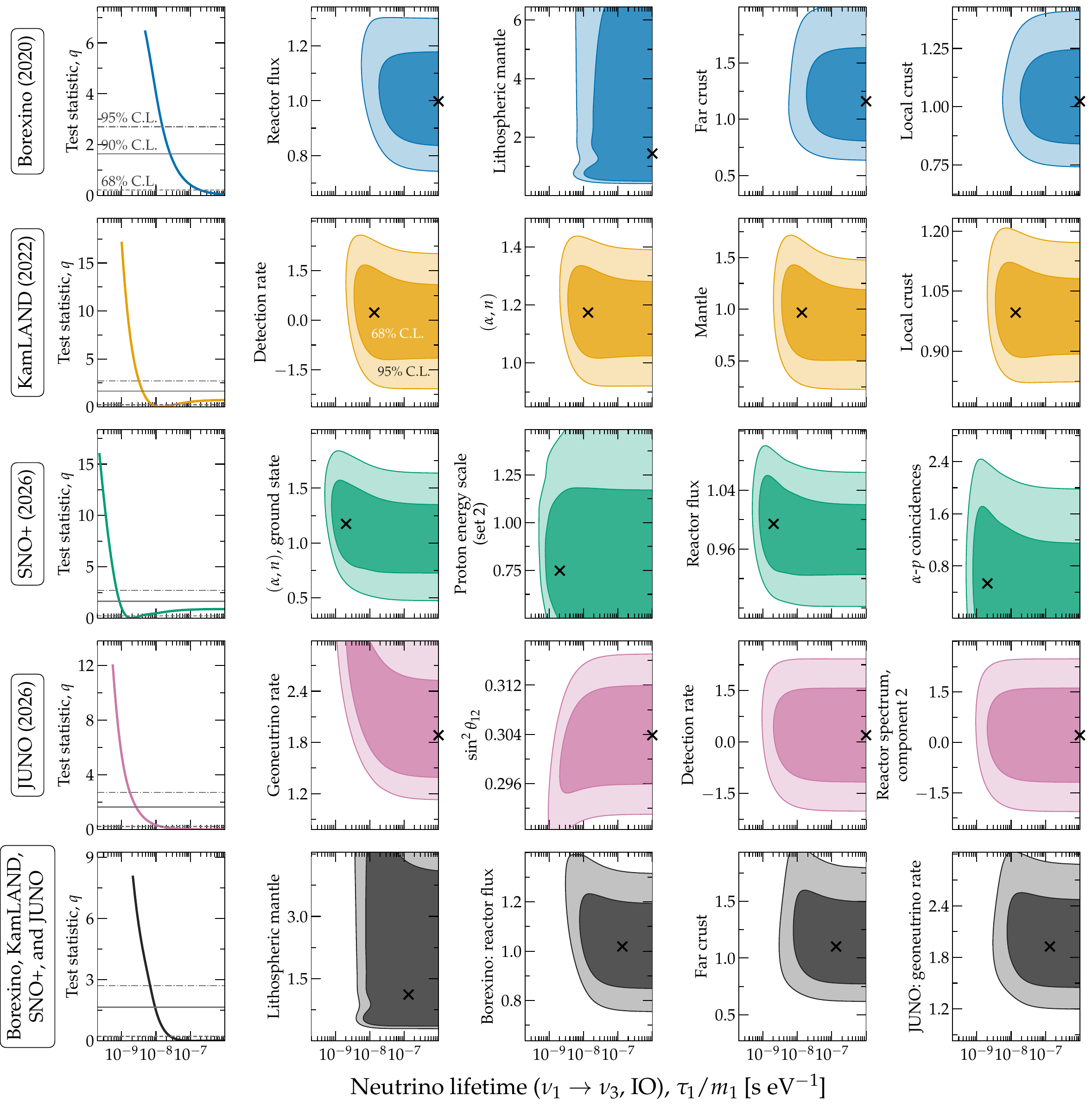}
 \caption{\textbf{How the lifetime trades off against the parameters that move most with it.} Each row is one fit to data, for $\nu_1 \to \nu_3$ in the inverted ordering on the geochemical Earth. \textit{Left:} the test statistic of $\tau_1/m_1$, with the levels at which its lower limits are read at 68\%, 90\%, and 95\%~C.L. \textit{Right:} the joint 68\% and 95\%~C.L. regions of $\tau_1/m_1$ and each of the four parameters that move most with it among those the data constrain, with every other parameter fitted. The cross marks the best fit; where the data prefer no decay, it sits on the right edge. Normalizations and the proton energy scales are relative to their nominal values; the energy scales, the detection rates, and the second component of JUNO's reference reactor spectrum are in units of their prior widths.}
 \label{fig:corner_rows}
\end{figure*}

\smallskip
\textit{Correlations.}---Figure~\ref{fig:corner_rows} shows, in its left column, the test statistic of $\tau_1/m_1$ in each fit. It rises steeply toward short lifetimes for Borexino and for the combination, but slowly for SNO+ and JUNO. Where it rises slowly, a small shift of a nuisance parameter moves the limit far. Therefore, single parameters move these two limits the most (\figu{impacts_experiments}). KamLAND's test statistic is lowest at $\tau_1/m_1 \approx 10^{-8}$~s~eV$^{-1}$, where its data favor decay, so its regions close at both ends. Where a region stays level, the parameter does not compete with decay.

\FloatBarrier
\section{Published limits on visible decay}
\label{app:published}

\tabl{published} lists the published limits that \figu{money} shows, with the statistics behind each. Three of them, from accelerator neutrinos, KamLAND's reactor antineutrinos, and SN1987A, are read at a test statistic of 2.71 for one parameter, the threshold of our 95\%~C.L. limits. The other four rest on other statistics: the limit from solar neutrino spectra on a region in two parameters, the one from IceCube on a Bayesian credible interval, and the two from solar antineutrinos on an upper limit on the number of detected events. For these two, their authors require that the antineutrinos produced by decay stay below the upper limit, which bounds the lifetime. Therefore, \figu{money} compares orders of magnitude, while Sec.~\ref{sec:comparison} compares limits read at the same threshold.

\begin{table*}[t!]
 \caption{\label{tab:published}\textbf{Statistics of the limits in \figu{money}.} For each limit, the data, the decay channel, how the limit is set, and the threshold it is read at. See Sec.~\ref{sec:comparison} for details.}
 \centering
 \begin{ruledtabular}
 \begin{tabular}{lllll}
  Data & Channel & Statistics & Threshold & Ref. \\
  \hline
  Accelerators: T2K, MINOS & $\nu_3 \to \nu_1$ (NO) & Frequentist, one parameter & $\Delta\chi^2 = 2.71$ & \cite{Gago:2017zzy} \\
  Reactors: KamLAND & $\nu_2, \nu_1 \to \nu_3$ (IO) & Frequentist, one parameter & $\Delta\chi^2 = 2.71$ & \cite{Porto-Silva:2020gma} \\
  Sun, $\bar\nu_e$ search: KamLAND & $\nu_2 \to \bar\nu_1$ (Majorana) & Event upper limit (Feldman--Cousins) & 90\%~C.L. & \cite{KamLAND:2003gfh} \\
  Sun, $\bar\nu_e$ spectra & $\nu_3 \to \bar\nu_1$ (NO, Majorana) & Event upper limit & 90\%~C.L. & \cite{Picoreti:2021yct} \\
  Sun, $\nu$ spectra & $\nu_2 \to \nu_1$ (Dirac) & Frequentist, two parameters & $\Delta\chi^2 = 6.18$ ($2\sigma$) & \cite{deGouvea:2023jxn} \\
  SN1987A & $\nu_1, \nu_2 \to \nu_3$ (IO), one lifetime & Frequentist, one parameter & $\Delta\chi^2 = 2.71$ & \cite{Ivanez-Ballesteros:2023lqa} \\
  High-energy astro.: IceCube & $\nu_1, \nu_2 \to \nu_3$ (IO) & Bayesian, marginalized & 90\% credible & \cite{Bustamante:2020niz} \\
  \hline
  Geoneutrinos (this work) & Every channel (\tabl{channels}) & Frequentist, one parameter, one-sided & $\Delta\chi^2 = 1.64$ (90\%) & --- \\
 \end{tabular}
 \end{ruledtabular}
\end{table*}

\section{Constraint on solar mixing parameters}
\label{app:nufit}

\begin{figure}[t!]
 \centering
 \includegraphics[width=\columnwidth]{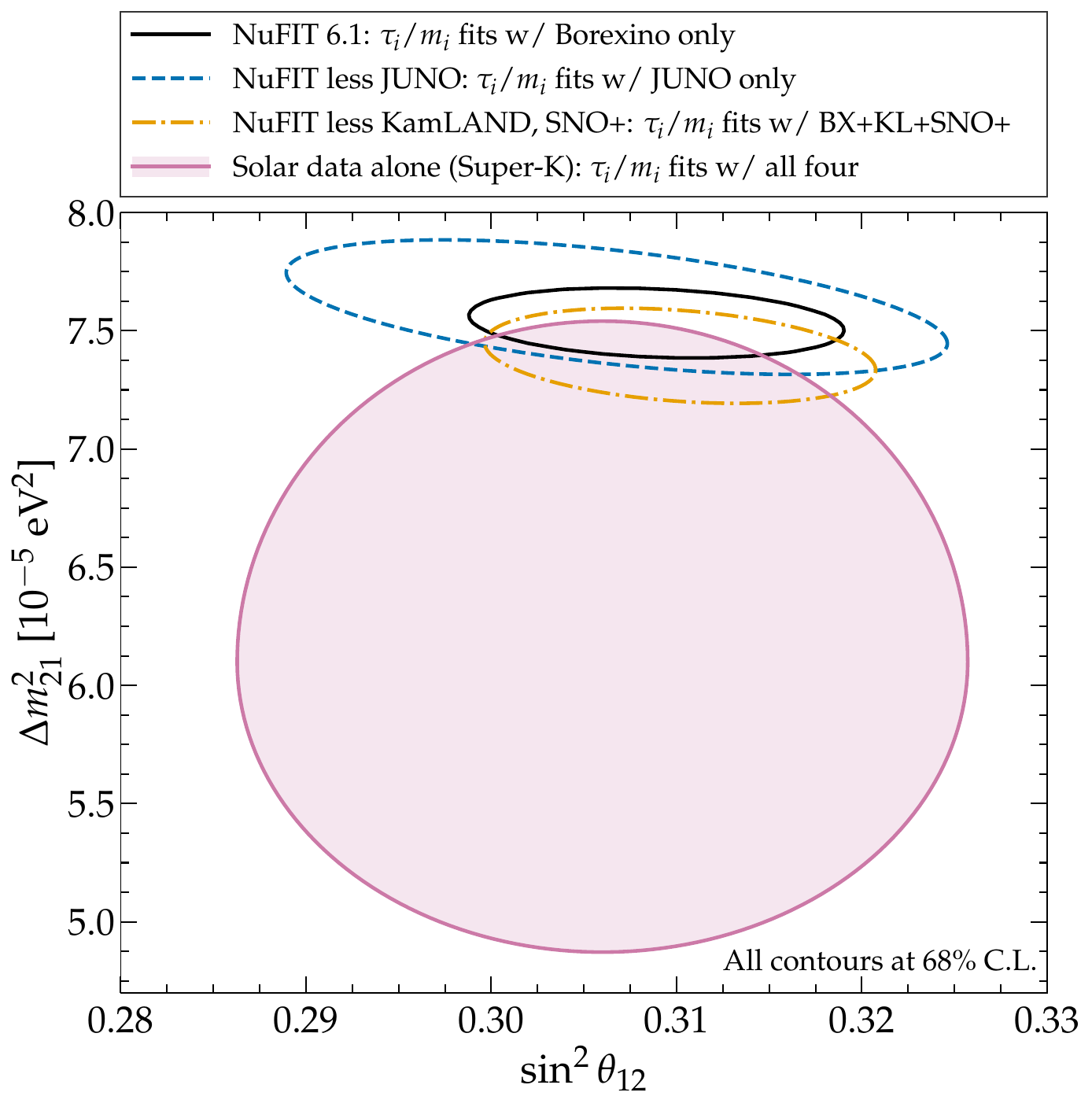}
 \caption{\label{fig:nufit}\textbf{Constraint on $\theta_{12}$ and $\Delta m^2_{21}$ that each fit uses.} The constraints are for the inverted ordering. The fits to KamLAND alone and to SNO+ alone use constraints close to NuFIT's, not shown. The solar data alone allow a far wider range of $\Delta m^2_{21}$, centered lower than the reactor data.}
\end{figure}

How much of the signal the decay of $\nu_1$ removes depends on the electron-flavor content of $\nu_1$, $|U_{e1}|^2 = \cos^2\theta_{12}\cos^2\theta_{13}$ (Sec.~\ref{sec:channels}). A fit of the neutrino lifetime with $\theta_{12}$ free could make up for the events that decay removes by moving $\theta_{12}$. In our tests, such fits set almost no limit on the lifetime, except at JUNO, whose reactor antineutrinos measure $\theta_{12}$ themselves. Therefore, we constrain $\theta_{12}$ with NuFIT~6.1~\cite{Esteban:2024eli, NuFIT61}, whose table gives it jointly with $\Delta m^2_{21}$. (NuFIT fixes $\cos^2\theta_{13}$ to 0.06\%, against 1\% for $\cos^2\theta_{12}$, and takes $\theta_{13}$ mostly from short-baseline reactor experiments absent from our fits.)

NuFIT obtains this constraint from the solar neutrino data together with the reactor antineutrinos of KamLAND, SNO+, and JUNO~\cite{NuFIT61}. Yet our fits to these three experiments contain the same reactor antineutrinos (Sec.~\ref{sec:reactors}). Used as the prior of such a fit, NuFIT's constraint would count them a second time, constraining $\theta_{12}$ more tightly than the data allow. With less room to move $\theta_{12}$, the fit could make up for less of what decay removes, so its limits could come out too strong. For JUNO, they would: with NuFIT's constraint as published, its limits at 90\%~C.L. would be 1.3 to 1.4 times stronger. The limit of all four from data would be 1.2 times stronger. The other fits change by at most 6.2\%, in either direction. Borexino needs no such correction: NuFIT contains its solar neutrinos, not its antineutrinos, so the fit to Borexino keeps NuFIT's constraint as published.

Figure~\ref{fig:nufit} shows the constraint that each fit uses. To avoid the double counting, each fit needs NuFIT's constraint without the reactor antineutrinos of the experiments that it contains. The information removed is not lost: each fit recovers it from its own reactor spectra. However, NuFIT does not tabulate such constraints~\cite{NuFIT61}, so we derive them from its global constraint ourselves.

Near its minimum, NuFIT's $\Delta\chi^2$ is close to a quadratic in $\sin^2\theta_{12}$ and $\log_{10}\Delta m^2_{21}$: the quadratic reproduces its $1\sigma$ widths to within 0.5\%. NuFIT's $\Delta\chi^2$ is approximately the sum of those of the data sets it combines. To remove an experiment, we subtract that experiment's own $\Delta\chi^2$, approximated by a Gaussian built from its published measurement~\cite{KamLAND:2013rgu, SNO:2025chx, JUNO:2025gmd}. A test supports this: removing JUNO from NuFIT~6.1 gives widths within 3\% of those of NuFIT~6.0, which predates JUNO's data~\cite{Esteban:2024eli}. For the fit of Borexino, KamLAND, and SNO+, we subtract the measurements of KamLAND and SNO+ together. (Our KamLAND lifetime fit uses only its spectrum below 2.7~MeV, so only part of KamLAND's measurement is counted twice. Subtracting all of it anyway moves our KamLAND limits by at most 0.2\%.)

For the fit of all four experiments, we would have to subtract KamLAND, SNO+, and JUNO together, leaving the solar data alone. But in $\Delta m^2_{21}$, the three published measurements together have an uncertainty of 1.2\%, smaller than the 1.3\% of NuFIT's global fit, which includes them and the solar data. Subtracting them would remove more than the global fit contains, leaving no valid constraint. Therefore, for the fit of all four, we use Super-Kamiokande's fit to all solar neutrino data: $\sin^2\theta_{12} = 0.306 \pm 0.013$ and $\Delta m^2_{21} = (6.10^{+0.95}_{-0.81})\times10^{-5}$~eV$^2$~\cite{Super-Kamiokande:2023jbt}. We take it as uncorrelated Gaussians, since no correlation is published.

\FloatBarrier
%

\end{document}